\documentclass[12pt,british]{article}
\usepackage[pdfstartview=FitH,colorlinks=true,linkcolor=black,anchorcolor=black,citecolor=black,urlcolor=black]{hyperref}
\usepackage{amsthm}
\usepackage{graphicx}
\usepackage{amsmath}
\usepackage{amssymb}
\usepackage{amsfonts}
\usepackage{appendix}
\usepackage{slashed}
\usepackage{afterpage}
\usepackage{cite}
\usepackage{pdfsync,hyperref}
\usepackage{amsthm,graphicx,amsmath,amssymb,amsfonts}
\usepackage{babel}
\usepackage{setspace}
\usepackage{textcomp}
\usepackage{esint}         
\usepackage{geometry}
\usepackage{breakurl}
\usepackage{bbold}
\usepackage{titling,authblk}
\newcommand{\M}{\mathcal M}
\newcommand{\A}{\mathcal A}
\newcommand{\Ds}{\slashed D}
\newcommand{\HH}{\mathcal H}
\newcommand{\sig}{{\boldsymbol{\sigma}}}
\newcommand{\ds}{\slashed \partial}
\newcommand{\sI}{\colorind I}
\newcommand{\sC}{\mathsf C}
\newcommand{\sD}{\mathsf D}
\newcommand{\sJ}{\colorind J}

\newcommand{\cinf}{{C^\infty({\cal M})}}
\newcommand{\one}{{\mathbb 1}}
\newcommand{\XX}{{\mathbb X}}
\newcommand{\I}{\mathbb 1}

\font\mybb=msbm10 at 12pt
\def\bb#1{\hbox{\mybb#1}}

\def\nn{\nonumber}
\newcommand{\tr}[1]{\:{\rm tr}\,#1}
\newcommand{\Tr}[1]{\:{\rm Tr}\,#1}

\def\be{\begin{equation}}
\def\ee{\end{equation}}
\def\bea{\begin{eqnarray}}
\def\eea{\end{eqnarray}}

\newcommand{\del}{\partial}

\newcommand{\ii}{{\mathrm i}}
\newcommand{\dd}{{\mathrm d}}

\newcommand{\spinind}[1]{\mathit{#1}}
\newcommand{\dotspinind}[1]{\mathit{\dot #1}}
\newcommand{\colorind}[1]{{\mathrm{#1}}}

\newcommand{\flavind}[1]{{\mathbf #1}}
\newcommand{\partind}[1]{{\sf #1}}

\numberwithin{equation}{section}
\numberwithin{figure}{section}

\theoremstyle{plain}

\newtheorem{prop}{Proposition}[section]

\newtheorem{df}[prop]{Definition}
\newtheorem{rem}[prop]{Remark}
  \newtheorem{example}[prop]{Example}

\usepackage{amsthm,graphicx,amsmath,amssymb,amsfonts,framed}
\usepackage{slashed}
\usepackage{afterpage}
\usepackage{cite}
\usepackage{color}
\usepackage{hyperref}

\definecolor{verde}{cmyk}{.83,.21,1,.08}

\def\alg{{\cal A}}

\font\mybb=msbm10 at 12pt
\def\bb#1{\hbox{\mybb#1}}

\def\nn{\nonumber}

\def\be{\begin{equation}}
\def\ee{\end{equation}}
\def\bea{\begin{eqnarray}}
\def\eea{\end{eqnarray}}

\newcommand{\la}{\label}
\newcommand{\ba}{\begin{eqnarray}}
\newcommand{\ea}{\end{eqnarray}}

\begin{document}

\begin{titlepage}

%
%
%
%
%
%
%
%
%

\title{ Spectral Noncommutative Geometry, Standard Model and all that\vspace{5pt}}

\renewcommand\Affilfont{\itshape}
\setlength{\affilsep}{1.5em}
\renewcommand\Authands{ and }

\author[1]{Agostino Devastato\thanks{astinodevastato@gmail.com}}
\author[1,2]{Maxim Kurkov\thanks{Max.Kurkov@na.infn.it}}
\author[1,2,3]{Fedele Lizzi\thanks{fedele.lizzi@na.infn.it}}

\affil[1]{INFN, Sezione di Napoli, Italy\vspace{5pt}}
\affil[2]{Dipartimento di Fisica ``Ettore Pancini'', Universit\`{a} di Napoli {\sl Federico~II}\vspace{5pt}, Napoli, Italy}
\affil[3]{Departament de F\'{\i}sica Qu\`antica i Astrof\'{\i}sica and Institut de C\'{\i}encies del Cosmos (ICCUB),
Universitat de Barcelona. Barcelona, Spain}

\date{}

\maketitle


\begin{abstract}
We review the approach to the standard model of particle interactions based on spectral noncommutative geometry. The paper is (nearly) self-contained and presents both the mathematical and phenomenological aspects. In particular the bosonic spectral action and the fermionic action are discussed in detail, and how they lead to phenomenology. We also discuss the Euclidean vs. Lorentz issues and how to go beyond the standard model in this framework.
\end{abstract}

Pacs {\sl 02.40 Gh, 11.10 Nx, 12-60.-i}.

\end{titlepage}

\tableofcontents

\newpage

\section{Introduction}
The aim of this review is to present the standard model of particle interaction as a particular geometry of spacetime, a \emph{noncommutative geometry}. The standard model is a very successful theory, and it seems to be working well beyond ``factory specifications''. The CERN measurement of the mass of the Higgs mass has been the crowing of several decades of confirmations of the model. Of course not everything is understood, and there are loose ends which may lead to significative advances. Yet little  is know about the \emph{origin} of it. Why among all possible Yang-Mills gauge theories nature has privileged a particular set of particles transforming under a particular representation? A full answer to these questions will of course not be given here, nevertheless we are convinced that a geometrical vision of the standard model may be a key to progress. There are other reasons to consider spacetime as a noncommutative geometry, mainly coming from quantum gravity, where the object to quantize is spacetime itself. 

The noncommutative geometry of the standard model presents a very mild version of noncommutativity. Spacetime itself is still an ordinary (possibly curved) manifold, but it is tensor multiplied times a noncommutative, but finite geometry: a matrix algebra. This geometric constructions correspond to the so called spectral triples, which we discuss in the present review.
Noncommutative geometry is then used to give a \emph{spectral} view of the geometry, its symmetries, and the action. There are several important features of this view. The first is that the Higgs particle is on the same footing of the other bosons, it represents the fluctuations of the internal geometry. It becomes another ``intermediate boson", except that its ``intermediate boson" nature refers to the internal space, and is therefore a scalar from the continuous spacetime point of view. The other important point is that not any Yang-Mills gauge theory can be described in this framework, on the contrary, if one makes the request that the internal finite dimensional space is the noncommutative equivalent of a manifold, then very few models survive, the standard model among them, or Pati-Salam, but none of the other Grand Unified Theories such as SU(5) or SO(10). 

The starting point of noncommutative geometry is based on the spectra of operators, and the action is also based on the spectral data of a generalised Dirac operator. Using the mass (or rather the Yukawa couplings) of known fermions as input in this generalised operator, it is possible to write the action of the standard model, in a curved gravitational background (which is not quantised). The action is composed of a scalar product and a trace properly regularised, and is given at a particular point in the renormalisation flow. In this point the three fundamental interaction have the same strength. It is known that such a point (in the absence of new physics) exists only approximatively, but this gives the possibility to calculate quantities such as the mass of the Higgs. In its bare formulation this number is not the correct one, but further refinements of the model make it compatible with experiment. 

We will start our review with a brief recap of the standard model coupled with gravity. This will help not only to set notations, but also to understand what we want to explain. The following section introduces noncommutative geometry from a spectral point of view, as pioneered by Connes. The two previous aspects come together in the fourth section, where the standard model as an almost commutative geometry is presented from a geometrical point of view. The connection with physics is further developed in the following two sections, dedicated to the description of the spectral action and the ``numbers'' that one can produce from these models. Sections 7 and 8 describe further developments which go beyond the standard model, while Section 9 discusses the issue of the Lorentzian vs. Euclidean signature of the model. A final section concludes the review and gives some pointers for further developments.

References to original work are given when the topic is introduced, but here we would like to refer to other, often complementary, reviews. One of us (FL) gave series of lectures on these topics at the Cimpa School \emph{Noncommutative Geometry and Applications to Quantum Physics} in Qui Nhon (Vietnam) in 2017, at the COST action training school in Corfu (2017) and at the Bose Research center in Kolkata (2018). A reduced and partial version of these lectures can be fount in~\cite{Corfu}. A complete reference, from a mathematician point of view is in a book by W.~Van Suijlekom~\cite{WalterBook}, see also his more recent review with Chamseddine~\cite{Chamseddine:2019fjq}.

We conclude this introduction with a little bit of history, without any pretence to be complete. 
The first works on the Higgs as a fluctuation of a noncommutative geometry are in the work of Dubois-Violette, Kerner and Madore~\cite{KernerDuboisMadore} and Connes and Lott~\cite{ConnesLott}. For an early discussion of the Higgs (before the spectral action) see also~\cite{Sitarz:1993zf}.
A lot of preparatory work was done by the Marseille group, notably Daniel Kastler, Bruno Iochum and Thomas Schucker and several young people, in a series of papers lasting over a decade. No review would be complete without mentioning their work. One of the latest review by T.~Schucker, which still has some fresh interest, is~\cite{Schucker}.  A more recent ``friendly'' overview is~\cite{Sitarz:2013yva}.

\section{The Standard Model and Gravity \label{se:standardmodel}} 
\setcounter{equation}{0}
Our understanding of fundamental interactions is based on the two pillars: the Standard Model of particle physics and the General Relativity.
The Standard Model (SM) is  a quantum field theory which addresses the electromagnetic,  weak and strong interactions of elementary particles. It had an impressive successes, \emph{predicting}, for example, the W, Z and Higgs bosons well before experiments demonstrated their existence.
General Relativity (GR) describes gravity in terms of differential geometry. It also is supported by a several experiments, starting from the perihelion precession of Mercury, up to modern astrophysical observations of binary pulsars dynamics.

Below we briefly review the action for these two theories. The aim of this microreview is twofold. On the one hand we set the notations, on the other we outline some relevant aspects of SM which are closely related to the spectral action principle. 
We use the ``recent version" of the Standard Model incorporating right handed neutrino. Then discuss some aspects
of the renormalization group flow. One of the aims of this review is to show \emph{how} the SM action can be obtained from the spectral Noncommutative Geometry (NCG), which treats gauge and gravitational degrees of freedom on an equal footing.  Therefore we introduce the SM action directly on a \emph{curved} gravitational background and after that discuss the gravitational action as a part of the bosonic action.

\subsection{The Classical Fermionic action of Standard Model\label{sub:SMclact}}
The fermions of the Standard Model are combined in multiplets according to their transformation properties upon the action 
of the gauge group $SU(3)\times SU(2) \times U(1)$. Note one important aspect: \emph{all the multiplets transform according to either  fundamental or  trivial representations of 
$SU(3)$ and $SU(2)$}. The action of the abelian subgroup $U(1)$ is defined by the hypercharge $Y$. These transformation properties are summarised in Tab.~\ref{Tab1}.
\begin{table}[htb]
\centering{}%
\begin{tabular}{cc|c|c|c|c|c|l}
\cline{2-7}
&  \multicolumn{1}{ |c| }{ $\boldsymbol{\mathrm{v}}_{\mathcal{R}}$  } & $ \boldsymbol{\mathrm{e}}_{\mathcal{R}}$ & $\boldsymbol{\mathrm{L}}_\mathcal{L}$ 
&  $\boldsymbol{\mathrm{u}}_{\mathcal{R}}$ & $\boldsymbol{\mathrm{d}}_{\mathcal{R}}$ & $\boldsymbol{\mathrm{Q}}_{\mathcal{L}}$\\ 
\cline{1-7}
\multicolumn{1}{ |c| }{SU(3)} & t & t & t & f & f & f &   \\ \cline{1-7}
\multicolumn{1}{ |c| }{SU(2)} & t & t & f & t &  t & f &   \\ \cline{1-7}
\multicolumn{1}{ |c| }{U(1)} & 0 & -2 & -1 & 4/3 & -2/3 & 1/3 &
\\ \cline{1-7}
\end{tabular}
\protect\caption{\sl Representations of the gauge subgroups on various fermionic multiplets of the Standard Model. For the nonabelian subgroups $SU(3)$ and $SU(2)$ the letters ``f" and ``t" indicate 
a fundamental and  
a trivial representations respectively. For the abelian subgroup $U(1)$ we specify the hypercharge $Y$.
}\label{Tab1}
\end{table}
Hereafter  $\boldsymbol{\mathrm{v}}$, $\boldsymbol{\mathrm{e}}$, $\boldsymbol{\mathrm{u}}$ and $\boldsymbol{\mathrm{d}}$ indicate neutrinos, electron-like leptons, upper  and down like quarks respectively. Moreover $\boldsymbol{\mathrm{Q}}_{\mathcal{L}}$ corresponds to the quark doublets $(\boldsymbol{\mathrm{u}}_{\mathcal{L}}, \boldsymbol{\mathrm{d}}_{\mathcal{L}})$ while $\boldsymbol{\mathrm{L}}_\mathcal{L} $ corresponds to the lepton doublets $(\boldsymbol{\mathrm{v}}_{\mathcal{L}}, \boldsymbol{\mathrm{e}}_{\mathcal{L}})$.
By boldface characters we indicate that the multiplets have to be replicated by three generations, for example 
$\boldsymbol{\mathrm{e}} = (e, \mu, \tau )$ and so on.  
The subscripts $\mathcal{L}$ and $\mathcal{R}$ indicate the \emph{chirality} of the fermions $\psi$ from the corresponding multiplets. By definition the left and the right handed fermions satisfy:
\be
\psi_{\mathcal{L}} = \frac{1}{2}\left(1-\gamma^5\right)\psi_{\mathcal{L}}, \quad\quad
\psi_{\mathcal{R}} = \frac{1}{2}\left(1+\gamma^5\right)\psi_{\mathcal{R}}. \label{chirdef}
\ee

The fermionic action of the Standard Model reads:
\begin{eqnarray} \label{fermionicactionusual}
 S^{\rm {M}}_F &=& \int d^4 x \sqrt{-g^{\rm M}} \left\{ \right.
 i  \overline{\left(\boldsymbol{\mathrm{u}}_{\mathcal R}\right)} \slashed{\nabla}_{\mathrm{M}}\boldsymbol{\mathrm{u}}_{\mathcal R}
+ i \overline{\left(\boldsymbol{\mathrm{d}}_{\mathcal R}\right)} \slashed{\nabla}_{\mathrm{M}} \boldsymbol{\mathrm{d}}_{\mathcal R}
+  i \overline{\left(\boldsymbol{\mathrm{Q}}_{\mathcal L}\right)} \slashed{\nabla}_{\mathrm{M}} \boldsymbol{\mathrm{Q}}_{\mathcal L} \nonumber\\
&+& i  \overline{\left(\boldsymbol{\mathrm{v}}_{\mathcal R}\right)} \slashed{\nabla}_{\mathrm{M}}\boldsymbol{\mathrm{v}}_{\mathcal R}
+ i \overline{\left(\boldsymbol{\mathrm{e}}_{\mathcal R}\right)} \slashed{\nabla}_{\mathrm{M}} \boldsymbol{\mathrm{e}}_{\mathcal R}
+  i \overline{\left(\boldsymbol{\mathrm{L}}_{\mathcal L}\right)} \slashed{\nabla}_{\mathrm{M}} \boldsymbol{\mathrm{L}}_{\mathcal L}  \nonumber \\
 &-& \left[ \overline{\left(\boldsymbol{\mathrm{Q}}_{\mathcal L}\right)} \left[\hat{y}^{\dagger}_u \otimes \tilde H\right] \boldsymbol{\mathrm{u}}_{\mathcal R}
+\overline{\left(\boldsymbol{\mathrm{Q}}_{\mathcal L}\right)}\left[\hat{y}_d^{\dagger} \otimes  H\right] \boldsymbol{\mathrm{d}}_{\mathcal R} + \right.
\nonumber\\
 &+& 
 \overline{\left(\boldsymbol{\mathrm{L}}_{\mathcal L}\right)}\left[\hat{Y}^{\dagger}_u \otimes \tilde H \right] \boldsymbol{\mathrm{v}}_{\mathcal R}
+ \overline{\left(\boldsymbol{\mathrm{L}}_{\mathcal L}\right)}\left[\hat{Y}^{\dagger}_d \otimes \ H\right] \boldsymbol{\mathrm{e}}_{\mathcal R} \nonumber\\
&+&\frac{1}{2} \overline{\left( C_{\rm M}\boldsymbol{\mathrm{v}}_{\mathcal R}\right)}\left[ \hat y_M^{\dagger}  \right] \mathrm{M}_R\boldsymbol{\mathrm{v}}_{\mathcal R}    
 \left.\left.  
+ \mbox{c.c.}\right]\right\}
 \la{SMink}.
\end{eqnarray}
Below we clarify the meaning of various quantities which enter in this formula, considering inputs of various nature separately.
Note that this structure, in addition to the fermions, contains \emph{all} other dynamical degrees of freedom: the vector gauge bosons, the scalar bosons and gravity  which appears via the vierbeins. In some sense this observation gives a heuristic basis for the spectral action principle, which we discuss  in Sec.~\ref{se:SpectralAction}.

\subsubsection*{Geometric input and spin structures.} 
The label ``${\rm {M}}$" in~\eqref{SMink} indicates the fact that we are dealing with the Minkowskian theory, i.e.\ the metric tensor $g^{\mathrm{M}}_{\mu\nu}$ has signature $(+,-,-,-)$.
As we will see in Sec.3, in the NCG spectral approach an important role is played by the Euclidean structures, which we will label by ``$\mathrm{E}$".
The quantity $g^{\rm M}$ is the determinant of the metric tensor. 
The expression $\slashed\nabla_{\mathrm{M}}$ is a contraction over the world index\footnote{We use $\mu,\nu,\rho,  ...$ as world indices and $a,b,c,...$ as flat (tangential) indices.}  of the $\gamma^{\mu}_{\mathrm{M}}$ and the covariant derivatives $\nabla_{\mu}^{\mathrm{M}}$.
These can be obtained from the \emph{flat} gamma matrices  $\gamma^{a}_{\mathrm{M}}$, which satisfy
\be
\left\{\gamma^a_{\mathrm{M}},\gamma^b_{\mathrm{M}}\right\} = \eta^{ab} \mathbb{1}_4, \label{gammaflatac}
\ee
 contracting them with the vierbeins $e_{a}^{\mu}$ over the flat (tangential) index $a$:
\be
 \gamma^{\mu}_{\mathrm{M}} = e^{\mu}_{a} \gamma^a_{\mathrm{M}}.
\ee
The quantities 
\be
\eta^{ab} = \mathrm{diag}(+1,-1,-1,-1) = \eta_{ab}
\ee
and $\mathbb{1}_4$, which enter in~\eqref{gammaflatac}, stand for the flat Minkowskian metric and for the unit four by four matrix in spinorial indexes respectively. The vierbeins 
satisfy
\be
e^{\mu}_{a} e^{\nu}_b \,g^{\mathrm{M}}_{\mu\nu}  = \eta_{ab}.
\ee
The symbol bar in~\eqref{SMink} indicates the Dirac conjugation of spinors according to the following rule
\be
\bar\psi = \psi^{\dagger}\gamma^{0}_{\mathrm{M}},
\ee
where $\dagger$ stands for the Hermitian conjugation and $\gamma^{0}_{\mathrm{M}}$ is the 0-th \emph{flat} Dirac matrix.
The covariant derivative $\nabla_{\mu}^{\mathrm{M}}$ has the structure
\be
\nabla_{\mu}^{\mathrm{M}} = \partial_{\mu} -  \ii \mathcal{A}_{\mu}  - \frac{\ii}{2} \omega_{\mu}^{\mathrm{M}}, \label{cdgen}
\ee
where $ \mathcal{A}_{\mu}$ stands for the gauge connection, which acts on various fermionic multiplets in a different way, 
and which we discuss in details below.
The spin-connection 
\be
\omega_{\mu}^{\mathrm{M}} = \left[\omega_{\mu}^{ab}\right]^{\mathrm{M}}\sigma^{\mathrm{M}}_{ab}, \la{MinkSpinConnection}
\ee
is  common for all fermionic multiplets.
In this formula
\be
\left[\omega_{\mu}^{ab}\right]^{\mathrm{M}}= e^a_{\nu}\, g_{\mathrm{M}}^{\nu\xi}\,\partial_{\mu} e^b_{\xi}
+ e^a_{\nu} \left[\Gamma^{\nu}_{\mu\xi}\right]^{\mathrm{M}}e^{b}_{\rho} \,g_{\mathrm{M}}^{\rho\xi},
\label{LCsDetails}
\ee
and
\be
\sigma^{\mathrm{M}}_{ab} = \frac{\ii \left[\gamma^c_{\mathrm{M}}, \gamma^d_{\mathrm{M}}\right]}{4}\, \eta_{ac} \eta_{bd},
\ee
stand for generators of the defining representation of $Spin(1,3)$. These four by four matrices act on the spinorial index of various multiplets.
 In~ \eqref{LCsDetails} the quantities 
$\left[\Gamma^{\nu}_{\mu\xi}\right]^{\mathrm{M}}$ are the Christoffel symbols of the second kind, which are computed with the Minkowskian metric
$g_{\mu\nu}^{\mathrm{M}}$. For a  generic metric tensor $g_{\mu\nu}$ these read
\begin{equation}
\Gamma_{\nu\rho}^{\mu}\left[g_{\mu\nu}\right]\equiv\frac{1}{2}\,g^{\mu\lambda}\left(\partial_{\rho}g_{\lambda\nu} + \partial_{\nu}g_{\lambda\rho} - \partial_{\lambda}g_{\nu\rho}\right). \label{Christ}
\end{equation}

In~\eqref{LCsDetails} and~\eqref{Christ} we used the metric tensor with two upper indexes and the vierbeins with lower world and upper flat indexes. For any given metric tensor $g_{\mu\nu}$  and a vierbein field $e_a^{\mu}$ seen as square matrices in their indexes these objects are nothing but the corresponding inverse matrices:
\be
g_{\mu\nu} g^{\nu\lambda} = \delta_{\mu}^{\lambda},\quad e_{\mu}^a e^{\mu}_b = \delta^{a}_b,\quad e_{\mu}^{a} e^{\nu}_a = \delta_{\mu}^{\nu}.
\ee
The spin connection renders the local ``pseudo-rotational" invariance of~ \eqref{SMink} upon simultaneous local $SO(1,3)$ pseudo-rotations of  vierbeins and the corresponding $Spin(1,3)$ transformations of  fermionic fields. 
In conclusion we notice  that the fermionic action, being a scalar, remains invariant upon local diffeomorhisms. We remind that  upon these transformations  fermionic fields transform  as \emph{scalars}.

\subsubsection*{Gauge input.} 
Let us take a closer look at the gauge connection $\mathcal{A}$, which enters in the covariant derivative~\eqref{cdgen}.
For various fermionic multiplets the gauge connection reads:
\begin{eqnarray}
\mathcal{A}^{\boldsymbol{\mathrm{v}}_{\mathcal{R}}}_{\mu} &=& g_1\,Y_{\boldsymbol{\mathrm{v}}_{\mathcal{R}}}\, B_{\mu}, \nonumber\\
\mathcal{A}^{\boldsymbol{\mathrm{e}}_{\mathcal{R}}}_{\mu} &=&  g_1\,Y_{\boldsymbol{\mathrm{e}}_{\mathcal{R}}}\, B_{\mu}  ,\nonumber\\
\mathcal{A}^{\boldsymbol{\mathrm{L}}_\mathcal{L}}_{\mu} &=&  g_1\,Y_{\boldsymbol{\mathrm{L}}_\mathcal{L}}\, B_{\mu}  +
g_2\, W_{\mu},\nonumber\\
\mathcal{A}^{\boldsymbol{\mathrm{u}}_{\mathcal{R}}}_{\mu} &=& g_1\,Y_{\boldsymbol{\mathrm{u}}_{\mathcal{R}}}\, B_{\mu}  
 + g_3\, G_{\mu}
,\nonumber\\
\mathcal{A}^{\boldsymbol{\mathrm{d}}_{\mathcal{R}}}_{\mu} &=& g_1\,Y_{\boldsymbol{\mathrm{d}}_{\mathcal{R}}}\, B_{\mu}  
 + g_3\, G_{\mu}  ,\nonumber\\
\mathcal{A}^{\boldsymbol{\mathrm{Q}}_\mathcal{L}}_{\mu} &=& g_1\,Y_{\boldsymbol{\mathrm{Q}}_\mathcal{L}}\, B_{\mu}  
 + g_2\, W_{\mu} + g_3\, G_{\mu},   \la{gaugeconnectdef}
\end{eqnarray}
where the superscripts indicate the corresponding multiplets. 
The quantities $g_{1}$, $g_{2}$ and $g_{3}$ stand for the $U(1)$, $SU(2)$ and $SU(3)$ gauge coupling constants respectively, $B_{\mu}$ is a real vector field, and
the abelian  hypercharges $Y$ for various multiplets are collected in Tab.~\ref{Tab1}. 
The matrix valued fields $W_{\mu}$ and $G_{\mu}$ are built from the real vector fields $W_{\mu}^{\alpha}$, $\alpha = 1,2,3$ and $G_{\mu}^{j}$,  $j = 1,\ldots,8$ according to the formulas
\bea
W_{\mu} &=& 
\frac{\sigma_{\alpha}}{2}\cdot W_{\mu}^{\alpha}, \nonumber\\
G_{\mu} &=& \frac{\lambda_{j}}{2} \cdot G_{\mu}^{j},
\eea
act on the Weak isospin index of left doublets and on the colour index of quark triplets.  The 2 by 2 matrices $\sigma_{\alpha}$ are Pauli matrices, 
and $\lambda_i$ stand for the 3 by 3 Gell-Mann matrices.

A presence of the gauge fields in the covariant derivatives guarantees a local $SU(3)\times SU(2) \times U(1)$ gauge invariance of the kinetic terms for fermions, which are given by the first two lines of~\eqref{SMink}.


\subsubsection{Scalars the Dirac mass terms.}
The third and the fourth lines of~ \eqref{SMink} define the Dirac mass terms. 
The quantities $\hat{Y}_{u}$, $\hat{Y}_{d}$, $\hat{y}_{u}$, $\hat{y}_{d}$ are arbitrary (dimensionless) complex 3 by 3 Yukawa matrices which act on the generation indices of fermionic multiplets.

According to our conventions the quantity 
\be
H = \left(\begin{array}{c} H^{\mathrm{up}} \\ H^{\mathrm{down}} \end{array}\right)
\ee
is a weak isospin doublet of scalar fields, which transforms according to the fundamental representation of the $SU(2)$ gauge subgroup and has the hypercharge $Y_H = 1$.
By definition the ``charged conjugated" scalar doublet equals to 
\be
\tilde{H} = \ii\sigma_2 H^*, \label{tHiggsdef}
\ee 
where * stands for the complex conjugation and  $\sigma_2$ is the second Pauli matrix. One can easily see 
that $\tilde{H}$ transforms as $H$ under $SU(2)$ transformations, however it has opposite hypercharge: 
$Y_{\tilde{H}} = -1$.  These transformation properties ensure gauge invariance of the Dirac mass terms.
We emphasise that $H$ and $\tilde{H}$ are contracted with fermionic multiplets over the $SU(2)$-representation indexes. 
To emphasize the fact that the Yukawa matrices act on different indices we have written in~\eqref{SMink} explicitly the tensor product $\otimes$, which is usually omitted. 

\subsubsection*{Charge conjugation and the Majorana mass terms.}
The  Majorana mass terms, which are given by the last line of~\eqref{SMink}. 
The quantity $ \hat{y}_{M}$ is a 3 by 3 complex matrix acting
 on a generation index,   $\mathrm{M}_R$ is the dimensionful constant, which  sets the Majorana mass scale for the right handed neutrinos, and the operation
\be
C_{\mathrm M}  = -\ii \gamma^2_{\mathrm{M}}\,\circ\, \mbox{complex conjugation}, \label{CMdef}
\ee
is the Minkowskian charge conjugation. 

The charged conjugated  spinor multiplet  and the original one transform in the same way upon the
the $Spin(1,3)$ transformations, but upon  local gauge transformations and  global $U(1)$  transformations (e.g.\ the lepton and baryon number accidental symmetries)  the charged conjugated multiplet transforms according to the complex conjugated representation. 
In particular the Majorana mass terms explicitly violate the lepton number conservation. 

\subsection{Bosonic action}
The bosonic action of SM bosons together with gravity is given by 
\be
S_{B}^{\mathrm{M}} = \int d^4 x\, \sqrt{-g^{\mathrm{M}}} \left(L_{\rm g} + L_{\rm s} + L_{\rm gr} \right),
\ee
where $L_{\rm g}$,  $L_{\rm s}$ and $L_{\rm gr}$ stand for the gauge, scalar and the gravitational Lagrangians respectively.
The gauge Lagrangian is a sum of the Yang-Mills Lagrangians for the nonabelian vector fields $G_{\mu}$ and $W_{\mu}$ and the Maxwell Lagrangian
for the abelian vector field $B_{\mu}$:
\begin{equation}
L_{\rm g}  =  -\frac{1}{4}G_{\mu\nu}^j G^{\mu\nu\,j} - \frac{1}{4} W_{\mu\nu}^{\alpha} W^{\mu\nu\,\alpha}
- \frac{1}{4}B_{\mu\nu} B^{\mu\nu}.  \la{GaugeLagr}
\end{equation}
here the quantities $G_{\mu\nu}^j$, $W^{\mu\nu\,\alpha}$ and $B_{\mu\nu}$ stand for the field-strength tensors associated
with the corresponding gauge connections:
\bea
G_{\mu\nu}^{i} &=& \partial_{\mu}G^i_{\nu} - \partial_{\nu}G^i_{\mu} + g_3 f^{i j k} G_{\mu}^{j} G_{\nu}^k, \nonumber\\
W_{\mu\nu}^{\alpha} &=& \partial_{\mu}W^{\alpha}_{\nu} - \partial_{\nu}W^{\alpha}_{\mu} + g_2 f^{\alpha \beta \gamma} W_{\mu}^{\beta} W_{\nu}^{\gamma} ,\nonumber\\
B_{\mu\nu} &=& \partial_{\mu}B_{\nu} - \partial_{\nu}B_{\mu}. \la{gaugestrength}
\eea
The Lagrangian for the Higgs scalar $H$, which is minimally coupled to gravity, reads: 
\begin{equation}
L_{\rm s} =  g^{\mu\nu}_{\mathrm{M}}\nabla_{\mu} H^{\dagger} \nabla_{\nu}H  - V(H). \la{ScLagr}
\end{equation}
According to the transformation properties of the Higgs doublet upon the action of the gauge group, which we discussed above,
 the covariant derivative, which appears in the kinetic terms, involves the $SU(2)$ and the $U(1)$ gauge connections:
\be
\nabla_{\mu} H = \left(\partial_{\mu}  - \ii g_1\, Y_H\,B_{\mu} - \ii g_2 W_{\mu}\right) H.
\ee
The potential term $V(H)$ is the famous Mexican hat potential, which plays a crucial role in the Higgs mechanism of a  spontaneous symmetry breaking: 
\be
V(H) =  - \mathfrak{m}^2 H^{\dagger} H +  \lambda\left(H^{\dagger} H\right)^2. \la{ScPot}
\ee
In this formula $\lambda$ is a positive dimensionless quartic coupling constant. The quantity $\mathfrak{m}$ is a real constant of the dimension of a mass - the so called ``mass parameter" (not to be confused with the actual mass!). 
This potential has  a family of nontrivial minima $\left(H^{\dagger}H \right)_0^2$, which are usually parametrised by the constant
\be
v = \sqrt{\left(H^{\dagger}H \right)_0^2}, \label{vevdef}
\ee
which is called the ``Higgs vacuum expectation value". Its experimental value is 246 GeV.

Let us fix the notations for the Riemann tensor, Ricci tensor and the scalar curvature for generic metric tensor $g_{\mu\nu}$ 
\bea
R^{\mu}_{~\nu\rho\sigma}\left[g_{\mu\nu}\right] &=& \partial_{\sigma}\Gamma^{\mu}_{\nu\rho} - \partial_{\rho}\Gamma^{\mu}_{\nu\sigma}
+ \Gamma^{\lambda}_{\nu\rho}\Gamma^{\mu}_{\lambda\sigma} - \Gamma^{\lambda}_{\nu\sigma}\Gamma^{\mu}_{\lambda\rho}. \nonumber\\ 
R_{\mu\nu}\left[g_{\mu\nu}\right] = R^{\sigma}_{~\mu\sigma\nu} &=&   \partial_{\nu}\Gamma^{\sigma}_{\mu\sigma} -\partial_{\sigma}\Gamma^{\sigma}_{\mu\nu}+
\Gamma^{\lambda}_{\mu\sigma}\Gamma^{\sigma}_{\lambda\nu}
- \Gamma^{\lambda}_{\mu\nu}\Gamma^{\sigma}_{\lambda\sigma}. \nonumber\\  
R\left[g_{\mu\nu}\right] &=& g^{\mu\nu}\left\{  \partial_{\nu}\Gamma^{\sigma}_{\mu\sigma} - \partial_{\sigma}\Gamma^{\sigma}_{\mu\nu}
 + \Gamma^{\lambda}_{\mu\sigma}\Gamma^{\sigma}_{\lambda\nu}- \Gamma^{\lambda}_{\mu\nu}\Gamma^{\sigma}_{\lambda\sigma} \right\}. \label{RS}
\eea
Now we introduce the  
the gravitational Einstein-Hilbert action:
\be
L_{\rm gr} = -\lambda_{\mathrm{c}} \, + \, \frac{M_{\rm Pl}^2}{16\pi}R\left[g_{\mu\nu}^{\rm M}\right].  \la{GraLagr}
\ee
The quantity $\lambda_{\mathrm{c}}$ is a real parameter of the dimension $[\mathrm{mass}^{4}]$,  which defines a cosmological term.
The quantity $M_{\rm Pl}$ is a  parameter of the dimension of a mass, whose experimentally observed value is of the order of $10^{19}$ GeV.

\subsection{Renormalization group flow: relevant aspects}
\label{RGF_section}
Upon quantisation various couplings of the Standard Model exhibit a dependence on the energy scale  - the normalisation point $\mu$.
Such a dependence is often called the renormalisation group (RG) flow or the RG running, and  it is 
described mathematically by the RG equations:  an autonomous system of the first order ordinary differential equations.
In the spectral approach to the SM an important role is played by the
RG flow of the gauge couplings.  As we will see in Sect.~\ref{se:comparison}, it provides us with a range of values for the UV cutoff $\Lambda$, which is needed to define the bosonic spectral
action, therefore let us focus on it. The RG equations for
 the  gauge coupling constants $g_{i}$ read:
\be
\mu\frac{\dd g_{i}}{\dd\mu}  =  \beta_{g_i} \label{gaugeRG}
\ee
where the righthand sides are called the \emph{beta}-functions of the corresponding couplings. These beta functions, generally speaking, 
depend on all the coupling constants, which are involved in the model and 
 can be computed
in a perturbative way via the loop expansion~\cite{Machacek1,Machacek2,Machacek3}.
\begin{figure}
\centering{}\includegraphics[scale=1.3]{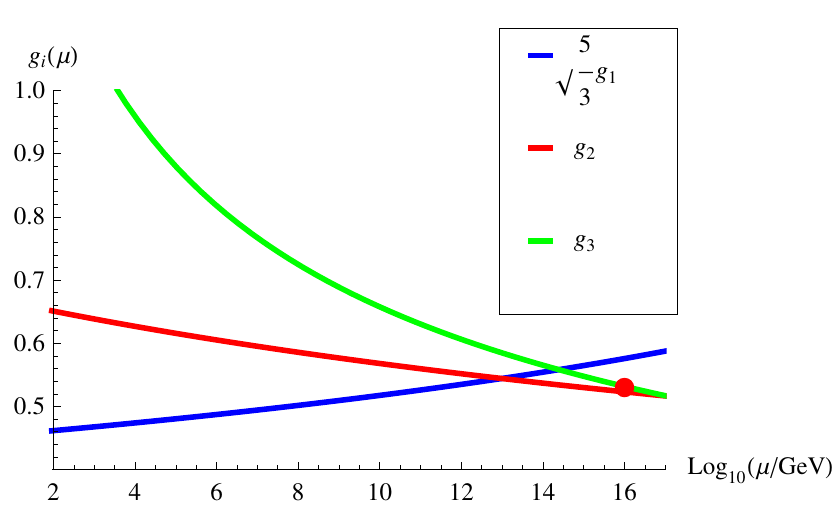}\protect\caption{The RG running of the gauge couplings $g_{i}$, $i=1,2,3$ in the Standard Model.}
\label{GRun}
\end{figure}
 At the one loop approximation the beta functions for the gauge couplings are given by
\begin{eqnarray}
\beta_{g_1} & = & \frac{g_{1}^{3}}{16\pi^{2}}\frac{41}{6},\nonumber \\
\beta_{g_2} & = &-\frac{g_{2}^{3}}{16\pi^{2}}\frac{19}{6},\nonumber \\
\beta_{g_3} & = & -\frac{g_{3}^{3}}{16\pi^{2}}7,
\end{eqnarray}
therefore at this approximation the system~\eqref{gaugeRG} is closed, and moreover the three equations are not coupled with each other. 
The initial conditions for these equations are can be taken from the experiment, in particular at $\mu = M_Z$:
\begin{equation}
g_{1}(M_Z)=0.3575\,,\,\, g_{2}(M_Z)=0.6514\,,\,\, g_{3}(M_Z)=1.1221.
\end{equation}
and the corresponding solutions are plotted on
 Fig.~\ref{GRun}.  

Using the results of~\cite{Machacek1,Machacek2,Machacek3} one can check that the two loop corrections that do not change this plot significantly.
Even though the gauge couplings do not exhibit an exact unification at a single scale $\mu$, there is a some sort of an approximate
unification when $\mu$ varies from $10^{14}$GeV to $10^{17}$GeV. This observation, as we will see, is very important for the spectral 
action principle.

\section{Geometry from Algebras}
 \label{se:spectraltriple}
\setcounter{equation}{0}

In this chapter we present an enlarged definition of geometry, which could encompass also spaces such as the quantum phase space of particles, or the standard model. Taking inspiration from position and momentum in quantum mechanics, we aim at describing geometries (both ordinary or noncommutative) using the spectral properties of a (possibly noncommuting) algebra of operators. We will introduce the necessary mathematical objects keeping the mathematical rigour at a minimum, and with examples mostly relating to ordinary geometry. More details for the construction, still for a nonmathematical audience are in~\cite{Corfu}. The reader interested to details may wish to look at one of the several monographs  on the subject, for example~\cite{Connesbook, Landi, Ticos, ConnesMarcolli, wss}. 

\subsection{Algebras}

An associative algebra $\mathcal{A}$ over the field of complex numbers $\mathbb{C}$
is a vector space on $\mathbb{C}$, so that every object like $\alpha a+\beta b$
with $\alpha,\beta\in\mathcal{\mathbb{C}}$ and $a,b\in\mathcal{A}$,
belongs to $\mathcal{A}$, equipped with a product $\mathcal{A}\times\mathcal{A\rightarrow A}$
which is associative, i.e.\ $(ab)c=a(bc)=abc$ and distributive over addition 
\begin{equation}
a(b+c)=ab+ac,\,\,\,(a+b)c=ac+bc,\,\,\,\forall a,b,c\in\mathcal{A}
\end{equation}
In the general case the product is not commutative so that 
\begin{equation}
ab\neq ba
\end{equation}
when $\mathcal{A}$ has a unit we call the algebra ``unital''. 

A $C^{*}-algebra$ admits
an antilinear involution $^{*}:\mathcal{A}\rightarrow\mathcal{A}$
so that 
\begin{equation}
a^{**}=a \ ; \ 
(ab)^{*}=b^{*}a^{*}\ ;\
(\alpha a+\beta b)^{*}=\bar{\alpha}a^{*}+\bar{\beta}b^{*}\,\,\forall a,b\in\mathcal{A}\,;\,\alpha,\beta\in\mathbb{C}
\end{equation}
with overbar denoting complex conjugation. Moreover the algebra has a norm
 with properties 
\begin{equation}
\Vert a\Vert\geq0,\,\Vert a\Vert=0\,\Leftrightarrow a=0\ ;\
\Vert\alpha a\Vert=\alpha\Vert a\Vert\ ;\
\Vert ab\Vert\leq\Vert a\Vert\Vert b\Vert\ ;\
\Vert a+b\Vert\leq\Vert a\Vert+\Vert b\Vert
\label{eq: propriet=00003D0000E0 norma}
\end{equation}
and moreover
\begin{equation}\label{cstarnorm}
\Vert a^{*}\Vert=\Vert a\Vert \; \ \Vert a^{*}a\Vert=\Vert a\Vert^{2} \,\forall a\in\mathcal{A}
\end{equation}
\begin{example}
\label{Un-esempio-di *Algebra Operatori}
An example of (noncommutative) $C^*$-algebra is that of bounded operators on a Hilbert space. The  involution $*$ is given
by the adjoint and the norm given by the operator norm
\begin{equation}
\|B\|=\mathrm{sup}\{\|B\chi\|_{\mathcal{H}}\,|\,\chi\in\mathcal{H},\,\|\chi\|_{\mathcal{H}}\leq1\}\,.
\end{equation}
as particular case there is the noncommutative algebra $\mathbb{M}_{n}(\mathbb{C})$
of $n\times n$ matrices $X$ with complex entries, with $X^{*}$
given by the Hermitian conjugate of $X$. The norm can also be written
as 
\begin{equation}
\|X\|=\mbox{positive square root of the largest eigenvalue of }X^{*}X\,\,.
\end{equation}
\end{example}

\begin{example}
\label{Esempio *Algebra, funzioni continue}An other example is $C_{0}(M)$
, the commutative algebra of continuous functions on a compact Hausdorff
topological space $M$ with {*} denoting complex conjugation and the
norm given by the supremum norm,
\begin{equation}
\|\,f\,\|_{\infty}=\mathrm{\underset{x\in M}{sup}}|f(x)|\,.\label{eq: Norma Funzioni}
\end{equation}
Note that square integrable functions are not a $C^*$ algebra since the norm does not satisfy~\eqref{cstarnorm}.
\end{example}

 A proper,
norm closed subspace of the algebra $\mathcal{A}$ is a right (resp.\ left) ideal  $\mathcal{I}$ if $a\in\mathcal{A}$
and $b\in\mathcal{I}\Rightarrow ba\in\mathcal{I}$
(resp.\ $ab\in\mathcal{I}$). A subspace which is both a left
and a right ideal is called a two-sided ideal. 
Each ideal
is automatically an algebra. Here we will only consider $*$-ideals, so that $\mathcal I=\mathcal I^*$.
If $\mathcal{A}$ is a $C^{\ast}$-algebra, then the quotient $\mathcal{A}/\mathcal{I}$
is also a $C^{*}$-algebra. 

\subsection{Commutative spaces: Gel'fand-Naimark theorem{\label{sec:-teorema di Gel'fand Naimark} }}

There is a complete duality between \emph{Hausdorff topological spaces} and \emph{commutative $C^{*}$-algebras.}
This is the content of the Gel'fand-Naimark theorem, more precisely, given any topological space $M$, it is possible to naturally associate to it a commutative $C^*$-algebra: that of complex valued function on $M$, vanishing on the frontier of $M$ in case the space is noncompact. Conversely
given any commutative $C^{*}$ -algebra $\mathcal{C}$, 
a Hausdorff topological space $M$ such that $\mathcal{A}$ is isometrically
$^{\ast}$-isomorphic to the algebra of continuous functions $C(M)$ can be reconstructed.
In other words, the study of any Hausdorff topological space is equivalent
to the study of the commutative $C^{*}$ -algebras.~\cite{Diximer,Doran}. 

Let us consider a commutative $\mathcal{C}^{*}-$algebra, $\mathcal{C}$.
We call $\mathcal{\hat{C}}$ the \textit{structure space} $\mathcal{C}$,
i.e.\ the space of equivalence classes of irreducible representations
of $\mathcal{C}$. Every irreducible representation of the commutative
$C^{*}$ -algebra $\mathcal{C}$ is one-dimensional, so that a (non-zero)
$\ast$-linear functional $\varphi:\mathcal{C}\rightarrow\mathcal{C}$
satisfies $\varphi(ab)=\varphi(a)\varphi(b)$, for any $a,b\in\mathcal{C}$.
It follows that, for unital algebras,  $\varphi(\mathbb{1})=1$, $\forall\varphi\in\mathcal{\hat{C}}$.
Any such multiplicative functional is also called a character of $\mathcal{C}$
and the space $\mathcal{\hat{C}}$ is then also the space of all characters
of $\mathcal{C}$. 
The structure space $\widehat{\mathcal{C}}$ becomes a topological
space called Gel'fand space endowed with the Gel'fand
topology, i.e.\ the topology of pointwise convergence on $\mathcal{C}$.
A sequence $\{\varphi_{n}\}\in\mathcal{\widehat{C}}$ converges to
$\varphi\in\widehat{\mathcal{C}}$ iff $\forall a\in\mathcal{C}$
the sequence $\{\varphi_{n}(a)\}$ converges to $\varphi(a)$ in the
topology of $\mathbb{C}.$ If the algebra $\mathcal{C}$ has a unit
then $\mathcal{\widehat{C}}$ is a compact Hausdorff space, otherwise
$\mathcal{\widehat{C}}$ is only locally compact. It is easy to see that $C(\widehat{\mathcal C})=\mathcal C$. 
 
\begin{example}
Consider the case of the algebra given by $n$ copies of complex numbers, $\mathcal{C}=\mathbb C^{n}$, which we may represent on the Hilbert space $\mathcal H=\mathbb C^{nm}$ as diagonal matrices:
\be
\mathcal C\ni\begin{pmatrix}
\lambda_1 \mathbb 1_{m_1} &&&\\
& \lambda_2 \mathbb 1_{m_1} &&\\
&&\ddots &\\
&&& \lambda_n \mathbb 1_{m_1} 
\end{pmatrix}
\ee
with $\sum m_i=m$
 A 1-dimensional representation $\pi:\,\mathcal{C\rightarrow}\mathbb{C}$
is
$
\pi(a)=\lambda_{1}.
$
Correspondingly an example of character $\mathcal{\varphi\in C}$ is
$
\phi(a)=\lambda_{1}.
$
Of course the same procedure can be repeated for all $i$, so that we have $n$ characters, corresponding to $n$ points, which are open and closed at the same time.
\end{example}

Let be $M$ a (locally) compact topological space. As we have shown
in Ex.~(\ref{Esempio *Algebra, funzioni continue}), we have a natural
$C^{\ast}$ -algebra $\mathcal{C}(M)$. On the other side, given a commutative algebra one reconstructs a space of characters with a topology.  We can recognise the points of $M$ via the \emph{evaluation map}, i.e.\ given a particular  character $\varphi=x_0$ (the choice of notation is not casual), we write the simple expression 
\be
x_0(a)=a(x_0)
\ee
both expression are a complex number. In the first expression this number is seen coming from a map which associates a number to every element of the algebra, in the second we stress that we can as well see it as a map from the algebra, seen as made of the functions of points, in $\mathbb C$.
The evaluation map is a homeomorphism
of $M$ onto $\widehat{\mathcal{C}(M)}$, and int his way we have seen the equivalence. This is the essence of the Gel'fand-Naimark theorem. For a more rigorous treatment see the cited literature.  

The noncommutative geometry studies therefore algebras, usually noncommutative. In some cases even for noncommutative algebras it is possible to associate an usual Hausdorff space, this is the case of matrix valued functions on some topological space $M$. This is captured by the concept of \emph{Morita equivalence} (see for example~\cite{Landi, Ticos}), in other cases, for instance the noncommutative algebra generated by the $p$ and $q$ of particle quantum mechanics, it is impossible to talk of `points', and we are left with a genuine noncommutative space.

\subsection{Spectral triples}
\label{sespec}
The basic device in the construction of noncommutative geometry is the \emph{spectral
triple} $\left(\mathcal{A},\mathcal{H},\mathcal{D}_0\right)$ consisting of 
a  {*}-algebra
$\mathcal{A}$ of bounded operators in a Hilbert space $\mathcal{H}$
 - containing the identity operator -
and a non-necessarily  bounded self-adjoint operator $\mathcal{D}_0$ on $\mathcal{H}$ with compact resolvent.  Namely $(\mathcal{D}_0-\lambda)^{-1}$ must be a compact operator when $\lambda$ is not in the spectrum of $\mathcal{D}_0$. In the case of ordinary Riemannian manifolds $\mathcal{D}_0$ is the Dirac operator, and we will often in the following call it as such, even if in some cases the operator will be very different from the one introduced by Dirac. We also require that $[\mathcal{D}_0,a]\equiv \mathcal{D}_0a-a\mathcal{D}_0\in\mathcal{B\left(H\right)}$ for a dense subalgebra of $\mathcal A$. In the case of finite dimensional $\mathcal H$ every operator has compact resolvent. Otherwise the condition implies that the eigenvalues of $\mathcal{D}_0$ have finite multiplicities and they grow to infinity.

There are two more operators which play a role, they are generalizations of chirality and charge conjugation of the ``canonical" spectral triple, see Sect.~\ref{canonical_triple} below. The spectral triple is said to be \emph{even} if  there is an operator $\Gamma$ on $\mathcal{H}$, $\Gamma=\Gamma^{*},\,\Gamma^{2}=1,$
such that 
\begin{equation}
\begin{array}{l}
\Gamma \mathcal{D}_0+\mathcal{D}_0\Gamma=0,\\
\Gamma a-a\Gamma=0,\forall a\in\mathcal{A}
\end{array}\label{eq: Propriet=0000E0 GAMMA5}
\end{equation}
If $\mathcal{H}$ is even then it is possible to separate
\begin{equation}
\mathcal{H=}\frac{(1+\Gamma)}{2}\mathcal{H\oplus}\frac{(1-\Gamma)}{2}\mathcal{H}=\mathcal{H}_{L}\oplus\mathcal{H}_{R}
\end{equation}

Finally, the spectral triple is said to be \emph{real} if  there is an antilinear isometry 
$\mathcal{J}$, called real structure by mathematicians. For physicists it is intimately related to the charge conjugation operator.  This operator implements an action of the opposite algebra\footnote{Identical to $\alg$ as a vector space, but with reversed
product: $a^{\circ} b^{\circ} = (ba)^{\circ}$.}
$\alg^{\circ}$ obtained by identifying
$
b^\circ = \mathcal{J}b^*\mathcal{J}^{-1},
$
and which commutes with the action of $\alg$, and of the generalized one-forms imposing the following conditions, often called zeroth and first order condition respectively:
\begin{eqnarray}\label{zerofirstorder}
[a, \mathcal{J}b\mathcal{J}^{-1}] &=& 0 \nonumber\\
{}[[\mathcal{D}_0,a], \mathcal{J}b\mathcal{J}^{-1}] &=& 0\quad \forall \; a,b\in\alg\end{eqnarray}
The operator $\mathcal{J}$ must obey three further properties:
\bea
\mathcal{J}^2 &=& \pm \mathbb 1 \nonumber\\
\mathcal{J}\mathcal{D}_0&=& \pm \mathcal{D}_0\mathcal{J}\nonumber\\
\mathcal{J}\Gamma&=&\pm\Gamma \mathcal{J} \label{KOcondition}
\eea 
with choice of signs determined by the  algebraic concept of dimension, called KO-dimension (see for example~\cite{ConnesMarcolli}). We will come back to the choice of signs later.

These  elements satisfy a set of properties allowing to prove the \textit{Connes reconstruction theorem}: 
given any spectral triple $\left(\mathcal{A},\mathcal{H},\mathcal{D}_0\right)$
  with commutative $\mathcal{A}$ satisfying the required
  conditions, then ${\mathcal A}\simeq  C^{\infty}(\M)$ for some
    Riemannian spin manifold $\M$, which we discuss next.

\subsection{Canonical triple over a Manifold \label{canonical_triple}}
An important example of a spectral triple is the \emph{canonical} triple over a compact  Riemannian spin manifold $(\mathcal{M},g^{\mathrm{E}}_{\mu\nu})$. From now on $g^{\mathrm{E}}_{\mu\nu}$ stands for the Euclidean metric tensor, defined on $\mathcal{M}$.
By construction the elements of the canonical spectral triple are: 
\be
\left(C(\mathcal{M}),\mathrm{sp}(\mathcal{M}),\slashed{D},\gamma^5, J \right). \la{cst}
\ee
 As we highlighted above, according to Connes's theorem this spectral triple allows to recover the Riemann space $(\mathcal{M},g^{\mathrm{E}}_{\mu\nu})$. It is convenient to consider the vierbeins but not the metric tensor as the independent geometric input. In particular for a given field of vierbeins $e_{\mu}^{a}$  the corresponding Euclidean metric tensor reads:
\be
 g^{\mathrm{E}}_{\mu\nu}  =   e_{\mu}^{a} e_{\nu}^b\, \delta_{ab}, \la{euclmetric}
\ee
whilst the metric tensor with the upper indices is defined by the relation 
\be
g^{\mathrm{E}}_{\mu\nu}g_{\mathrm{E}}^{\nu\lambda} = 
\delta_{\mu}^{\lambda}.
\ee Below we give a detailed description of various ingredients of the canonical spectral triple~\eqref{cst}, restricting ourselves to the physically relevant case $\dim(\mathcal{M}) = 4$. 

The Hilbert space $\mathcal{H}$ is the space of square integrable spinors $\mathrm{sp}(\mathcal{M})$. 
The algebra $\mathcal{A}$ is the commutative infinite dimensional algebra $C(\mathcal{M})$ of continuous functions on  $\mathcal{M}$. 
The elements $f\in\mathcal{A}$ act as multiplicative operators on $\mathcal{H}$, 
\begin{equation}
(f\,\psi)(x)\equiv f(x)\psi(x)\,,\,\,\forall f\in\mathcal{A},\psi\in\mathcal{H}.
\end{equation}

The Dirac operator $\mathcal{D}_0$ is the standard one, which we denote through $\slashed{D}$:
\be
\slashed{D} = \ii \gamma^\mu_{\mathrm{E}} \nabla_\mu.
\ee
In this formula the Euclidean Dirac matrices $\gamma^\mu_{\mathrm{E}} \equiv e^\mu_a \gamma^a_{\mathrm{E}}$ are selfadjoint and the \emph{flat} gamma matrices $\gamma^a_{\mathrm{E}}$ satisfy the anti-commutation relation
\be
\left\{\gamma^a_{\mathrm{E}},\gamma^b_{\mathrm{E}}\right\} = \delta^{ab} \mathbb{1}_4. \label{gammaflatacE}
\ee
The covariant derivative on the spinor bundle over $\mathcal{M}$ 
\be
 \nabla_\mu =\partial_{\mu} - \frac{\ii}{2} \omega_{\mu}^{\mathrm{E}}
\ee
contains the Euclidean spin connection (which is different from the Minkowskian one in~\eqref{MinkSpinConnection}):
\be
\omega_{\mu}^{\mathrm{E}} = \left[\omega_{\mu}^{ab}\right]^{\mathrm{E}}\sigma^{\mathrm{E}}_{ab}, \la{EuclSpinConnection}
\ee
where
\be
\left[\omega_{\mu}^{ab}\right]^{\mathrm{E}}= e^a_{\nu}\, g_{\mathrm{E}}^{\nu\xi}\,\partial_{\mu} e^b_{\xi}
+ e^a_{\nu} \left[\Gamma^{\nu}_{\mu\xi}\right]^{\mathrm{E}}e^{b}_{\rho} \,g_{\mathrm{E}}^{\rho\xi},
\label{LCsDetailsE}
\ee
and
\be
\sigma^{\mathrm{E}}_{ab} = \frac{\ii \left[\gamma^a_{\mathrm{E}}, \gamma^b_{\mathrm{E}}\right]}{4},
\ee
stand for generators of the defining representation of $Spin(4)$. The Christoffel symbols, which enter in~\eqref{LCsDetailsE}, are constructed with
the Euclidean metric tensor $g^{\mathrm{E}}_{\mu\nu}$ (c.f.~\eqref{euclmetric}), using the relation~\eqref{Christ}.
The grading $\Gamma$ is the chirality matrix $\gamma^5$ i.e.\ the usual  product of all four Dirac's $\gamma^\mu_{\mathrm{E}}$
It is identical to the chirality matrix, which we mentioned in the previous section in the Minkowskian context, c.f. Eq.~\eqref{chirdef}.
The real structure $\mathcal{J}$ of the canonical spectral triple, which we denote through $J$,  is defined as:
 \be
J =   \ii\gamma^0_{\mathrm{E}}\gamma^2_{\mathrm{E}}\circ \mbox{complex conjugation}. 
  \ee
This is  the Euclidean charge conjugation: the spinor $J\psi$ transforms in the same way as $\psi$ upon the action of the $Spin(4)$ transformations, while upon the unitary transformations of the spinor $\psi$, which do not effect spinorial index\footnote{e.g. the local $U(1)$ transformations}, the field $J\psi$ transforms according to the complex conjugated representation. The operation $J$ looks very similar to the
  charge conjugation~\eqref{CMdef}, which we introduced in the Minkowskian context before. There is, however, a substantial difference between the two: whilst the former preserves chirality, the latter changes. We will come back to this important point in the next section in the context of the ``Lorentzianisation" of the Euclidean NCG.

\subsection{Noncommutative Manifolds}
In this section we will discuss how to characterise manifolds with the algebraic data of the previous section\footnote{We keep calling the set a ``triple'', even if it is composed by five elements in the real even case.}.

Connes~\cite{Connesmanifold} has shown that the following seven ``axioms'' characterise spin manifolds in the commutative case, and generalise to the noncommutative one. We will give the list for completeness, even if some will not be discussed further since they play no role central in the following, although they are of course important for other aspects of geometry.
\begin{enumerate}

\item{\bf Dimension.} The dimension of the manifold can be read from the rate of growth of the eigenvalues of the Dirac operator. Consider the ratio of the number ${N_\omega}$ of eigenvalues smaller than a value $\omega$, divided by $\omega$ itself. Then
$
{\lim_{\omega\to\infty} N_\omega/\omega^{\frac d2}}
$
does not diverge or vanishes for a single value of $d$, which defines the dimension.

\item{\bf Regularity.} For any $a\in{\mathcal A}$ both $a$ and
    $[\mathcal{D}_0,a]$ belong to the domain of $\delta^k$ for any integer
    $k$, where $\delta$ is the derivation given by $\delta(T) =
    [|\mathcal{D}_0|,T]$. In other words there is a sufficient number of ``smooth'' functions.

\item{\bf Finiteness.} The space $\bigcap_k {\rm Dom} (\mathcal{D}_0^k)$
    is a finitely generated projective left $\mathcal A$ module. Not discussed, plays no role.

{\item {\bf Reality.} There exist  $\mathcal{J}$ with the commutation
    relation fixed by the number of dimensions with the property

\begin{enumerate}
\item{\emph{Commutant, also called order zero}.} $ [a, \mathcal{J}b^*\mathcal{J}^{-1}] = 0, \forall
    a,b$

\item{\emph{First order.}} $ [[\mathcal{D}_0,a], \mathcal{J}b^*\mathcal{J}^{-1}] =
    0~,\forall  a,b $
\end{enumerate}
}
\item{\bf Orientation} There exists a Hochschild cycle $c$
    of degree $n$ which gives the grading $\Gamma$ , This
condition gives an abstract volume form.

\item{\bf Poincar\'e duality} A Certain intersection form
    determined by $\mathcal{D}_0$ and by the K-theory of $\mathcal A$ and its
    opposite is nondegenerate. Not discussed, plays no role.
\end{enumerate}

These structures, abstract as they may seem, will be put to work in the next sections for a description of the standard model of particle interaction.

\newcommand{\gz}{\cdot}
\section{Almost Commutative Geometry and fermionic action of the Standard Model \label{se:SpectralAction}}

In this section we discuss how the fermionic action of the standard model~\eqref{SMink} can be obtained from a particular kind of noncommutative geometry: an \emph{almost commutative geometry}. By this we mean the product of an ordinary geometry, namely the canonical triple for a manifold described in Sect.~\ref{canonical_triple}, times a finite dimensional triple, i.e.\ a triple described by a finite dimensional algebra. The latter algebra is  represented on a Hilbert space also finite dimensional, and the Dirac operator is just a Hermitean matrix. 

\subsection{Finite spectral triple.}
Here we describe the \emph{noncommutative} finite spectral triple $\left(\mathcal A_F,\mathcal{H}_F,D_F,\gamma_F,J_F\right)$.\subsubsection*{The Hilbert space $\mathcal{H}_F$}
The finite dimensional Hilbert space, which is needed to reconstruct the Standard Model within the NCG approach is:
\be
\mathcal{H}_F = \mathbb C^{96}.
\ee
Let us show where the number 96 comes from.  
By construction the basic elements of $\mathcal{H}_F$  are labeled by the independent chiral fermions of the Standard Model and the corresponding charge conjugated fermions, 
therefore  $\mathrm{dim}\,\mathcal{H}_F$ equals to a number of the independent chiral fermions of the Standard Model times two.
The fact that the NCG approach treats the charge conjugated fermions as independent entities is a peculiar feature of the formalism, and we will come back to it below in the context of the fermionic quadrupling and Lorentz symmetries in Sects.~\ref{se:Wicketc} and~\ref{sec:twistlorentz}. 

Let us count how many chiral fermions contribute to the action of the Standard Model~\eqref{SMink}. We remind that for each generation we have a lepton left doublet 
$\boldsymbol{\mathrm{L}}_\mathcal{L}$ plus two right handed singlets $\boldsymbol{\mathrm{v}}_{\mathcal{R}}$ and  
$\boldsymbol{\mathrm{e}}_{\mathcal{R}}$, and a doublet $\boldsymbol{\mathrm{Q}}_{\mathcal{L}}$ and two singlets for quarks
$\boldsymbol{\mathrm{u}}_{\mathcal{R}}$,  $\boldsymbol{\mathrm{d}}_{\mathcal{R}}$
 times three colours. Since we are dealing with three generations we arrive to 48 independent chiral fermions, times two to take antiparticles into account: $\mathrm{dim}\,\mathcal{H}_F = 96$.

Hereafter we  label the elements of $\mathcal{H}_F$
in the following way:
\be
( \boldsymbol \nu_R, \boldsymbol e_R, \boldsymbol L_L,\boldsymbol u_R, \boldsymbol{d_R}, \boldsymbol Q_L, \boldsymbol \nu_R^c, \boldsymbol e_R^c, \boldsymbol L_L^c,
\boldsymbol u_R^c, \boldsymbol{d_R^c}, \boldsymbol Q_L^c ) \label{HfStruct}
\ee
{The symbols here must confronted with the ones used, e.g.\ in~\eqref{fermionicactionusual}. Superficially they are the same, and correspond to the same set of particles. But in~ \eqref{fermionicactionusual} the unslanted $\boldsymbol{\mathrm{Q}}_{\mathcal L}$,  $\boldsymbol{\mathrm{u}}_{\mathcal R}$ etc.\ are spacetime fields. In~\eqref{HfStruct} the slanted $\boldsymbol Q_L$, $\boldsymbol{u}_R$ etc.\ correspond to elements of a finite dimensional Hilbert space. Also the chiral indices, $\mathcal L, \mathcal R$ vs. $L,R$ are different, because they are eigenvectors of two different gradings. In particular $\mathcal L,\mathcal R$ (c.f.~\eqref{chirdef}) refer to chirality $\gamma^5$, while $L,R$ to $\gamma_F$.}
 With the superscript $c$  we indicate the elements of $\mathcal{H}_F$ which correspond to the charge conjugated SM multiplets.  
From now on, whenever it does not create confusion, we address the basic elements of $\mathcal{H}_F$  introduced by Eq.~\eqref{HfStruct}, which do not carry the superscript $c$ as ``particles", whilst the basic elements, which are labelled by the superscript $c$, we call the ``antiparticles". 

Unfortunately in this context we have no explanation for the presence of three generations, with identical quantum numbers, except for the different masses of the fermions. An extension of the model involving the Jordan algebra of Hermitean octonionic matrices  might give an explanation, as discussed in~\cite{Dubois-Violette:2016kzx, Carotenuto:2018zxd}.

\subsubsection*{The Dirac operator $D_F$}

Since we are in finite dimensions the Dirac operator will be a finite 96 by 96 matrix. We shall see later that the finite dimensional Dirac operator introduces the mass terms in the product-geometric fermionic action Eq.~\eqref{SFE}. This in the end  leads to the physical fermionic action~\eqref{SMink}, therefore it is natural to have it carry the information about the Yukawa couplings {$\hat{Y}_{u}$, $\hat{Y}_{d}$, $\hat{y}_{u}$, $\hat{y}_{d}$, 
 and also the Majorana mass  $ \hat{y}_{M}$ and $\mathrm{M}_R$} . In the basis~\eqref{HfStruct} it has the following form (for graphical reasons we substitute the block of zeros by a dot):
{\small 
\begingroup
\setlength{\arraycolsep}{10.5pt}
\renewcommand{\arraystretch}{1.2}
\be
D_F=\left[\!\begin{array}{cccccc|cccccc}
\gz & \gz &\boldsymbol  \Upsilon_\nu\!\!\! &  \gz & \gz & \gz & \boldsymbol {{\Upsilon}}^{\dagger}_R\!\!\! & \gz & \gz & \gz & \gz & \gz \\
\gz & \gz &\boldsymbol  \Upsilon_e\!\!\! & \gz &   \gz & \gz &  \gz & \gz & \gz & \gz & \gz & \gz \\
\boldsymbol \Upsilon_\nu^\dag\!\!\! & \boldsymbol \Upsilon_e^\dag\!\!\! & \gz & \gz & \gz &  \gz & \gz & \gz & \gz &  \gz & \gz & \gz \\
 \gz & \gz & \gz & \gz & \gz & \boldsymbol \Upsilon_u\!\!\! & \gz & \gz &  \gz & \gz & \gz & \gz \\
\gz &  \gz & \gz & \gz & \gz & \boldsymbol \Upsilon_d\!\!\! & \gz & \gz & \gz & \gz & \gz & \gz \\
\gz & \gz &  \gz &\boldsymbol  \Upsilon_u^\dag\!\!\! &\boldsymbol  \Upsilon_d^\dag\!\!\! & \gz & \gz & \gz & \gz & \gz & \gz & \gz \\
\hline
\boldsymbol \Upsilon_R\!\!\! &  \gz & \gz & \gz & \gz & \gz & \gz & \gz & \boldsymbol {\Upsilon}^*_\nu\!\!\! &  \gz & \gz & \gz \\
 \gz & \gz & \gz & \gz & \gz & \gz & \gz & \gz &\boldsymbol{ \Upsilon}^*_e\!\!\! & \gz &  \gz & \gz \\
\gz & \gz & \gz &  \gz & \gz & \gz &\boldsymbol  \Upsilon_\nu^t\!\!\! & \boldsymbol \Upsilon_e^t\!\!\! & \gz & \gz & \gz & \gz \\
\gz & \gz &  \gz & \gz & \gz & \gz &  \gz & \gz & \gz & \gz & \gz & 
{\boldsymbol \Upsilon}^*_u \\
\gz & \gz & \gz & \gz & \gz & \gz & \gz & \gz & \gz & \gz & \gz & {\boldsymbol \Upsilon}^*_d \\
\gz & \gz & \gz & \gz & \gz & \gz & \gz & \gz & \gz &\boldsymbol  \Upsilon_u^t\!\!\! & \boldsymbol \Upsilon_d^t\!\!\! & \gz
\end{array}\!\right].  \label{DFbig}
\ee
\endgroup
}
where 
\bea
\boldsymbol  \Upsilon_{\nu} &=& \hat{Y}_{u} \otimes \tilde{h}_{\nu}^{\dagger} \nonumber\\
\boldsymbol  \Upsilon_{e} &=& \hat{Y}_{d} \otimes h_{e}^{\dagger} \nonumber\\
\boldsymbol  \Upsilon_{u} &=&  \hat{y}_{u} \otimes \tilde{h}_{u}^{\dagger} \nonumber \\
\boldsymbol  \Upsilon_{d} &=&  \hat{y}_{d} \otimes {h}_d^{\dagger}\nonumber\\
\boldsymbol {\Upsilon_R^{\dagger}} &=& \hat{y}_{M}\otimes\mathrm{M}_R.  \label{choice}
\eea
{In these formulas  
the two component columns $h_{\nu,e,u,d}$ (in the Weak isospin indexes) are chosen as follows (hereafter $v$ is the Higgs vacuum expectation value, introduced in~\eqref{vevdef}):} 
\be
h_{\nu} = \left(\begin{array}{c} v \\ 0   \end{array}\right), 
\quad h_{e} = \left(\begin{array}{c} 0 \\ v   \end{array}\right),
\quad h_{u} = \left(\begin{array}{c} v \\ 0   \end{array}\right), 
\quad h_{d} = \left(\begin{array}{c} 0 \\ v   \end{array}\right).
\ee
{We remind,} the tilde in~\eqref{choice} indicates charge conjugated weak isospin doublets e.g. $\tilde{h}_{\nu} = \sigma_2 {h}^*_{\nu}$, where $\sigma_2$ stands for the second Pauli matrix {(c.f.~\eqref{tHiggsdef}).}

\subsubsection*{The noncommutative algebra $A_F$}
Under assumptions on the representation $-$ irreducibility and existence of
a separating vector $-$ it is possible to show~\cite{Chamseddine:2008uq} that the most general finite algebra in~\eqref{produ} satisfying all conditions for the noncommutative space to be a manifold is
\begin{equation}
\mathcal{A}_F=\mathbb{M}_{a}(\mathbb{H})\oplus\mathbb{M}_{2a}(\mathbb{C})\quad\quad
a\in \mathbb{N}^*. \label{genericalgebra}
\end{equation}
This algebra acts on an Hilbert space of dimension $2(2a)^2$~\cite{Chamseddine:2007fk,Chamseddine:2008uq}.\
To have a non trivial grading on $\mathbb{M}_{a}(\mathbb{H})$ the integer 
$a$ must be at least 2, meaning the simplest possibility is 
\be
{\cal A}_F= 
\mathbb{M}_{2}(\mathbb{H})\oplus\mathbb{M}_{4}(\mathbb{C}). \label{bigalgebra}
\ee
The grading condition $[a,\Gamma]=0$, with $\Gamma$ given in (\ref{produ}) below\footnote{The chosen internal grading just considers left/right particles to have eigenvalue $\pm1$.},
reduces the algebra to the left-right algebra:
\be
\mathcal{A}_{LR}=\mathbb{H}_L\oplus\mathbb{H}_R\oplus\mathbb{M}_{4}(\mathbb{C}).
 \label{repa2}
\ee
This is basically a Pati-Salam model~\cite{PatiSalam}, one of the not
many models allowed by the spectral action~\cite{Constraints}. 
The order one condition reduces further the algebra
(for a review see also~\cite{Walterreview})
\be
{\cal A}_{sm}=\mathbb C\oplus \mathbb H\oplus \mathbb M_3(\mathbb C), \label{smalgebra}
\ee
where $\mathbb H$ are the quaternions, which we represent as $2\times
2$ matrices, and $\mathbb M_3(\mathbb C)$ are $3\times 3$ complex
valued matrices. ${\cal A}_{sm} $ is the algebra of the standard model, that is the one whose unimodular group is
U(1)$\times$SU(2)$\times$U(3).
The details of these reductions can be found in~\cite[appendix A]{coldplay}.

In the basis~\eqref{HfStruct} an element of the algebra $a=(\lambda,h,m)$ with  $\lambda\in\mathbb C, h\in\mathbb H$ and $m\in\mbox{Mat}_3(\mathbb C)$ is represented by the matrix\footnote{Here and in the following we omit the unit matrices like $\otimes\mathbb 1_3$ when for example 
a complex number act on a quark, and likewise for doublets etc.} :
{\small 
\begingroup
\setlength{\arraycolsep}{11.5pt}
\renewcommand{\arraystretch}{1.2}
\be
a=\left[\!\begin{array}{cccccc|cccccc}
\lambda  & \gz & \gz &  \gz & \gz & \gz & \gz & \gz & \gz & \gz & \gz & \gz \\
\gz & \lambda^* & \gz & \gz &   \gz & \gz &  \gz & \gz & \gz & \gz & \gz & \gz \\
\gz & \gz & h & \gz & \gz &  \gz & \gz & \gz & \gz &  \gz & \gz & \gz \\
 \gz & \gz & \gz & \lambda & \gz & \gz & \gz & \gz &  \gz & \gz & \gz & \gz \\
\gz &  \gz & \gz & \gz & \lambda^* & \gz & \gz & \gz & \gz & \gz & \gz & \gz \\
\gz & \gz &  \gz & \gz & \gz & h & \gz & \gz & \gz & \gz & \gz & \gz \\
\hline
\gz &  \gz & \gz & \gz & \gz & \gz & \lambda & \gz & \gz &  \gz & \gz & \gz \\
 \gz & \gz & \gz & \gz & \gz & \gz & \gz & \lambda & \gz & \gz &  \gz & \gz \\
 \gz & \gz & \gz &  \gz & \gz & \gz &\gz & \gz & \lambda & \gz & \gz & \gz \\
\gz & \gz &  \gz & \gz & \gz & \gz &  \gz & \gz & \gz & m & \gz & \gz \\
\gz & \gz & \gz & \gz & \gz & \gz & \gz & \gz & \gz & \gz & m & \gz \\
\gz & \gz & \gz & \gz & \gz & \gz & \gz & \gz & \gz & \gz & \gz & m
\end{array}\!\right].  \label{algebraconproiettori}
\ee
\endgroup
}
\subsubsection*{The real structure $J_F$ }
This exchange particles with antiparticles, and performs a complex conjugation (it is an antiunitary operator). It is a bloc  diagonal operator which can be expressed as:
\be
J_F=\left[\begin{array}{cc} \mathbb 0 & \mathbb 1_{48}\\ \mathbb 1_{48} & 0 \end{array}\right]
\circ cc.  \label{JF}
\ee
We emphasise that this operation is antiunitary in $\mathcal{H}_F$.
\subsubsection*{The grading $\gamma_F$}
The last ingredient of the finite dimensional spectral triple is defined in the basis~\eqref{HfStruct} as follows:
{\small \begingroup
\setlength{\arraycolsep}{11.5pt}
\renewcommand{\arraystretch}{1.2}
\be
\gamma_F=\left[\!\begin{array}{cccccc|cccccc}
-\mathbb 1  & \gz & \gz &  \gz & \gz & \gz & \gz & \gz & \gz & \gz & \gz & \gz \\
\gz & - \mathbb 1 & \gz & \gz &   \gz & \gz &  \gz & \gz & \gz & \gz & \gz & \gz \\
\gz & \gz & \mathbb 1 & \gz & \gz &  \gz & \gz & \gz & \gz &  \gz & \gz & \gz \\
 \gz & \gz & \gz &  - \mathbb 1 & \gz & \gz & \gz & \gz &  \gz & \gz & \gz & \gz \\
\gz &  \gz & \gz & \gz & -\mathbb 1 & \gz & \gz & \gz & \gz & \gz & \gz & \gz \\
\gz & \gz &  \gz & \gz & \gz & \mathbb 1 & \gz & \gz & \gz & \gz & \gz & \gz \\
\hline
\gz &  \gz & \gz & \gz & \gz & \gz &  \mathbb 1 & \gz & \gz &  \gz & \gz & \gz \\
 \gz &  \gz & \gz & \gz & \gz & \gz & \gz & \mathbb 1 & \gz & \gz &  \gz & \gz \\
 \gz & \gz & \gz &  \gz & \gz & \gz &\gz & \gz & - \mathbb 1 & \gz & \gz & \gz \\
\gz & \gz &  \gz &  \gz & \gz & \gz &  \gz & \gz & \gz & \mathbb 1 & \gz & \gz \\
\gz & \gz & \gz & \gz &  \gz & \gz & \gz & \gz & \gz & \gz & \mathbb 1 & \gz \\
\gz & \gz & \gz & \gz & \gz & \gz & \gz & \gz & \gz & \gz & \gz & -\mathbb 1
\end{array}\!\right]  \label{gammaF}
\ee
\endgroup
}
Note that the signs of  unities on the diagonal correspond to the chiralities of the corresponding fermionic multiplets, which are equal to plus one for the left-handed particles and right-handed anti-particles and to the minus one for the right-handed particles and the left-handed antiparticles.

\subsection{The product geometry.}
Almost commutative spectral triple $\left(\mathcal A,\mathcal H,\mathcal D_0,\Gamma,\mathcal J\right)$ is defined as a product of the infinite dimensional canonical commutative spectral triple $\left(C(M),\mathrm{sp}(M),\slashed{D},\gamma^5, J \right)$ and the finite dimensional noncommutative spectral triple $\left(\mathcal A_F,\mathcal{H}_F,D_F,\gamma_F,J_F\right)$ according to the following rule:
\bea
\mathcal A &=&C(M)\otimes \mathcal A_F, \nonumber \\
\mathcal H &=& \mathrm{sp}(M)\otimes \mathcal \mathcal{H}_F, \nonumber\\
\mathcal D_0 &=&  \slashed{D} \otimes 1_F + \gamma^5 \otimes {D}_F \label{D0}, \nonumber\\
\Gamma &=& \gamma^5\otimes \gamma_F, \nonumber\\
\mathcal J&=&J\otimes J_F, \label{produ}\label{almostspectraltriple}
\eea
where, we emphasise, the real structure $\mathcal J$ is antiunitary in $\mathcal{H}$. The $\gamma^5$ in the definition of $\mathcal D_0$ in the second term is necessary. Otherwise, for a general product of two spectral triples, the resulting operator could not have a compact resolvent~\cite{DabrowskiDossena}.  The choice of putting the chirality in the first or second addend is irrelevant, the operator $\slashed{D} \otimes \gamma + 1 \otimes {D}_F$ is unitarily equivalent to the one defined above.  There are still some ambiguity in this definition, to cite just one, the product we have defined does not generalize in a natural way to a further product of triples, the order of product matters and we would have that the product of triples is nonassociative. Fortunately all problems are neatly solved considering a \emph{graded} product of triples~\cite{Boyle:2016cjt, Farnsworth:2016qbp, Bizi:2016lbv}.

In what follows we parametrise the elements of the Hilbert space $\mathcal{H}$ as follows:
\be
\Psi =  (\boldsymbol{\mathrm{v}}_R, \boldsymbol{\mathrm{e}}_R, \boldsymbol{\mathrm{L}}_L, 
           \boldsymbol{\mathrm{u}}_R, \boldsymbol{\mathrm{d}}_R, \boldsymbol{\mathrm{Q}}_L,   
           \boldsymbol{\mathrm{v}}_R^c, \boldsymbol{\mathrm{e}}_R^c, \boldsymbol{\mathrm{L}}_L^c,          
            \boldsymbol{\mathrm{u}}_R^c, \boldsymbol{\mathrm{d}}_R^c, \boldsymbol{\mathrm{Q}}_L^c             
             )^{\mathrm{T}},\label{HStruct}
\ee
This  parametrisation  looks very similar to the parametrisation~\eqref{HfStruct} of the elements of $\mathcal{H}_F$, but the change of typeface  indicates that the elements of $\mathcal{H}$ are spinors, no longer complex numbers. In these notations 
$\boldsymbol{\mathrm{u}}_R$ is a collection of 4-component spinors which transforms upon the action of the gauge group  as the right handed quarks 
{$\boldsymbol{\mathrm{u}}_{\mathcal{R}}$}, $\boldsymbol{\mathrm{u}}_R^c$  is an \emph{independent} collection of  4-component spinors which transforms upon the action of the gauge group  as the charge conjugated right handed quark {$J\boldsymbol{\mathrm{u}}_{\mathcal{R}}$} field and so on. These spinors are \emph{non-chiral}, in the sense that they are not eigenvectors of $\gamma^5$.

This almost commutative spectral triple is a central ingredient and basically the starting point of the NCG approach to the Standard Model. The forthcoming discussion is devoted to description of how to construct the classical action of the Standard Model {\eqref{SMink}} using this spectral triple. Nevertheless
the product structure exhibits a very peculiar feature, which is known as a ``fermionic quadrupling". 
Below we describe, what the problem is. Before we proceed we outline some important aspects of the general structure of the Dirac operator.

{\subsubsection*{Constraints on the Dirac Operator\footnote{We thank Latham Boyle and Shane Farnsworth for discussions and correspondence on this issue.}}

The Dirac operator $D_F$ defined in~\eqref{DFbig} correctly reproduces the Yukawa coupling of fermions including a possible Majorana mass for the neutrinos. One of the important aspects of of this approach is the fact that the mathematical framework on NCG singles out its structure, modulo a few caveats which we will discuss. As we have seen in Sect.~\ref{sespec} there are constraints on the commutation of $D$ with the elements of the algebra, $J$ and $\Gamma$, which in turn impose limitations on $D_F$. The condition~\eqref{KOcondition} depends on a choice of signs, which in turn depend on the number of dimensions. While for the continuous part the choice is unambiguous, for the discrete algebra the definition of dimension is less clear. The dimension stemming from the growth of eigenvalues is ill defined for finite dimensional operators. One might think that it is zero, but this choice leads to an unphysical Dirac operator. Mathematically however, there is a different definition of dimension~\cite{Connesbook, Connes:1996fu} based on K-homology, which indeed dictates the choice of signs.
It turns out~\cite{AC2M2} that the proper choice for the number of dimension which reproduces the correct couplings is 6 (mod 8). This choice of signs remarkably is also the one one would obtain if one were to use Minkowskian rather than Euclidean quantities~\cite{Barrett}. 

But we are not yet there, even with these provisions there still are spurious couplings. These can be eliminated imposing a condition called the `massless photon condition''~\cite{AC2M2}. This is the requirement of commutation of $D_F$ with elements of the algebra of the kind$(\lambda,\lambda\, \mathbb 1_2, \mathbb 0_3)$, which in the representation~\eqref{algebraconproiettori} are matrices which in the first nine blocks are a multiple of the identity, and vanish in the remaining three blocks. This condition does not impose any restriction in the strong force sector, but it eliminates unwanted couplings and keeps an unbroken U(1) symmetry. The condition has originally been imposed by hand, as it has no evident geometric meaning. Moreover it must be imposed on the unfluctuated $\mathcal D_0$, the procedure would not work for the covariant Dirac operator of~\eqref{fluctu}.

Some solutions have been proposed to this. The unwanted couplings also disappear if one imposes~\cite{Boyle:2016cjt}, in the finite part, a ``second order condition'', i.e.\ the generalization of~\eqref{zerofirstorder}:
\be
[[D_F,a],[D_F,b]]=0 \quad ; \forall a,b\in\mathcal A_F
\ee
A similar condition can be imposed~\cite{Brouder:2015qoa} for the full triple only up to ``junk forms'', i.e.\  higher forms which appear spuriously when commuting the Dirac operator with the element of the algebra, and have to be quotiented out to reobtain the usual de Rham cohomolgy\footnote{A proper treatment of junk form is beyond the scopes of this review, a terse description of them can be found in~\cite{Landi, Ticos}.}. Unlike the massless photon condition, the second order condition has a mathematical origin. In~\cite{Farnsworth:2013nza, Boyle:2014wba, Farnsworth:2014vva, Shanethesis} an extension of noncommutative geometry to the nonassociative case has been introduced. The standard model, and its restriction including the second order conditions, emerges imposing associativity constraints to this generalised geometry. In particular in~\cite{Boyle:2016cjt} the algebra of the spectral triple is enlarged to a superalgebra ($\mathbb Z_2$ graded) with differential form and the Hilbert space. Within this scheme the second order comes naturally solving the junk form issue.

The Dirac operator can have some extra, not experimentally excluded, coupling between particles if one considers alternative internal grading operators. The choice of $\gamma_F$ having eigenvalue +1 and -1 on left and right particles respectively may seem natural, but is not necessary. In~\cite{FrancescoLudwik} a different grading, imposed by a noncommutative generalization of Clifford symmetry. The new grading is related to the old one by
\be
\gamma_F = \left(\mathrm{Q} - \mathrm{L} \right)\gamma^{\mathrm{st}}_F,
\ee
where $\mathrm{Q}$ and $\mathrm{L}$ stand for the projectors of the ``quark" and ``leptonic" subspaces of $\mathcal{H}_F$ respectively. The structure of $D_F$ in this case changes, and more couplings are allowed. They have been investigated in~\cite{DeepPurple}. The theory allows for extra bosonic fields, with some couplings which disappear after elimination of the spurious degrees of freedom, which are present due to the fermionic quadrupling discussed below. The extra terms are compatible with known physics. A complete phenomenological analysis has not been yet performed, but their role for the renormalization flow is studied in~\cite{Aydemir:2019txw}.

\subsubsection*{The fermion quadruplication problem.}
 Since the Dirac spinor has four components, the element $\Psi$ of the full Hilbert space $\mathcal{H}$ is described by  384 independent complex valued functions, whilst the fermionic action of the Standard Model {\eqref{SMink} depends just on 96. As we have seen at the beginning of this section:  48 chiral fermions, which have just two independent components}. Clearly there is some overcounting, called for historical reasons \emph{fermion doubling}~\cite{LMMS, BrunoPepeThomas}. Let us explain the origin of the overcounting.

For the forthcoming discussions it is convenient to split the Hilbert space $\mathcal{H}$ of the almost commutative geometry as follows:
\be
\mathcal{H} = \mathrm{sp}(M)\otimes \mathcal{H}_F = \mathcal{H}_L \oplus \mathcal{H}_R \oplus \mathcal{H}_L^c \oplus \mathcal{H}_R^c, \label{Hprod}
\ee
The subscripts $L$ and $R$ indicate the transformation properties of the corresponding Dirac multiplets upon the action of the gauge group. In other words $\mathcal{H}_L$ consist of the multiplets of the  \emph{nonchiral} 4-component spinors
corresponding to left particles and right antiparticles:
\be
 (\boldsymbol{\mathrm{v}}_R, \boldsymbol{\mathrm{e}}_R,  
           \boldsymbol{\mathrm{u}}_R, \boldsymbol{\mathrm{d}}_R,                       
             )^{\mathrm{T}} \in\mathcal{H}_R,
             \quad (\boldsymbol{\mathrm{L}}_L, 
            \boldsymbol{\mathrm{Q}}_L         
             )^{\mathrm{T}} \in \mathcal{H}_L,
\ee
and analogously
\be
            ( \boldsymbol{\mathrm{v}}_R^c, \boldsymbol{\mathrm{e}}_R^c,         
            \boldsymbol{\mathrm{u}}_R^c, \boldsymbol{\mathrm{d}}_R^c
             )^{\mathrm{T}} \in\mathcal{H}_R^c,
             \quad (
           \boldsymbol{\mathrm{L}}_L^c,          
           \boldsymbol{\mathrm{Q}}_L^c             
             )^{\mathrm{T}} \in \mathcal{H}_L^c.
\ee
 On the other side the Standard Model
is described by the  multiplets of the \emph{chiral} fermions, which are the eigenvectors of the left and right chiral projectors which we indicate with a different typeface, {see
\eqref{chirdef}}.
 In other words apart from the particle with the correct chirality $\mathcal{H}$ contains a particle with the same quantum numbers but with the opposite chirality - the mirror particle.
Folllowing~\cite{DireStraits} we will call this doubling the ``mirror doubling". 

Another doubling has the following origin. With the physical Lorentz signature the fermionic action of the Standard Model~\eqref{SMink} does not contain any independent variables with the index ``c", which indicates the charge conjugated field:  the charge conjugated spinor is obtained from the original one via the the charge conjugation operation 
\eqref{CMdef}
i.e.\ they are not independent variables. 

This second doubling is called in~\cite{DireStraits} the ``charge conjugation doubling". We notice that the Minkowskian charge conjugation operation~\eqref{CMdef}
changes chirality, whilst the Euclidean charge conjugation $J$ (i.e.\ the real structure of the canonical spectral triple) does not. This fact is very important and results in the necessity to carry out the (anti) Wick rotation to the Lorentzian signature in order to eliminate this doubling. 
\subsection{Covariant Dirac operator.}
The Dirac operator $\mathcal{D}_0$, which enters in the almost commutative spectral triple is not sufficient to build the fermionic action~\eqref{SMink}, since the latter  contains the gauge and the scalar fields as well. Nevertheless, in analogy with the introduction of the covariant derivative,  there is an elegant way to introduce these fields in the game, which is based on the spectral triple only: one has to consider the so called fluctuated Dirac operator
\be
\mathcal{D} = \mathcal D_0 +\sum_i a_i [\mathcal D_0,b_i] +\sum_i   \mathcal J \,a_i [\mathcal D_0,b_i] \mathcal J^\dagger   \label{fluctu},
\ee
for generic elements $a_i, b_i\in\mathcal A$. Both gauge and scalar fields in the spectral approach derive from these fluctuations. {Mathematically these fluctuations are nothing but the addition to the Dirac operator of a connection one form, which in the spectral approach to geometry are seen as operators themselves.}

There are two sources, which give nonzero contributions to the commutators $[\mathcal D_0,b_i]$. The first is the noncommutativity of $\slashed{D}$, which enters in the first term of $\mathcal{D}_0$, with the elements $\mathcal{C}(M)$  i.e.\ the infinite dimensional (tensor) factor of $\mathcal{A}$. The second is the noncommutativity of $D_F$, which enters in the second term of $\mathcal{D}_0$, with the elements of $A_{sm}$ i.e.\ the finite dimensional (tensor) factor of $\mathcal{A}$.

Considering the fluctuations of the Dirac operator~\eqref{fluctu} one can show (see~\cite{AC2M2} for details) that the former class of the fluctuations recovers exactly the gauge fields, which correspond to the gauge group of the Standard Model $U(1)\times SU(2)\times SU(3)$, whilst  the latter class of the fluctuations leads to the Higgs scalars. More precisely the fluctuated Dirac operator has the following structure:
\be
\mathcal{D} = \ii\gamma^{\mu}_{\mathrm{E}}\nabla_{\mu}^{\mathrm{E}} + \gamma^5 \otimes M, \label{fluctu2}
\ee 
where the covariant derivative $\nabla_{\mu}^{\mathrm{E}}$ contains both the Euclidean Levi-Civita spin-connection and the $U(1)\times SU(2) \times SU(3)$ gauge connection\footnote{The explicit form of $\nabla_{\mu}^{\mathrm{E}}$ is presented in the next section in~\eqref{EuclCovDiv}.}, and the Matrix $M$ is obtained from $D_F$ via the replacement of the constant 2 component columns (c.f.~\eqref{choice}) by the two component complex scalar field $H$ according to the following rule:
\bea
\tilde h_{\nu} &\longrightarrow& \tilde{H}, \nonumber\\
 h_{e}   &\longrightarrow&  H, \nonumber \\
\tilde  h_{u}       &\longrightarrow& \tilde{H}, \nonumber\\
   h_{d}     &\longrightarrow& H 
  \label{rules}.
\eea
Note that upon the fluctuations of the Dirac operator the entry $\mathrm{M}_R$ remains a constant i.e.\ it does not transform into a field. The fluctuated Dirac operator~\eqref{fluctu} transforms in a covariant manner 
upon the simultaneous transformation of the gauge and the scalar fields, in particular the combination
\be
\left(\mathcal{J}\Psi\right)^{\dagger}\mathcal{D}\,\Psi,\quad \Psi\in\mathcal{H}
\ee 
is gauge invariant, and can be considered to be a natural candidate for the fermionic action.
Nevertheless such a fermionic action depends on four times more degrees of freedom than the fermionic action of the Standard Model due to the fermionic quadrupling, which we discussed above.

In order to get rid of the mirror doubling one has to extract the subspace $\mathcal{H}_{+}$ of $\mathcal{H}$ which contains just the fermions with correct chiralities, which has the following structure:
\be
\mathcal{H}_{+}  = \left(H_L\right)_{\mathcal{L}} 
\oplus \left(\mathcal{H}_R\right)_{\mathcal{R}} \oplus \left(\mathcal{H}_L^c\right)_{\mathcal{R}} \oplus \left( \mathcal{H}_R^c\right)_{\mathcal{L}}. \label{Hp}
\ee
In the original paper~\cite{AC2M2} such an extraction was presented in the form
\be
P_+ \mathcal{H}_{+} = \mathcal{H}_{+}
\ee
 where the  projector $P_+$ is defined via the grading $\Gamma$ of the almost commutative geometry as follows:
\be
P_{+} = \frac{1}{2}\left(\mathbb 1_{\mathcal{H}} + \Gamma\right). 
\ee
The Euclidean fermionic action introduced in~\cite{AC2M2}, which is free of the mirror doubling reads:
\be
{S}_F^{\mathrm{E}} = \frac{1}{2}\int d^4x \sqrt{g^{\mathrm{E}}}\left(\mathcal{J}\Psi_{+}\right)^{\dagger}\mathcal{D}\Psi_{+}, \quad \Psi_+ \in \mathcal{H}_+. \label{SFE}
\ee
This action still suffers of the charge conjugation doubling, and it is Euclidean, however some progress is there:  it  correctly\footnote{Since the eigenvalues of the Dirac operator grow indefinitely, the expression is however still formal. A regularisation is needed.}  recovers the action in the form of a Pfaffian. 
Another issue is the fact that we are in the Euclidean context. We turn to this issue.

\subsection{The physical fermionic action. \label{se:Wicketc}}
In this section we describe the (anti) Wick rotation to the Lorentzian signature and explain how to get rid of the charge doubling. We will see that the two operations are actually connected.

\subsubsection*{ Wick rotation: general remaks.}
Upon the Wick rotation we mean a procedure, which allows to connect the action  of a Minkowskian quantum field theory $S^{\rm M}[{\rm fields}, g_{\mu\nu}^{\rm M}] $ with the action $S^{\rm E}[{\rm fields}, g_{\mu\nu}^{\rm E}] $ of a Euclidean quantum field theory. The path integrands must transform in a proper way:
\begin{equation}
 \exp{\left(-S^{\rm E}[{\rm fields}, g_{\mu\nu}^{\rm E}] \right)} 
\longleftrightarrow \exp{\left(\ii S^{\rm M}[{\rm fields}, g_{\mu\nu}^{\rm M}] \right)}.  \label{WickGen}
\end{equation}
In a flat space-time in the Cartesian coordinates the imaginary time formalism, based on the replacement $t\longrightarrow \ii t$, is usually used. This prescription is too naive. One can see~\cite{Visser} that it may fail for spacetimes for which the choice of ``time'' depends on coordinates. 
To go from an Euclidean to a Lorentzian theory in a self consistent manner we proceed differently.
 Each expression $F$ which involving vierbeins $ e_{\mu}^a$ is transformed according to:
\be
\mbox{Wick:} \quad F\left[e_{\mu}^0, e_{\mu}^j \right]
\longrightarrow \quad F\left[\ii e_{\mu}^0, e_{\mu}^j\right],
\,\, \, j = 1,2,3. 
\label{bosWickvier}
\ee 
This procedure perfectly works for the bosonic fields in scalar, vector and gravitational sectors.
Let us consider the bosonic actions, which, as we will see in the next chapter, come out from the bosonic spectral action,
\bea
S_{\rm scal}^{\rm E}\left[g_{\mu\nu},A_\mu,  \phi\right] &=& \int d^4 x \sqrt{g^{\rm E}}
\left\{ \sum_{j=1}^N \left(g^{\mu\nu}_{\rm E} \nabla_{\mu} \phi_j^{\dagger} \nabla_{\nu}\phi_j  -  \frac{1}{6}R\left[g_{\mu\nu}^{\rm E}\right] 
\phi_j^{\dagger} \phi_j\right) + V(\phi)
\right\}, \nonumber\\ 
S^{\rm E}_{\rm gauge} \left[F_{\mu\nu}\right]&=&  \int d^4 x \sqrt{g^{\rm E}} \, g^{\mu\alpha}_{\rm E} g^{\nu\beta}_{\rm E} \,{\rm tr}\, F_{\mu\nu} F_{\alpha\beta}, \nonumber\\
S^{\rm E}_{\rm grav}\left[g_{\mu\nu}^{ \rm E}\right] &=& \int d^4 x \sqrt{g^{\rm E}} \left(\lambda  +  \frac{M_{\rm Pl}^2}{16\pi}R\left[g_{\mu\nu}^{\rm E}\right] 
+ a  C_{\mu\nu\alpha\beta}\left[g_{\mu\nu}^{\rm E}\right] C^{\mu\nu\alpha\beta}\left[g_{\mu\nu}^{\rm E}\right]\right), \label{eactions}
\eea
and their Minkowskian counterparts:
\bea \label{SFL}
 S_{\rm scal}^{\rm M} \left[g_{\mu\nu}^{ \rm M}, \phi_j\right]  &=& \int d^4 x \sqrt{-g^{\rm M}}
\left\{ \sum_{j=1}^N \left(g^{\mu\nu}_{\rm M} \nabla_{\mu} \phi_j^{\dagger} \nabla_{\nu}\phi_j  -  \frac{1}{6}R\left[g_{\mu\nu}^{\rm M}\right] 
\phi_j^{\dagger} \phi_j\right) - V(\phi)
\right\}, \nonumber\\
 S_{\rm gauge}^{\rm M} \left[g_{\mu\nu}^{ \rm M}, A_{\mu}\right]  &=& \int d^4 x  \sqrt{-g^{\rm M}} \,\left(- g^{\mu\alpha}_{\rm M} g^{\nu\beta}_{\rm M} \,{\rm tr}\, F_{\mu\nu} F_{\alpha\beta}\right), \\
 S_{\rm grav}^{\rm M} \left[g_{\mu\nu}^{ \rm M}\right] &=&
\int d^4 x \sqrt{-g^{\rm M}} \left(-\lambda \, + \, \frac{M_{\rm Pl}^2}{16\pi}R\left[g_{\mu\nu}^{\rm M}\right] 
- a  C_{\mu\nu\alpha\beta}\left[g_{\mu\nu}^{\rm M}\right] C^{\mu\nu\alpha\beta}\left[g_{\mu\nu}^{\rm M}\right]\right),  \nonumber\label{mactions}
\eea
where $\phi$ is a multicomponent scalar field, $V$  - its potential, $A_{\mu}$ is the vector potential and $F_{\mu\nu}$ stands for the corresponding field-strength tensor. The quantities 
$R[g_{\mu\nu}]$ and $C_{\mu\nu\eta\xi}[g_{\mu\nu}]$ are the scalar curvature and the Weyl tensor, which are build\footnote{See~\cite{DireStraits} for details.} from the metric tensor $g_{\mu\nu}$;  the cosmological constant and the Planck mass are denoted through $\lambda$ and $M_{\rm Pl}$ respectively; the quantity $a$ is the dimensionless constant. 

After the Wick rotation~\eqref{bosWickvier} the Euclidean bosonic actions~\eqref{eactions} map onto the Minkowskian version~\eqref{mactions} exactly in a way, which fits the general prescription~\eqref{WickGen}:
\begin{equation}
\mbox{Wick:}\quad \exp{\left(-S_{\mathrm{bos}}^{\rm E}[{\rm fields}, g_{\mu\nu}^{\rm E}] \right)} 
\longrightarrow \exp{\left(\ii S_{\mathrm{bos}}^{\rm M}[{\rm fields}, g_{\mu\nu}^{\rm M}] \right)} , \quad \mathrm{bos} = {\rm scal},\,{\rm gauge},\,{\rm grav}.\label{cond}
\end{equation}
The fermionic case is subtle, since the doubling plays a role.

\subsubsection*{The fermionic case.}
Applying the Wick rotation of the vierbeins~\eqref{bosWickvier} to the NCG Euclidean fermionic action~\eqref{SFE} one obtains:
\be
\mathrm{Wick}:\quad \exp\left( -S_F^{\mathrm{E}}[\mathrm{spinors}, e^a_{\mu}]\right) 
 \longrightarrow \exp\left(\ii{S}_F^{\mathrm{M\,\, doubled}}[\mathrm{spinors}, e^a_{\mu}]\right),
\ee
where the fermionic action ${S}_F^{\mathrm{M\,\, doubled}}$ is Lorentz invariant, however it depends on twice more independent variables and it is not real. 
The former means that the classical configuration space of the theory, which is described by this action, is twice bigger than needed. This implies that canonical quantisation, needed, in particular, to describe the asymptotic states, see~\cite{DireStraits} for discussions. Both issues can be resolved via the elimination of the charge doubling, which is rendered as
the following identification of the variables in the action ${S}_F^{\mathrm{M\,\, doubled}}$ from the subspaces $\mathcal{H}_L^c$ and $\mathcal{H}_R^{c}$ with the variables from $\mathcal{H}_L$ and $\mathcal{H}_R$:
\be
\mathrm{step \,\,1}: 
\begin{cases}
\;\left(\psi_L^{c}\right)_{\mathcal R} \in  \underbrace{ \left(H_L^c \right)_{\mathcal R}  }_{\subset H_+}
\quad\mbox{has to be identified with}\quad
 C_{\mathrm M}\left(\psi_L\right)_{\mathcal L} , \quad \left(\psi_L\right)_{\mathcal L}\in\underbrace{\left(H_L\right)_{\mathcal L}}_{\subset H_+} \\
\;\left(\psi_R^{c}\right)_{\mathcal L} \in  \smash{\underbrace{ \left(H_R^c \right)_{\mathcal L}  }_{\subset H_+}}
\quad\mbox{has to be identified with}\quad
 C_{\mathrm M}\left(\psi_R\right)_{\mathcal R} , \quad \left(\psi_R\right)_{\mathcal R}\in\smash{\underbrace{\left(H_R\right)_{\mathcal R}}_{\subset H_+}}
\rule{0pt}{16pt}
\end{cases} \label{fWick2}.\vspace{14pt}
\ee
We emphasise that the identification~\eqref{fWick2} makes sense \emph{after} the Wick rotation to Lorentzian signature: since the quantities to be identified transform in the same way under the Lorentzian $SO(1,3)$ transformations rather than Euclidean $SO(4)$ rotations. This establishes a deep connection between the resolution of the two naively-thinking independent issues: the Euclideness of the approach and a presence of the charge conjugation doubling.
Since there is no risk of confusion anymore, hereafter we simplify the notations:
\be
\mbox{change of notations}:\quad 
\left(\psi_L\right)_{\mathcal L} \longrightarrow \psi_{\mathcal L}, \quad 
\left(\psi_R\right)_{\mathcal R} \longrightarrow \psi_{\mathcal R}, 
\ee

One more step has to be done to complete the discussion. The result of the application of the procedure~\eqref{fWick2} to the 
${S}_F^{\mathrm{M\,\, doubled}}$ leads to the Lorentz invariant fermionic action, which is  real and depends on the correct number of the degrees of freedom. Nevertheless all the mass terms involve
the chirality $\gamma^5$ matrix, what implies that the discrete symmetries of this theory do not coincide with the ones of the Standard Model, in particular the outcoming QED sector contains the axial mass terms, which break P-invariance! In order to resolve this final issue one has to carry out the axial transformation of all the spinors
\be
\mathrm{step \,\,2}: \quad\psi \longrightarrow e^{-\frac{\ii\pi}{4}\gamma^{5}} \psi \label{axtrans}.
\ee
It is very important, that this last step must be performed \emph{before} the quantisation: otherwise one will get an additional Pontrtyagin gauge action which comes out from the abelian axial anomaly.
Summarising all together we see that:
\be
\mbox{Wick rotation + step 1 + step 2}: \quad \exp{\left( -S_F^{\mathrm{E}} \right)}\longrightarrow  \exp{\left( \ii S^{\rm {M}}_F \right)},
\ee
{where, we remind,  $S^{\rm {M}}_F $ has been introduced in~\eqref{SMink}.}

\section{Bosonic spectral Action}

In this section we discuss how to define the action for bosons using the spectral data. First we present the original
cutoff-based definition of the bosonic spectral action (BSA). After that we discuss the heat kernel expansion, which on the one side allows to compute it in the low-energy approximation, whilst on the other side naturally suggests to define the ``asymptotically expanded BSA". Then we demonstrate another  computation of the quadratic part of the bosonic spectral action, valid at all energy scales.This calculation shows that the ultraviolet behaviour of the original BSA drastically differs from its asymptotically expanded version. A recent book describes in detail the spectral action~\cite{EcksteinIochum}.

In the presentation here we postulate the spectral action. This action is  natural from the spectral geometry point of view, which is the theme behind noncommutative geometry. It is however possible to connect it to other structures. A precursor was the \emph{finite mode regularization} introduced in~\cite{AndrianovBonora1, AndrianovBonora2, Fujikawa} in QCD. In~\cite{AndrianovLizzi,AndrianovKurkovLizzi,AndrianovKurkovLizzi2,KurkovLizziHiggs,Kurkov:2013gma} it is argued that a structure similar to the bosonic spectral action can emerge from anomalies. A  different regularization, based on the $\zeta$ function, also gives rise to the action~\cite{zeta}. It is also possible to see the fermionic action as ``spectral'', and in this case neutrinos play a fundamental role~\cite{Sitarz:2008qv,Sakellariadou:2019ihz}.

\subsection{Formal Definition}
We start from the fluctuated Dirac operator $\mathcal{D}$, that we introduced in Eq.~\eqref{fluctu2}, i.e.\ the Dirac operator, which enters in the Euclidean fermionic action~\eqref{SFE}.
It is remarkable, that one can define the action for bosons in terms of this object as well. By definition\footnote{From now on ${\rm Tr }\,$ stands for the functional trace on $L^2$. } 
\begin{equation}
S_B := {\rm Tr }\, \chi \left(\frac{\mathcal{D}^2}{\Lambda^2}\right),  \la{BSAdef}
\end{equation}
where $\chi$ is a cutoff function, and $\Lambda$ is the cutoff scale. The former is assumed to be an arbitrary function such that the trace is well defined\footnote{Recall that here we assume spacetime, to be compact, Euclidean and without a boundary.}. A natural choice of this function is  the characteristic function of the unit interval or its smooth approximation. In such a case the right hand side has a clear meaning: bosonic spectral  action is the number of the eigenvalues of the Dirac operator $\mathcal{D}$, which are smaller than the UV cutoff scale $\Lambda$. We shall see from the discussion of the predictive power of the BSA in the next section, there will be a natural range of values for the parameter $\Lambda$ between $10^{14}$GeV and $10^{17}$GeV.

Even though the definition~\eqref{BSAdef} is quite simple and natural, it is absolutely not obvious that this structure correctly reproduces a bosonic action of the Standard Model. In the next subsection we demonstrate \emph{how} to extract the information from the definition~\eqref{BSAdef}.

\subsection{The Heat Kernel Expansion.}
The traditional approach to BSA is based on the heat kernel expansion. Presenting the generic cutoff function $\chi(z)$ as a superposition of decreasing exponents,
\be
\chi(z) = \int_{0}^{\infty} dt\,e^{-tz} \hat{\chi}(z), \la{LT}
\ee 
where $\hat{\chi}$ stands for the inverse Laplace transform (which we assume exists) of $\chi$, we see that it is sufficient to study the case
\be
\chi(z) = \exp{(-z)}. \la{expcutoff}
\ee 
The important point is the fact that the the square of the Dirac operator $\mathcal{D}^2$, which enters in the definition of the BSA~\eqref{BSAdef}, is a \emph{Laplace-type} operator. 

\subsubsection*{Laplace-type operators}
The literature on Laplace-type operators is sterminate. Here we present some facts on these operators,  necessary and hopefully sufficient  to understand  the forthcoming discussion. After we introduce the 
quantities  relevant for our  case $L = \mathcal{D}^2$.
In general (see e.g.~\cite{Vassilevich}) the Laplace type operator, is an operator, which acts on smooth sections of some vector bundle $V$
over the Riemannian manifold $\mathcal{M}$, and
which has the following structure: 
\be
L=-(g^{\mu\nu}_{\mathrm{E}}\nabla^{\mathrm{E}}_\mu\nabla^{\mathrm{E}}_{\nu} +E). \la{LaplGen}
\ee
In this formula $E$ is some endomorphism of $V$, 
and the covariant derivative:
\be
\nabla^{\mathrm{E}}_{\mu} = \partial_{\mu} + \sigma_{\mu}, \la{EuclCovDiv}
\ee
is defined by some connection $\sigma_{\mu}$ on $V$. In other words the Laplace-type operator is uniquely defined by the three entries: the Euclidean metric  $g^{\mu\nu}_{\mathrm{E}}$, the endomorphism $E$ and the connection $\sigma_{\mu}$. 
We also notice that the combination $g^{\mu\nu}_{\mathrm{E}}\nabla^{\mathrm{E}}_{\mu}\nabla^{\mathrm{E}}_{\nu}$ in~\eqref{LaplGen} contains the Christoffel symbols associated with $g^{\mu\nu}_{\mathrm{E}}$ (c.f.~\eqref{Christ}), since the first covariant derivative $\nabla^{\mathrm{E}}_{\mu}$ acts on a quantity carrying one coordinate index. One can say that the sections of the vector bundle over $\mathcal{M}$ are the multicomponent fields on  $\mathcal{M}$, whilst the endomorphism $E$ is a matrix valued function which acts on these multicomponent fields.

In our case $L = \mathcal{D}^2$, and the vector bundle $V$ is chosen so that its basic space 
is our (Euclidean) four dimensional ``spacetime" manifold $\mathcal{M}$, whilst its smooth sections
are elements of $\mathcal{H}$, 384-component fields on $\mathcal{M}$. 
The connection $\sigma_{\mu}$ is the one in the covariant derivative~ \eqref{fluctu2}. It contains the Euclidean spin-connection and 
the $SU(2)\times SU(2)\times U(1)$ gauge connection. In the basis~\eqref{HStruct} the connection $\sigma$ is given by 384 by 384 matrix valued function:
\be
\sigma_{\mu} = -\frac{\ii}{2}\omega_{\mu}^{\mathrm{E}}\otimes 1_{96} - 
   \ii\mathcal{A}^{\mathcal{H}}_{\mu}, \la{oursigma}
\ee
where $\omega_{\mu}^{\mathrm{E}}$ is the Euclidean spin-connection~\eqref{EuclSpinConnection},
whilst the gauge connection is 
\be
\ii \mathcal{A}^{\mathcal{H}}_{\mu} =  \ii \mathcal{A}_{\mu}\oplus  (\ii\mathcal{A}_{\mu})^*, \la{AHdef}
\ee
where $\mathcal{A}_{\mu}$ is the gauge connection of the Standard Model, which we introduced 
in Sect.~\ref{se:standardmodel}. 
The superscript ``$\mathcal{H}$" indicates that the action of the gauge connection $\mathcal{A}_{\mu}$ 
is promoted to the subspace $\mathcal{H}^c_R \oplus\mathcal{H}^c_L$, whose presence manifests the anti-charge doubling.
By definition $\mathcal{A}_{\mu}$ acts on $\mathcal{H}_R\oplus\mathcal{H}_L$
 as a 192 by 192 matrix valued function:
 \be
\mathcal{A}_{\mu}  = \mathcal{A}^{\boldsymbol{\mathrm{v}}_{\mathcal{R}}}_{\mu}
  \oplus \mathcal{A}^{\boldsymbol{\mathrm{e}}_{\mathcal{R}}}_{\mu}
  \oplus \mathcal{A}^{\boldsymbol{\mathrm{L}}_\mathcal{L}}_{\mu}
  \oplus \mathcal{A}^{\boldsymbol{\mathrm{u}}_{\mathcal{R}}}_{\mu}
  \oplus \mathcal{A}^{\boldsymbol{\mathrm{d}}_{\mathcal{R}}}_{\mu}
  \oplus \mathcal{A}^{\boldsymbol{\mathrm{Q}}_\mathcal{L}}_{\mu},
\ee
where various blocks are defined by Eq.~\eqref{gaugeconnectdef}. 
In this formula the superscripts indicate which subspaces are affected by the corresponding blocks:
for example\footnote{Since the gauge connection acts nontrivially only on the gauge indices, its action
 is insensitive to a presence or absence of the mirror doubling, which has to do with the spinorial chiral structure.
 Therefore a mentioning of $\mathcal{R}$ and $R$ (and  $\mathcal{L}$ and $L$) in this context together can not create any confusion.}
 $\mathcal{A}^{\boldsymbol{\mathrm{e}}_{\mathcal{R}}}_{\mu}$
 acts on $\boldsymbol{\mathrm{e}}_R$, $\mathcal{A}^{\boldsymbol{\mathrm{L}}_\mathcal{L}}_{\mu}$ acts
 on  $\boldsymbol{\mathrm{L}}_L$ and etc.
 Defining the ``gauge curvature" via
\be
\mathcal{F}^{\mathcal{H}}_{\mu\nu} = \partial_{\mu}\mathcal{A}^{\mathcal{H}}_{\nu} - \partial_{\nu}\mathcal{A}^{\mathcal{H}}_{\mu} + \left[\mathcal{A}^{\mathcal{H}}_{\mu},\mathcal{A}^{\mathcal{H}}_{\nu}\right], \la{gaugecurv}
\ee
we present the 384 by 384 matrix valued function (viz. endomorphism)  $E$:
\bea
E = -(\ii\gamma^{\mu}_{\mathrm{E}}\otimes 1_{96}) [\nabla_{\mu}^{\mathrm{E}},\gamma^5\otimes M ]- M^2 
- \frac{R}{4}\otimes \one_{\mathcal{H}} + \frac{\ii}{4}[\gamma^{\mu}_{\mathrm{E}}\otimes 1_{96},\gamma^{\nu}_{\mathrm{E}}\otimes 1_{96}]
\mathcal{F}_{\mu\nu}^{\mathcal{H}}. \la{EE}
\eea
We remind, the 384 by 384 matrix $M$, which contains all the scalar fields, is defined
after Eq.~\eqref{fluctu2}.  Hereafter in this section the Riemann tensor $R_{\sigma\rho\mu\nu}$, the Ricci tensor $R_{\mu\nu}$ and the scalar curvature $R$
are computed with the metric $g_{\mu\nu}^{\mathrm{E}}$ according to~\eqref{RS} and~\eqref{Christ}. 

\subsubsection*{Heat kernel trace and its asymptotic expansion}
With the choice of the exponential cutoff function~\eqref{expcutoff} the bosonic spectral action~\eqref{BSAdef} is the  trace of the heat operator associated with $L =\mathcal{D}^2$, or simply the heat kernel trace. For a generic Laplace-type operator~\eqref{LaplGen}, the heat operator
$K(t) = \exp\left(-tL\right)$ solves the initial value problem 
\be
\left\{\begin{array}{lcl}
\partial_t K(t) &=& -L K, \\
K(0) &=&  \mathbb{1}_{V}
\end{array}
\right.
\ee 
for the heat equation, where the parameter $t$, which for our purposes has to be identified with $\Lambda^{-2}$, is called for historical reasons ``proper time". The quantity $\mathbb{1}_{V}$ is the unit matrix in the bundle.
In our case $L = \mathcal{D}^2$, the unity $\mathbb{1}_V$  coincides with $\mathbb{1}_{\mathcal{H}} = \mathbb{1}_{384}$.

The following plays a key role: for generic Laplace-type operator $L$ and $d$-dimensional compact manifold $\mathcal{M}$ without boundary the heat kernel trace is well defined, and the following asymptotic \emph{heat kernel expansion} holds~\cite{Vassilevich} at arbitrary order $N$:
\be
{\mathrm{Tr}}\,\exp{\left(-tL\right)} \simeq \sum_{n=0}^{N} t^{n-\frac{d}{2}} a_{2n}\left(L\right) + \mathcal{O}\left(t^{N-\frac{d}{2}+1}\right), \la{hcexp}
\ee
where the quantities $a_{2n}(L)$ are the even\footnote{On manifolds without boundaries the odd heat kernel coefficients vanish.} heat-kernel (or DeWitt-Seeley-Gilkey) coefficients.  These coefficients are local polynomials of the Riemann tensor $R_{\mu\nu\eta\xi}$, the endomorphism $E$,  the ``curvature" $\Omega_{\mu\nu}$, which is defined in terms of the connection $\sigma_{\mu}$:
\be
\Omega_{\mu\nu}  = \partial_{\mu}\sigma_{\nu} - \partial_{\nu}\sigma_{\mu} + \left[\sigma_{\mu},\sigma_{\nu}\right],
\ee
and their covariant derivatives.  For our connection $\sigma$ given by~\eqref{oursigma} the ``curvature" $\Omega_{\mu\nu}$, with $\mathcal{F}^{\mathcal{H}}_{\mu\nu}$ defined by~\eqref{gaugecurv}, reads: 
\bea
\Omega_{\mu\nu} &=& \ii\mathcal{F}^{\mathcal{H}}_{\mu\nu} - \frac{1}{4}\gamma^{\sigma}_{\mathrm{E}}\gamma^{\rho}_{\mathrm{E}} 
R_{\sigma\rho\mu\nu} \otimes 1_{96}, \la{Oomega}
\eea

\begin{rem}The asymptotic heat kernel expansion~\eqref{hcexp} is valid at small proper time~$t$ only,  therefore we are in a low-energy  approximation. Otherwise, any finite ansatz of the expansion~\eqref{hcexp} would be a poor approximation of the heat kernel trace. At the end of this section (Eqs.~\eqref{HIGH} and~\eqref{LOW}) we present an explicit example illustrating this.
\end{rem}

\begin{rem}
Apart from the signature issue, we note that our construction necessitates an elliptic operator with \emph{discrete} spectrum, i.e.\ a compact space. Infrared compactification is usually a mere device necessary for the correct definition of operator. This may be naive, the infrared is being understood to play a fundamental role in general in field theory (see for example~\cite{Strominger, EOMAS, Scent}. Also in noncommutative field theory the issue is non trivial~\cite{mixing, Canfora:2015nsa,Vitale:2017nkp}, and the presence of boundaries for the Heat Kernel expansion results in a whole bunch of novelties, starting from parity anomalies~\cite{ConnesChamseddineboundary1, ConnesChamseddineboundary2, Vassilevichboundary, Kurkov:2017cdz, Kurkov:2018pjw}.
\end{rem}

The first three nontrivial heat kernel coefficients for a $d$-dimensional manifold without boundary are~\cite{Vassilevich}:
\begin{eqnarray}
a_0(L)&=&\frac{1}{(4\pi)^{d/2}}\int_{\mathcal{M}}\dd^d x\, \sqrt{g^{\mathrm{E}}}\,\,
\tr\one_{V},\nonumber\\
a_2(L)&=&\frac{1}{(4\pi)^{d/2}}\int_{\mathcal{M}}\dd^d x\, \sqrt{g^{\mathrm{E}}}\,\,
\tr\left(-\frac R6\, \one_{V}+{E}\right),\nonumber\\
a_4(L)&=&\frac{1}{(4\pi)^{d/2}}\frac{1}{360}\int_{\mathcal{M}}\dd^d x\, \sqrt{g^{\mathrm{E}}}\,\,
\tr( 5R^2 \,\one_{V} -2R_{\mu\nu}R^{\mu\nu} \,\one_{V}\nonumber\\
&&+2R_{\mu\nu\sigma\rho}R^{\mu\nu\sigma\rho}\,\one_{V}-60R{E}+180{E}^2+30\Omega_{\mu\nu}\Omega^{\mu\nu}), \label{spectralcoeff}
\end{eqnarray}
where ``$\tr$" stands
for the trace over bundle indices.
The heat kernel coefficients $a_n$ are universal: all the integrands in~\eqref{spectralcoeff} are \emph{not} sensitive to a particular shape of the manifold $\mathcal{M}$.
In our case $d=4$, hence we set $N = 2$ in the asymptotic expansion~\eqref{hcexp}. This way the contributions of the higher heat kernel coefficients will be suppressed by inverse powers of the UV cutoff $\Lambda$ in the low-energy regime. By ``low-energy regime" we mean the following. The bosonic background is chosen so that various bosonic fields  and their derivatives are much smaller than the corresponding powers of  $\Lambda$, for the Higgs field this results in $H^{\dagger}H \ll \Lambda^2$, $D_{\mu}H^{\dagger}D^{\mu}H \ll \Lambda^4$, etc.

In conclusion we see that if one uses a generic cutoff function $\chi$, instead of the exponential cutoff,  the low-energy asymptotic expansion
is still valid, albeit in a slightly different form.
Substituting the heat kernel expansion~\eqref{hcexp} in the decomposition~\eqref{LT} one finds that at $d = 4$ the bosonic spectral action~\eqref{BSAdef} exhibits the following asymptotic 
low-energy expansion in the inverse powers of the cutoff scale $\Lambda$
\ba
S_B \simeq \sum_{n=0}^{N} f_n\,\Lambda^{4-2n} a_{2n}\big(\mathcal{D}^2\big) + \mathcal{O}\left(\Lambda^{-2N +2}\right), \la{SBexp}
\ea
where the $f_n$ are the momenta of $\hat\chi$, defined by~\eqref{LT}:
\begin{eqnarray}
f_0&=&\int_0^\infty \dd x\, x \, \chi(x),\nonumber\\
f_2&=&\int_0^\infty \dd x  \,  \chi(x),\nonumber\\
f_{2n+4}&=&(-1)^n \del^{n}_x \chi(x)\big|_{x=0} \ \ n\geq 0.
\label{fcoeff}
\end{eqnarray}
In the case  $\chi = e^{-z}$ all these numbers, obviously, are equal to one.

\subsection{Asymptotically expanded Bosonic Spectral Action.}
The bosonic action of the Standard Model derives from the first three nontrivial heat kernel coefficients. Setting $N = 2$ in the expansion~ \eqref{hcexp} and truncating the remaining part 
we define the \emph{asymptotically expanded} BSA as the \emph{finite} ansatz of the asymptotic expansion~\eqref{SBexp}
\be
S_{B,4} := \Lambda^4 f_0 \,a_0(\mathcal{D}^2) +\Lambda^2 f_2\, a_2(\mathcal{D}^2) 
+ \Lambda^0 f_4\, a_4(\mathcal{D}^2), \label{anz}
\ee
where the subscript $4$ indicates that we took into account just the heat kernel coefficients up to $a_4$.  The terminology ``asymptotically expanded spectral action" was introduced  in~\cite{WvS1,WvS2}. Strictly speaking all the phenomenological studies~\cite{spectralaction,AC2M2,resilience,CCVPati-Salam} of the spectral action were devoted exactly to this asymptotically expanded BSA. Note that it is not necessary to truncate the expansion~\eqref{hcexp} at $N=2$: the contribution of a larger but \emph{finite} number of the higher heat kernel coefficients was also
considered~\cite{Nomadi}. 

Applying the general formulas~\eqref{spectralcoeff} to $L = \mathcal{D}^2$  one finds:
\bea
a_0\left(\mathcal{D}^2\right) &=& \frac{24}{\pi^2}\, \int_{\mathcal{M}} d^4x\,\sqrt{g^{\mathrm{E}}},  \nonumber\\
a_2\left(\mathcal{D}^2\right)&=& \frac{1}{\pi^2}\int_{\mathcal{M}} d^4x\,\sqrt{g^{\mathrm{E}}} \,\left(2R - 3y_{1} H^{\dagger}H 
- \frac{y_{2}}{2} \mathrm{M}_R^2 \right), \nonumber \\
a_4\left(\mathcal{D}^2\right) &=&\frac{1}{2\pi^2}\int_{\mathcal{M}} d^4x\,\sqrt{g^{\mathrm{E}}} \left({3 y_{1}}\left(D_{\mu}H^{\dagger}D^{\mu}H - \frac{R}{6}H^{\dagger}H\right)  
-\frac{ y_{2} }{12} R\,\mathrm{M}_R^2 \right.
\nonumber\\ 
& +& 3 z_{1}\left(H^{\dagger}H\right)^2  
+ \frac{z_{2}}{2}\,\mathrm{M}_R^4 + 2 z_{3}\left(H^{\dagger}H\right)\mathrm{M}_R^2   \nonumber\\
 &+&  g_3^2 G_{\mu\nu}^j G^{\mu\nu\,j} + g_2^2 W_{\mu\nu}^{\alpha} W^{\mu\nu\,\alpha}
+ \frac{5}{3}g_1^2 B_{\mu\nu} B^{\mu\nu} \nonumber\\
&+& \left. \frac{11}{30}\, \mathrm{GB} - \frac{3}{5}C_{\mu\nu\eta\lambda}C^{\mu\nu\eta\lambda}
\right), \la{hcres}
\eea
where $\mathrm{GB}$ denotes the Gauss-Bonnet density:
\be
\mathrm{GB}\equiv \frac{1}{4}\epsilon^{\mu\nu\rho\sigma}\epsilon_{\alpha\beta\gamma\delta}R^{\alpha\beta--}_{--\mu\nu}
R^{\gamma\delta--}_{--\rho\sigma}, \la{GB}
\ee
and 
\be
C_{\mu\nu\eta\lambda}C^{\mu\nu\eta\lambda} = R_{\mu\nu\eta\lambda}R^{\mu\nu\eta\lambda} - 2 R_{\mu\nu}R^{\mu\nu} - \frac{1}{3}R^2
\ee
stands for the square of the Weyl tensor.
The quantities $G_{\mu\nu}^j$, $W_{\mu\nu}^{\alpha}$ and  $B_{\mu\nu}$, defined by~\eqref{gaugestrength}, stand for the field strength of the $SU(3)$, $SU(2)$ and $U(1)$ gauge fields respectively.
The numbers $y_r$, $z_s$, $r=1,2$, $s=1,2,3$ are combinations of the Yukawa couplings:
\begin{eqnarray} 
{y_{1}} &{\equiv}& {{\rm tr} \left(\left[\hat y_u \hat y_u^{\dagger}\right] + \left[\hat y_d \hat y_d^{\dagger}\right]
+\frac{1}{3}\left[\hat Y_u \hat Y_u^{\dagger}\right] + \frac{1}{3}\left[\hat Y_d \hat Y_d^{\dagger}\right] \right)}, \nonumber\\
{y_{2}} & {\equiv}& { {\rm tr} \left(\hat y_M \hat y_M^{\dagger}\right)}, \nonumber
\\
%
{z_{1} }&{\equiv}&{ {\rm tr} \left(\left[\hat y_u \hat y_u^{\dagger}\right]^2 + \left[\hat y_d \hat y_d^{\dagger}\right]^2
+\frac{1}{3}\left[\hat Y_u \hat Y_u^{\dagger}\right]^2 + \frac{1}{3}\left[\hat Y_d \hat Y_d^{\dagger}\right]^2 \right)},\nonumber\\
{z_{2}} &{\equiv}&{ {\rm tr} \left(\hat y_M \hat y_M^{\dagger}\right)^2}, \nonumber\\
{z_{3}} &{\equiv}&{ {\rm tr} \left(\hat y_M \hat y_M^{\dagger}\right)\left(\hat Y_u \hat Y_u^{\dagger}\right)}.
\end{eqnarray} 
Substituting~\eqref{hcres} in~\eqref{anz} we arrive at the following expanded BSA:
\bea
S_{B,4} &=& \int_{\mathcal{M}} d^4 x\,\sqrt{g^{\mathrm{E}}}\,\left(\alpha_1 + \alpha_2 R + \alpha_3 C_{\mu\nu\eta\lambda}C^{\mu\nu\eta\lambda} + \alpha_4 \mathrm{GB} \right. \nonumber\\
&+&  \alpha_5 G_{\mu\nu}^j G^{\mu\nu\,j} + \alpha_6 F_{\mu\nu}^{\alpha} F^{\mu\nu\,\alpha}
+ \alpha_7 B_{\mu\nu} B^{\mu\nu} \nonumber\\
&+&\left.\alpha_8\left(D_{\mu}H^{\dagger}D^{\mu}H - \frac{R}{6}H^{\dagger}H\right)+ \alpha_9\, H^{\dagger}H   + \alpha_{10}\left(H^{\dagger}H\right)^2 \right) . \la{BSAans}
\eea
This action contains everything the standard model may wish, and even more: on the one side the actions for the gauge fields~ \eqref{GaugeLagr}, and for the Higgs scalar~\eqref{ScLagr} and~\eqref{ScPot}) are there (the second and the third lines respectively), but also the gravitational action (the first line) emerges from this formalism as well.
We emphasise that all the constants $\alpha_1$,...,$\alpha_{10}$ \emph{are fixed}, in terms of the ``fermionic input", which enters through the Dirac operator $\mathcal{D}$ via the gauge couplings, the Yukawa matrices and the $\mathrm{M}_R$ scale. This is in particular true for the coefficients of the Higgs field. Moreover, by construction (Eq.~\eqref{BSAdef}, Eq.~\eqref{anz} ) the answer~\eqref{BSAans} depends on the UV cutoff scale $\Lambda$ and  the cutoff function $\chi$ via the first three
momenta of its inverse Laplace transform~\eqref{fcoeff}.

The dimensionful constants
\bea
\alpha_1  &=& \frac{1}{\pi^2} \left( 24 f_0 \Lambda^4 - \left(\frac{y_2}{2}\right)f_2\Lambda^2 \mathrm{M}_R^2 + \left(\frac{z_4}{4}\right)f_4\mathrm{M}_R^4\right),
\nonumber\\ 
\alpha_2 &=& \frac{1}{\pi^2} \left(2f_2 \Lambda^2 - \frac{1}{24}y_2 f_4 \mathrm{M}_R^2\right), \la{c12}
\eea
fix the structure $\alpha_1 + \alpha_2 R$, which is nothing but the Einstein-Hilbert action (c.f.~Eq.~\eqref{GraLagr}) for gravity with the cosmological constant. The dimensionless constants
\bea
\alpha_3 &=& -\frac{1}{\pi^2}\frac{3}{10}\,f_4,
\nonumber\\
\alpha_4 &=& \frac{1}{\pi^2}\frac{11}{60} \,f_4, \la{c3_4}
\eea
fix the quadratic gravitational terms $\alpha_3 C_{\mu\nu\eta\lambda}C^{\mu\nu\eta\lambda} + \alpha_4 \mathrm{GB}$. Note that the Gauss-Bonnet term being topological does not affect the classical equations of motion, while the cosmological consequences of the Weyl square term  were studied in the spectral action context in~\cite{Mairi,Mairireview}.

The constants $\alpha_5$,..., $\alpha_{10}$, given by
\bea
\alpha_5 &=& \frac{g_3^2}{2\pi^2}\,f_4, \nonumber\\
\alpha_6 &=& \frac{g_2^2}{2\pi^2}\,f_4, \nonumber\\
\alpha_7 &=& \frac{5}{3}\frac{g_1^2}{2\pi^2}\,f_4, \nonumber\\
\alpha_8 &=& \frac{3y_1}{2\pi^2} f_4, \nonumber\\
\alpha_9 &=& \frac{1}{\pi^2}\left( -3y_1 f_2 \Lambda^2 + z_3f_4\mathrm{M}_R^2 \right),\nonumber\\
\alpha_{10} &=& \frac{3 z_1}{2\pi^2}f_4, \la{c5_10}
\eea 
constrain  the bosonic action of the standard model. In Sec.~\ref{se:comparison} we shall see that these constrains restrict the parameters of the Standard Model, in particular the Higgs quartic coupling will not be the independent parameter, but will come out from the spectral data. Moreover the constraints point at the interpretation of the spectral action: it has to be identified with the classical action of the standard model at the unification scale\footnote{As we shall discuss later, since experimentally there is no precise unification, one may use  a range of scales $10^{14}$GeV - $10^{17}$GeV.}. To obtain predictions the renormalisation group flow must be considered. 
6 we will demonstrate how to Wick rotate the asymptotically expanded BSA to the Lorentzian signature.

The asymptotically expanded BSA plays an important role as far as the low-energy phenomenology is concerned. Nevertheless the UV behaviour of the original BSA~\eqref{BSAdef} substantially differs from its asymptotically expanded version~\eqref{anz},  and it is quite interesting from the pointless geometry perspective. Therefore, before going to the phenomenological consequences of the spectral action and its predictive power, we pause to take a closer look at the original BSA~\eqref{BSAdef}.

\subsection{Beyond the low momenta approximation \la{BLMapprox}}
Here, following~\cite{Kuliva}, we apply less common but more sophisticated technique in order to extract the information from the bosonic spectral action~ \eqref{BSAdef}. The results regarding the gauge sector 
were obtained and studied in~\cite{ILV1,ILV2}.
We will show that this object 
exhibits two qualitatively different behaviours, with transition scale is given by  $\Lambda$.
While the low momenta regime of the BSA reproduces the asymptotically expanded BSA, which describes the Standard Model non minimally coupled with gravity,
the high momenta behaviour appears to be drastically different. We will see that exchange of high momenta bosons is impossible. 
The latter, due to the uncertainty principle, makes impossible  measurement of  distances smaller than the inverse cutoff scale $\Lambda^{-1}$,  
pointing to a scale in which {there may be present} a transition to a different geometric phase.

A generic contribution to the expansion~\eqref{hcexp} to order $ t^n = \Lambda^{-2n}$ has the following structure:
\be
\Lambda^{-2n}(\mbox{contribution}) = \int d^4 x \,\sqrt{g^{\mathrm{E}}}\,\,\left(\frac{\mbox{powers of fields, powers of $\partial$} }{\Lambda^{2n}}\right)
\ee
where powers of the cutoff scale $\Lambda$ in denominator are compensated by powers of fields and their derivatives in numerator.
Higher heat kernel coefficients contain higher derivatives of fields, and higher powers of fields, but  at low energies their contribution {can be neglected. In this way the BSA recovers Standard Model bosonic Lagrangian.}

We need to distinguish the notions of low/high momenta and low/high energy regimes, and we focus on the \emph{momenta} dependence. By momenta we mean to momenta of Fourier modes of various bosonic fields.  The low and high energy regimes are understood here comparing \emph{various} dynamical quantities of the positive energy dimension with the corresponding powers of the cutoff scale  $\Lambda$. In particular the high energy regime can be achieved in various ways: one can either consider highly oscillating fields (i.e.\ high momenta) or
 large amplitudes of gauge and scalar fields\footnote{These fields have the same energy dimension as  $\Lambda$. The gravitational field $g_{\mu\nu}$ is dimensionless. }, without requiring rapid oscillations. To avoid confusions, we emphasise that in the present  consideration  the word ``energy" is used exactly in the sense, described above, in particular, it  has nothing to do with the 0-th component of the 4-momenta.

We want to study the propagation of \emph{free} bosons in the spectral approach at arbitrary, in particular, high momenta. For this purpose we should compute the 
BSA up to quadratic order in fields, \emph{summing all derivatives}.
Barvinsky and Vilkovisky~\cite{Barvinsky:1990up} obtained a resummation of the heat
kernel expansion, that allows to derive linear equations of motion, valid for both high and low momenta regions.

\subsubsection*{Barvinsky-Vilkovisky expansion}
For a generic Laplace-type operator $L$ both the heat kernel expansion~ \eqref{hcexp} and the Barvinsky and Vilkovisky expansion involve the same ingredients: the proper time
$t$ and the ``curvatures" $\mathcal{Q} = (E, \Omega_{\mu\nu}, R_{\mu\nu\lambda\sigma})$. There is, however, a substantial difference.
Whilst the former is an expansion in powers of the proper time $t$, the latter
is an expansion in powers of the ``curvatures" $\mathcal{Q}$:
\begin{eqnarray}
&&\Tr\,{\exp{\left(- t L\right)}} \simeq \frac{1}{(4\pi t )^2} \int d^4x \sqrt{g^{\mathrm{E}}}\, {\rm tr}\, \Big[ 1 + t P
+ t^2\bigl( 
R_{\mu\nu} f_1\left(- t{\nabla^2}\right) R^{\mu\nu} +   Rf_2\left(-t{\nabla^2}\right)R
\nn\\
&& 
 + \,Pf_3\left(-t{\nabla^2}\right)R +P\,f_4\left(-t{\nabla^2}\right) P 
+ \Omega_{\mu\nu} f_5\left(-t{\nabla^2}\right)\Omega^{\mu\nu}
\bigr) \Big]  
+ \mathcal{O}\left(\mathcal{Q}^3\right)
.
 \la{BV}
\end{eqnarray}
In this formula $P\equiv E+\tfrac 16 R$,  $f_1,...,f_5$ are known functions:
\begin{eqnarray}
&&f_1(\xi)=\frac{h(\xi)-1+\tfrac 16 \xi}{\xi^2}\,,\qquad f_5(\xi)=-\frac {h(\xi)-1}{2\xi}\,,\nn\\
&&f_2(\xi)=\tfrac 1{288} h(\xi) -\tfrac 1{12} f_5(\xi) - \tfrac 18 f_1(\xi)\,,\qquad
f_3(\xi)=\tfrac 1{12} h(\xi) - f_5(\xi)\,,\nn\\
&&f_4=\tfrac 12 h(\xi)\, , 
\end{eqnarray}
\\
and
\begin{equation}
h(z):=\int_0^1 d\alpha \,e^{-\alpha (1-\alpha)\,z}\,. \label{h} 
\end{equation}

For illustrative purposes in what follows we will discuss a simplified BSA corresponding to a single fermion, interacting with abelian gauge, real scalar and gravitational fields.
We will compute for this special case the righthand side of~\eqref{BV} and confront the low and high momenta behaviours.
 
\subsubsection*{Simplified model}
Below we consider the simplified BSA with the exponential cutoff 
\be
S_{b} =\mathrm{Tr}\, \exp{\left(-\frac{\mathcal{D}^2}{\Lambda^2}\right)} \la{BSAtoy}
\ee
where the Dirac operator 
\be
\mathcal{D}  = i\gamma^{\mu}\nabla_{\mu} + \gamma_5 \phi .\la{Dtoy}
\ee
is a simplified version of the fluctuated Dirac operator~\eqref{fluctu2}.
 In order to understand \emph{why} the real BSA has the announced nontrivial behaviour this simplified BSA is sufficient. 
In~\eqref{Dtoy} $\phi$ stands for the real scalar field - the ``simplified" version of the Higgs field.  
The connection~\eqref{oursigma}
\be
\sigma_{\mu} = -\frac{\ii}{2}\omega_{\mu}^{\mathrm{E}} - 
   \ii {A}_{\mu}\cdot 1_4, \la{simpoursigma}
\ee
which enters in the covariant derivative~ \eqref{EuclCovDiv} contains both the spin-connection
and the abelian gauge connection $A_{\mu}$.
Defining  (c.f.\ Eq.~\eqref{gaugecurv})
\be
F_{\mu\nu}  =  \partial_{\mu}A_{\nu} - \partial_{\nu}A_{\mu} ,
\ee
one can easily check that in such a setup
\be
E = \frac{i}{4}\left[\gamma^{\mu},\gamma^{\nu}\right]
F_{\mu\nu} - \frac{R}{4}\,1_4 - \phi^2\,1_4 
 - i\gamma^{\mu}\gamma_5\phi_{;\mu}, \la{E}
\ee
and
\be
\Omega_{\mu\nu} =iF_{\mu\nu}\,1_4 - \frac{1}{4}\gamma^{\sigma}\gamma^{\rho}R_{\sigma\rho\mu\nu}. \la{Omega}
\ee
These formulas are nothing but the simplified versions of~\eqref{EE} and~\eqref{Oomega} correspondingly.
Substituting the expressions~\eqref{E} and~\eqref{Omega} for  $E$ and   $\Omega$ 
in the righthand side of~\eqref{BV} we obtain 
\begin{gather}
S_{b} \simeq \frac{1}{32\pi^2}\int\, d^4x\,\sqrt{g^{\mathrm{E}}}\left\{\right. 
 8\Lambda^4 - \Lambda^2\left(\frac{2}{3}R + 8\phi^2\right)\nn \\
 + R\left[8 f_2 \left(-\frac{\nabla^2}{\Lambda^2}\right) - \frac{2}{3}f_3 \left(-\frac{\nabla^2}{\Lambda^2}\right)
+ \frac{1}{18}f_4 \left(-\frac{\nabla^2}{\Lambda^2}\right)\right]R \nn \\
+ 8 \,R_{\mu\nu}  f_1 \left(-\frac{\nabla^2}{\Lambda^2}\right) R^{\mu\nu} - 
R_{\mu\nu\rho\sigma}\,f_5 \left(-\frac{\nabla^2}{\Lambda^2}\right) R^{\mu\nu\rho\sigma}\nn \\
+\phi^2\,\left[ - 8f_3 \left(-\frac{\nabla^2}{\Lambda^2}\right)+\frac{4}{3}f_4 \left(-\frac{\nabla^2}{\Lambda^2}\right)\right] R 
  + 8\, \phi\left[ - \nabla^2 f_4 \left(-\frac{\nabla^2}{\Lambda^2}\right)\right]\phi  
 + 8\, \phi^2 f_4 \left(-\frac{\nabla^2}{\Lambda^2}\right)\phi^2 \nn \\
\left. + F_{\mu\nu}\left[4f_4 \left(-\frac{\nabla^2}{\Lambda^2}\right) 
- 8 f_5 \left(-\frac{\nabla^2}{\Lambda^2}\right)\right]F^{\mu\nu} \right\}  + \mathcal{O}(\mathcal{Q}^3).\la{genres}
\end{gather}
This is the main result of an application of the Barvinsky-Vilkovisky expansion to the BSA~\eqref{BSAtoy}. 
Dependences on all possible momenta is ``captured" by the form factors $f_1,..,f_5$.
Let us take a closer look at the low and high momenta regimes.

Expanding at small $\xi$ the formfactors $f_1(\xi),..,f_5(\xi)$, which stand in~\eqref{genres},  what corresponds to the low-momenta regime, we arrive to  
\ba
S_{b}
&\simeq 
& \frac{1}{32\pi^2}\int\,d^4x\,\sqrt{g}\left\{
8\Lambda^4 - \Lambda^2\left(  8\phi^2 - \frac{2 R}{3}\right)\right. \nn\\
& +& 4\phi\left(-\nabla^2 - \frac{R}{6}\right)\phi + 4\phi^4 + \frac{4}{3}F_{\mu\nu}F^{\mu\nu} \nn \\
&-&\left. 
   \frac{1}{10}C_{\mu\nu\rho\sigma}C^{\mu\nu\rho\sigma} + \frac{1}{60}\mathrm{GB}
 \right\}
. \la{LOW}
 \ea
One can check that the anszatz~\eqref{LOW} correctly 
reproduces the heat kernel anszatz~\eqref{anz} up to the Gauss-Bonnet term~\eqref{GB}\footnote{In~\cite[Sect.~8]{Barvinsky:1990up} it is explained, 
that this Gauss-Bonnet term is actually $\mathcal{O}(\mathcal{Q}^3)$, therefore there is no contradiction with the heat kernel expansion.}. Nevertheless,
the Gauss-Bonnet term, being topological, does not affect classical equations of motion, therefore at low momenta~\eqref{genres} reproduces correctly all the classical dynamics, which comes out from the asymptotically expanded BSA $S_{B,4}$. 
In particular the free propagators of all the bosonic fields: they are the standard ones, i.e.\ the
low momenta bosons propagate in a standard way.

Consider now the high momenta regime of~\eqref{genres}.
Since we are interested in the propagators of the free bosonic fields we consider the quadratic part of the action only.
For the gravitational field we
consider fluctuations of the metric tensor over the flat background,
imposing the transverse and traceless gauge fixing condition 
\be
g_{\mu\nu} = \delta_{\mu\nu} + h_{\mu\nu}, \quad h^{\mu}_{\mu}\equiv \delta^{\mu\nu}h_{\mu\nu}=0,\quad \partial_{\mu}h^{\mu\nu} = 0.
\ee
For the gauge field we
impose the transversal gauge fixing condition
\be
\partial_{\mu}A^{\mu} = 0.
\ee
 Expanding the formfactors $f_1(\xi),..,f_5(\xi)$ at large $\xi$ we obtain for the quadratic part of the BSA
\be
S_{b}^{(2)}  \simeq  
\frac {\Lambda^4}{16\pi^2 } \int d^4x \left[ - \frac{3}{2}\,  h_{\mu\nu}h^{\mu\nu}
+ 8 \phi \frac 1{-\partial^2} \phi + 8 F_{\mu\nu} \frac 1{(-\partial^2)^2} F^{\mu\nu} \right],  \la{HIGH}
\ee
The low and high momenta regimes of BSA, are completely different. While the low momenta regime leads to
the standard propagators, at high momenta
the action does not contain positive powers of derivatives: in this regime high momenta particles do not propagate.

In conclusion we notice that the content of this subsection is relevant to the dynamics of the classical bosonic fields. The  nontrivial UV behaviour~\eqref{HIGH} implies that bosonic propagators do not decay at high momenta, therefore the quantum theory is not well defined, for once is not renormalizable, even if one treats the gravitational sector at the classical level only, see the discussion in~\cite{CCVPati-Salam} and~\cite{zeta}. Some sort of the UV completion may necessary, or a more drastic change of order parameter.  Another important aspect is the Wick rotation to the Lorentzian signature. It is easy to ``Wick rotate" any finite number of the heat kernel        coefficients (i.e.\ ``the asymptotically expanded BSA"), however it is not clear what to do with the complete BSA~\eqref{BSAdef}.

\begin{rem}
Presenting a generic smooth cutoff function as a superposition of decreasing exponents weighted with its inverse Laplace transform\footnote{We assume, of course, that the cutoff function is chosen in a way, that such a representation exists.}, we see that the high momenta behaviour~\eqref{HIGH} holds also in this case: high momenta bosons do not propagate.
This nontrivial behaviour can be seen as a resummation of the asymptotic expansion~\eqref{SBexp}, and it gives a qualitatively different behaviour from than any finite anzatz of this
expansion: the kinetic term of bosonic spectral action vanishes at high momenta. 
On the other side, in the case of the sharp cutoff the expansion contains just three nonzero terms, therefore
it does not make sense to discuss any resummation. Such an anszatz, obviously, grows up at large momenta.
Nevertheless, a careful analysis in~\cite{ILV1}  shows that this anszatz describes the behaviour of the left-hand side of~\eqref{SBexp} at small momenta compared to $\Lambda$ only, at high momenta the kinetic terms of the bosonic spectral action vanish, see the discussion is~\cite[Sect.~4.2]{ILV1}.
\end{rem}

\subsection{Comments on the non-compact case.}

Even thought in the spectral approach the manifold $\mathcal{M}$ is assumed to be compact, the physical applications definitely require a \emph{non-compact} space time. The non-compactness creates an infrared problem.
Let us clarify the origin of this issue.

For the sake of simplicity we restrict ourselves to the simplified model \eqref{BSAtoy}, assuming that the Riemann curvature\footnote{In particular we assume that the metric tensor has the structure $g_{\mu\nu}^{\mathrm{E}} = \delta_{\mu\nu} + h_{\mu\nu}(x)$, where the fluctuations $ h_{\mu\nu}$ over the flat background together with their derivatives fall off sufficiently fast at infinity.} $R^{\mu\nu\lambda\xi}$, the gauge curvature $F^{\mu\nu}$ and the scalar field $\phi$ decay fast at infinity. Unfortunately, even in such a simple setup, the right-hand side of  \eqref{BSAtoy} does not exist.
The heat kernel expansion \eqref{hcexp} points at the source of the problem: the 0-th coefficient (c.f. Eq.~\eqref{spectralcoeff}) is nothing but the volume of $\mathcal{M}$, which is infinite for the non-compact manifold. 
On the other side, the higher heat kernel coefficients, being local polynomials of $R^{\mu\nu\lambda\xi}$, $F^{\mu\nu}$ and  $\phi$, are well defined quantities.
In~\cite{ILV1} the following IR regularisation of the bosonic spectral action has been proposed:
\be
S_{b} =\mathrm{Tr}\,\left( \exp{\left(-\frac{\mathcal{D}^2}{\Lambda^2}\right)} - \exp{\left(-\frac{\mathcal{D}_0^2}{\Lambda^2}\right)} \right), \la{BSAtoyBis}
\ee
with $\mathcal{D}_0  := \ii \delta^{\nu}_A \gamma^{A}_{\mathrm{E}}\partial_{\nu}$. In such a construction the unpleasant IR divergences in the two terms cancel each other. 
It is remarkable, that the second  ``IR-regularising term" in \eqref{BSAtoyBis} does not affect the equations of motion. 
In conclusion we notice that the generalisation of \eqref{BSAtoyBis} for the arbitrary cutoff function $\chi$ is obvious: one has to replace $\exp(...)$ by $\chi(...)$.

\section{Physical constraints from Noncommutative Geometry \label{se:comparison}}

Here we discuss  the how the spectral action principle can restrict possible phenomenological values of physical quantities..
All the parameters of the bosonic spectral action~\eqref{BSAans} are not arbitrary numbers, but obey the constraints~\eqref{c12},~\eqref{c3_4} and~\eqref{c5_10}. Let us clarify what these constrains physically mean.

First of all let us normalise the gauge contribution to~\eqref{BSAans}, 
in a canonical way and Wick rotate it to the Lorentzian signature.
Performing the rescaling of the gauge potentials
\begin{equation}
G_{\mu}^{j} \longrightarrow\frac{1}{2} \frac{1}{\sqrt{\alpha_5}}G_{\mu}^j,
\quad W_{\mu}^{\alpha} \longrightarrow \frac{1}{2} \frac{1}{\sqrt{\alpha_6}}W_{\mu}^{\alpha},
\quad B_{\mu}\longrightarrow \frac{1}{2} \frac{1}{\sqrt{\alpha_7}} B_{\mu},
\end{equation}
together with the Wick rotation, which we discussed in Sect.\ref{se:Wicketc}, we arrive to the gauge Lagrangian of the Standard Model\begin{equation}
L_{\rm gauge}  =  -\frac{1}{4}G_{\mu\nu}^j G^{\mu\nu\,j} - \frac{1}{4} F_{\mu\nu}^{\alpha} F^{\mu\nu\,\alpha}
- \frac{1}{4}B_{\mu\nu} B^{\mu\nu}.
\end{equation}
Since both the fermionic and the bosonic actions come out from the same Dirac operator $\mathcal{D}$, the same rescaling must be performed in the vector-spinor couplings of the  fermionic action, what is equivalent to setting in~\eqref{gaugeconnectdef}:
\begin{equation}
g_3 = \frac{\pi}{2} \sqrt{\frac{2}{f_4}}, \quad g_2  = \frac{\pi}{2} \sqrt{\frac{2}{f_4}},
\quad g_1 = \frac{\pi}{2} \sqrt{\frac{3}{5}} \sqrt{\frac{2}{f_4}}, \la{unif}
\end{equation}

The relation~\eqref{unif} tells us that the gauge couplings must unify, while we know from the experiment, that the three interactions of the Standard Model: the strong, the electro-weak and the electro-magnetic have quite different strength at the energy scales, which are accessible  for the accelerators. On the other side we also know, that the the gauge couplings ``run" with the growth of energy according to the RG equations, which at one loop have the following form:
\be
\frac{\dd g_i(\mu)}{\dd \log\mu} =  \frac{1}{16\pi^2}\beta_i(g_1,g_2,g_3) , \quad i = 1,2,3,
\ee
where $\beta_i$ stand for the beta functions of the gauge couplings, whose explicit form can be found in~\cite{Machacek1, Machacek2, Machacek3}. It is remarkable that at the one loop level, which is usually used in the NCG context, these beta functions depend on on the gauge couplings only, not on the Yukawa or Higgs self-interaction constants.

From the experimental data we know that  some sort of the ``approximate unification"  occurs at the energies $10^{14}-10^{17}$ GeV, as shown in Fig.~\ref{GRun}. Therefore the relation~\eqref{unif} hints us how to \emph{postulate} the physical interpretation of the spectral action principle. This identifies the scale (or at least the range of scales) in which the BSA is written\footnote{Strictly speaking  the cutoff scale $\Lambda$ and the normalisation scales have different origin, and a priori there is no necessity to identify them. Nevertheless the philosophy of the spectral approach suggests to minimize the number of parameters, therefore we identify these two scales \emph{by construction}.}.

From now on we follow this paradigm and rewrite~\eqref{unif} as the \emph{initial} condition for the RG equations~\eqref{gaugeRG} at the scale $\Lambda$.
\begin{equation}
g_2^2\left(\Lambda\right) = g_3^2\left(\Lambda\right) = \frac{5}{3}g_1^2\left(\Lambda\right) = \frac{1}{f_4} \frac{\pi^2}{2}.
\label{unif1}
\end{equation}
Although the exact unification does not take place for the present version of the asymptotically expanded BSA~\eqref{anz}, one can consider the higher heat kernel coefficients, as it was done in~\cite{Nomadi}. These alter the RG flow and upon a proper (fine) tuning of the parameters may result in a precise unification. Note that~\eqref{unif1} imposes a constraint between the gauge couplings at the unification scale and the ``cutoff-function input" $f_4$~\eqref{fcoeff}.

Consider now the scalar sector of the model. Rescaling the Higgs field
\begin{equation}
H \longrightarrow \frac{1}{\sqrt{ \alpha_8}}\, H, \label{Hresc}
\end{equation} 
and performing the Wick rotation to the Lorentzian signature we arrive at the Higgs Lagrangian of the Standard Model 
\begin{equation}
L_H =  D_{\mu} H^{\dagger} D^{\mu}H - \frac{R}{6} H^{\dagger} H + \mu^2 H^{\dagger} H - \lambda\left(H^{\dagger} H\right)^2,
\end{equation}
where the Higgs mass parameter $\mu^2$ and the quartic coupling constant $\lambda$ are given by
\begin{equation}
\mu^2 \left(\Lambda\right) =    \frac{\alpha_9}{\alpha_8} \quad ; \quad
\lambda \left(\Lambda\right) = \frac{\alpha_{10}}{\alpha_8^2}. \la{quartcons}
\ee

It is important, that the rescaling~\eqref{Hresc} has to be performed in the scalar-spinor coupling of the fermionic action~\eqref{SFE} as well. 
Since according to the spectral action principle all the coefficients
in (properly normalized) spectral action have to be identified with the corresponding running constants taken at the scale
$\Lambda$, we arrive to the following relations
\begin{equation}
\hat{y}_{u}\left(\Lambda\right) = \frac{1}{\sqrt{ \alpha_{8}}}\cdot \hat{y}_{u}\left(\Lambda\right),\ \hat{y}_{d}\left(\Lambda\right) = \frac{1}{\sqrt{ \alpha_{8}}}\cdot \hat{y}_{d}\left(\Lambda\right),
\ \hat{Y}_{u}\left(\Lambda\right) = \frac{1}{\sqrt{ \alpha_{8}}}\cdot \hat{Y}_{u}\left(\Lambda\right), \ \hat{Y}_{d}\left(\Lambda\right) = \frac{1}{\sqrt{ \alpha_{8}}}\cdot \hat{Y}_{d}\left(\Lambda\right),
\end{equation}
which makes sense iff
\begin{equation}
\alpha_8 = 1. \la{simprel}
\end{equation}
Using the explicit expression for $\alpha_8$ and taking into account the constraint~\eqref{unif1} between the gauge couplings and the parameter $f_4$ we arrive to the following
``unification" relation between the Yukawa and the gauge couplings:
\begin{eqnarray}
{\rm tr}\left[
 \hat{y}_{u}^{\dagger}\left(\Lambda\right)\hat{y}_{u}\left(\Lambda\right)
+\hat{y}_{d}^{\dagger}\left(\Lambda\right)\hat{y}_{d}\left(\Lambda\right)
+\frac{1}{3}\hat{Y}_{u}^{\dagger}\left(\Lambda\right)\hat{Y}_{u}\left(\Lambda\right)
+\frac{1}{3}\hat{Y}_{d}^{\dagger}\left(\Lambda\right)\hat{Y}_{d}\left(\Lambda\right)
\right] \ = \frac{4}{3} g_2^2\left(\Lambda\right) 
\la{unif2}
\end{eqnarray}
In conclusion we notice that using the formula~\eqref{simprel} we can simplify~\eqref{quartcons}:
\be
\lambda\left(\Lambda\right)  = \frac{z_1}{y_1} = \frac{{ {\rm tr} \left(\left[\hat y_u \hat y_u^{\dagger}\right]^2 + \left[\hat y_d \hat y_d^{\dagger}\right]^2
+\frac{1}{3}\left[\hat Y_u \hat Y_u^{\dagger}\right]^2 + \frac{1}{3}\left[\hat Y_d \hat Y_d^{\dagger}\right]^2 \right)}}{ {{\rm tr} \left(\left[\hat y_u \hat y_u^{\dagger}\right] + \left[\hat y_d \hat y_d^{\dagger}\right]
+\frac{1}{3}\left[\hat Y_u \hat Y_u^{\dagger}\right] + \frac{1}{3}\left[\hat Y_d \hat Y_d^{\dagger}\right] \right)}}. \la{unif3}
\ee

The constrains~\eqref{unif1},~\eqref{unif2} and~\eqref{unif3} for the initial data for the RG flow is the crucial predictive power of the spectral action principle. Constraining the quartic coupling $\lambda$ at the scale $\Lambda$ one finds its value at the TeV scale, solving the RG equations\footnote{A complete set of the one loop  RG equations for the Standard Model can be found in~\cite{Machacek1, Machacek2, Machacek3}}.

Comparing the result with the known Higgs vacuum expectation value one \emph{predicts} the mass of the Higgs boson. The value predicted is between 167 and 172~GeV~\cite{WalterBook}, with the mainly uncertainty due to ambiguity of the choice of choice of the unification point. This value was a genuine prediction made in~\cite{spectralaction, AC2M2} \emph{before} LHC measured it to be smaller. While the prediction if clearly not satisfied, it is nevertheless remarkable that a physical theory, based on first mathematical principles, obtains a result not too distant from  the experimental one. Note in addition that the prediction is  made at one loop, and under the assumption that there is no new physics between the scale probed at LHC and the unification scale. In the next sections we will take the constructive point of view that the experimental results must a stimulus to improve the model, still keeping the mathematical roots.

There have been
several proposals in this sense, and some of them are reviewed in~\cite{Schucker}. In particular in~\cite{Stephan} it was proposed  that the presence of an extra scalar field, corresponding to the breaking of a extra U(1) symmetry, can bring down the mass of the Higgs to 126~GeV. This model however  contains extra fermions. Earlier examples of extensions are in~\cite{PrisSchucker,PaschkeScheckSitarz,SchuckerZouzou,Stephan:2005uj,Stephan:2007fa,Squellari:2007zr,Stephan:2009vm}. What is common to these extensions is the enlargement of the Hilbert space, i.e.\ the introduction of extra fermions. While this in not in principle a negative aspect, it is preferable to keep the number of type of fermions to be the presently known ones. Their intricate choice of quantum numbers, with the highly nontrivial cancellation of anomalies, suggests a fundamental role of the structure. Moreover, as we will see, they naturally fit into a Pati-Salam kind of structure. Another reason to keep the present fermionic structure is the fact that there is a (generalised) Hodge duality which play an important role in the understanding of the symmetries of the Dirac operator~\cite{FrancescoLudwik, Dabrowski:2017dct, Dabrowski:2017hfh, Dabrowski:2018ugg}.

Shortly after the original measurement of the Higgs mass, in~\cite{resilience} it was noticed that the presence of an extra field, present in the Dirac operator in the position occupied by the neutrino Majorana mass, could lower the mass of the Higgs, thus rendering the model compatible with experiment, at the price of a loss of predictive power, since a scale should be given to this field. 

This is done ruling out the hypothesis
of the ``big desert'' and considering an additional scalar field $\sigma$
that lives at high energies and gives mass to the Majorana
neutrinos. Explicitly $\sigma$ is obtained in~\cite{resilience} by turning (inside the finite dimensional part
  $D_F$ of the Dirac
  operator) the constant-entry $y_R$ of the
  Majorana matrix $\M_R$ into a field:
\begin{equation}
  \label{eq:68}
  y_R\to y_R\sigma(x)
\end{equation}
The extra field changes the renormalization flow, which will be not dictated by a Lagrangian which contains also the couplings of $\sigma$ with the Higgs\footnote{Different versions of the extension have Lagrangian qualitatively equal, but with possibly different coefficients.}:
\begin{align}\label{SBconsigma}S_{\mathrm{B}} & ={\displaystyle \frac{24}{\pi^{2}}f_{0}\Lambda^{4}\int d^{4}x\sqrt{g}}\nonumber\\
 & \,\,\,-\frac{2}{\pi^{2}}f_{2}\Lambda^{2}\int d^{4}x\sqrt{g}\left(R+\frac{1}{2}a\overline{H}H+\frac{1}{4}c\sigma^{2}\right)\nonumber\\
 & \,\,\,{\displaystyle +\frac{1}{2\pi^{2}}f_{4}\int d^{4}x\sqrt{g}\left[\frac{1}{30}\left(-18C_{\mu\nu\rho\sigma}^{2}+11R^{*}R^{*}\right)+\frac{5}{3}g_{1}^{2}B_{\mu\nu}^{2}+g_{2}^{2}(W_{\mu\nu}^{\alpha})^{2}+g_{3}^{2}(V_{\mu\nu}^{m})^{2}\right.}\nonumber\\
 & \,\,\,+\left.\frac{1}{6}aR\overline{H}H+b(\overline{H}H)^{2}+a|\nabla_{\mu}H_{a}|^{2}+2e\overline{H}H\sigma^{2}+\frac{1}{2}d\sigma^{4}+\frac{1}{12}cR\sigma^{2}+\frac{1}{2}c(\partial_{\mu}\sigma)^{2}\right]
\end{align}
With the new field the mass of the Higgs is lowered, and made compatible with the experiment, at the price of the loss of predictive power because there is a new parameter which is loosely constrained. A detailed analysis of the phenomenology is beyond the scope of this review, but we signal that it will be similar to Pati-Salam, with input coming from NCG, and we refer to~ \cite{CCvS, CCVPati-Salam, Chamseddine:2015ata} for a specific analysis,  and to~\cite{Aydemir:2013zua, Aydemir:2014ama, Aydemir:2015nfa, Aydemir:2016xtj,Aydemir:2018cbb} for a more general view.

The origin of the field $\sigma$ is quite different from the Higgs. The
latter, like the other bosons, are components of the gauge potential
$\mathbb A$. They are obtained from the commutator of $D_F$.
 with
the algebra: $D_F$ has
constant components, that is without
manifold dependence, but when these numbers
are commuted with elements of the algebra they give rise to the
desired bosonic fields. One could hope to obtain $\sigma$ in a
similar way, by considering $y_R$ as a Yukawa coupling. As explained
in~\cite{coldplay}, the problem is that in taking the
commutator with elements of the algebra ${\cal A}_{sm}$, the
coefficient $y_R$ does not contribute to the
potential because of the first
  order condition. This forced the authors
of~\cite{resilience} to ``promote to a field'' only the entry $y_R$, in a somewhat arbitrary
way. Indeed the components of $D_F$ cannot all be fields to start with, otherwise the model
would loose its predictive power, in that all Yukawa couplings would
be fields, and the masses of all fermions would run independently,
thus making any prediction impossible.
In the following (sections \ref{se:tripleforsm} and
\ref{grandbreaking}) we show that there is a way to obtain the field
$\sigma$ from $y_R$ by a fluctuation of the metric, provided one starts with
an algebra larger than the one of the standard model. Another problem is that  the  mass of this extra field should be of the order of $10^{11}$~GeV. On the other side, for the authors of~\cite{resilience} perform the renormalization group flow in an interval, $10^2-10^{10}$~GeV, where the extra field is decoupled. These considerations called for a rethinking of the problem. Another field could solve the problem, but it should emerge more naturally, and the renormalization group flow should take into account the possibility of having extra symmetries. There is a further way to introduce the field, using outer automorphism~\cite{Farnsworth:2014vva}. This is less ad hoc, but we will not discuss it. We will instead pass to the description of changes of the symmetry to accomodate the extra field as a connection.

\section{The grand symmetry model \label{se:grandalgebra}}

In this section we will present a method of finding the correct mass of the Higgs, based on an extension of the symmetries of the model first presented in~\cite{coldplay, confgrand}.   In the next section we will discuss an alternative scheme based on the violation of the order one condition.

\subsection{Mixing spinorial and internal degrees of freedom} 
\label{mixing}

The total Hilbert space ${\cal H}$ of the almost commutative geometry~\eqref{almostspectraltriple} is the tensor product of four
dimensional spinors by the 96-dimensional elements of $\mathcal
H_F$. Any of its element is a $\mathbb C^{384}$-vector valued
function on $\M$. From now on, for simplicity, we work with $N=1$ generation only, and
consider instead $384/3 = 128$ components vector. The total Hilbert space
can thus be written - at least in a local trivialization - in two ways:
\begin{equation}
\HH  = \mathrm{sp}(\mathcal{M}) \otimes \mathcal H_F = L^2(\M)\otimes \mathsf H_F
\label{eq:10}
\end{equation}
where $\sf H_F\simeq \mathbb C^{128}$ takes  into account both external (i.e.\ spin) and internal (i.e.\ particle) degrees of
freedom. 

For the purposes of this section it is useful to introduce another notation which can take into account the various representations of the groups in compact way 
We label
 the basis of $\sf H_F$ with four indices as $s\dot s\sC\sI\alpha$ according to the following:
\begin{itemize}
\item[$\spinind s ,\dotspinind s$]  describe the Dirac spinor with a double index notation: $\spinind{s=r,l}$ runs over the right, left
  parts  and $\dotspinind{s=\dot 0, \dot 1}$ over
  the particle, antiparticle parts of the spinors. 

\item[$\partind {C}$] differentiates 
  ``particles'' ($\partind{C=0}$) from ``antiparticles''  ($\partind{
    C=1}$).

\item[$\sI$] is a ``lepto-colour'' index: $\sI = 0$ identifies leptons while $\colorind
  I=1,2,3$ are the three colours of QCD.

\item[$\flavind \alpha$] is the flavour index. It runs over the set $u_R,d_R,u_L,d_L$ when $\colorind{I=1,2,3}$, and $\nu_R,e_R,\nu_L,e_L$ when $\colorind{I=0}$. 
\end{itemize}
As we said we suppress a further generation index, as it will not play any role in the following. It can easily be reinstated in the end.

We will represent a vector $\Psi\in\HH$ explicitly as a multi index-spinor as:
$\Psi^{\mathsf C \colorind I}_{\spinind s  \dotspinind s
  \flavind{\alpha}}\in L^2(\M)$. The position of the indices is arbitrary: $\Psi$ evaluated
at $x\in\M$ is a column vector, so all the
indices are row indices. An element $A$ in ${\cal B}(\cal H)$ is a
$128\times 128$ matrix whose coefficients are function of $M$, and
carries the indices
\begin{equation}
  \label{eq:97}
  A = A_{\sD s \sJ \dot s \alpha}^{\sC t \sI \dot t \beta}
\end{equation}
where $\sD, t, \sJ, \dot t, \beta$ are column indices with the same
range as $\sC, s, \sI, \dot s, \alpha$.

This choice of indices yields the chiral basis for the Euclidean
Dirac matrices:{\footnote{The multi-index $st$ after the closing parenthesis is
      to recall that the block-entries of the $\gamma$'s matrices
      are labelled by indices $s,t$ taking values in the set
      $\left\{l,r\right\}$. For instance the $l$-row, $l$-column block
      of $\gamma^5$ is $\mathbb I_2$. Similarly the entries of the
      $\sigma$'s matrices are labelled by $\dot s, \dot t$ indices
      taking value in the set $\left\{\dot 0, \dot 1\right\}$: for
      instance ${\sigma^2}^{\dot 0}_{\dot 0} ={\sigma^2}^{\dot 1}_{\dot 1} = 0$. }}
\begin{equation}
  \label{eq:3}
\gamma^\mu_E=\left(\begin{array}{cc}  0_2
  & {\sigma^\mu} \\ {\tilde\sigma^\mu}&  0_2
\end{array}\right)_{st},\quad
\gamma^5_E=\gamma^1_E\gamma^2_E\gamma^3_E\gamma^0_E =\left(\begin{array}{cc}  \mathbb I_2
  & 0_2\\ 0_2& - \mathbb I_2
\end{array}\right)_{st},
\end{equation}
where for $\mu = 0,1,2,3$ one defines
\begin{equation}
  \label{eq:1}
  \sigma^\mu =\left\{\mathbb I _2, -i\sigma_i, \right\},\quad
  \tilde\sigma^\mu =\left\{\mathbb I_2, i\sigma_i\right\}
\end{equation}
with $\sigma_{i}$, $i=1,2,3$ the Pauli matrices. Explicitly,
\begin{equation}
\nonumber
 \label{eq:43}
\sigma^0 =\mathbb I_2,\; \sigma^1 = - i \sigma_1 =  \left(\begin{array}{cc} 0 & -i \\ - i & 0
\end{array}\right)_{\dot s \dot t},\;
\sigma^2 = - i \sigma_2 =  \left(\begin{array}{cc} 0 & -1 \\ 1 & 0
\end{array}\right) _{\dot s \dot t},\;
\sigma^3 = - i \sigma_3 =  \left(\begin{array}{cc} -i & 0 \\ 0 & i
\end{array}\right) _{\dot s \dot t}.
\end{equation}

 The free Dirac operator $\slashed D$ extended  to ${\cal H}$
by  $\slashed D \otimes{\mathbb I}_F$ acts as {\footnote{We use Einstein summation on alternated up/down
    indices. For any $n$ pairs of indices $(x_1,y_1)$, $(x_2, y_2)$, ...
    $(x_n, y_n)$, we write
$\delta_{x_1x_2... x_n}^{y_1y_2... y_n}$ instead of
$\delta_{x_1}^{y_1}\delta_{x_2}^{y_2}... \delta_{x_n}^{y_n}$. For the
tensorial notation to be coherent, $\ds$ and $\gamma^\mu_E$ should carry lower
$s\dot s$ and upper $t \dot t$ indices. We systematically omit them to
facilitate the reading.}} 
 \begin{equation}
   \label{eq:3-bis}
\slashed D \otimes{\mathbb I}_F=\,\delta^{\sC\sI\beta}_{\sD\sJ\alpha}\slashed D =
 i\left(\begin{array}{cc}\delta^{\sI\beta}_{\sJ\alpha}\,\gamma^\mu_E\nabla_\mu^S&
     0_{64}\\0_{64}&
 \delta^{\sI\beta}_{\sJ\alpha}\,\gamma^\mu_E\nabla_\mu^S\end{array}\right)_{\sC\sD}.
 \end{equation}
 while indicating $D_{M_R}$ the Majorana part of the Dirac operator:
 
 \begin{equation}
\label{majoranadiracoperator}
D_{M_R}:= \gamma^5_E D_{F_{(Majorana)}} = \eta_s^t\,\delta_{\dot s}^{\dot t} \Xi_{\sJ\alpha}^{\sI\beta} \left(\begin{array}{cc} 0 & y_M \\ \bar y_M& 0\end{array}\right)_{\sC\sD}, \text{  with  }  \Xi :=\left(\begin{array}{cc}
1 &0 \\ 0 &0_3 \end{array}\right)
\end{equation}
 
In tensorial notation, the charge conjugation
operator is
\begin{equation}
J= i\gamma^0_E\gamma^2_E cc=i\left(
\begin{array}{cc} 
  {{\overline\sigma}^2}&0_2\\ 
  0_2 & {\sigma^2}
\end{array}\right)_{s t} \, cc = -i\eta_s^t \tau_{\dot s}^{\dot t}\, cc,
\label{eq:2}
\end{equation}
while
\begin{equation}
J_{F}=\left(\begin{array}{cc}
0 & \mathbb{I}_{16}\\
\mathbb{I}_{16} & 0
\end{array}\right)_{\sC\sD} cc,
\end{equation}
hence
\begin{equation}
({\cal J}\Psi)^{\partind C\colorind I}_{s\dotspinind s
  \flavind{\alpha}} = - i \eta^t_s\,\tau^{\dot t}_{\dot s}\, \xi^\sC_\sD \, \delta^{\sI\beta}_{\sJ\alpha}\,
\bar \Psi^{\partind D \colorind J
}_{t \dotspinind t \flavind{\beta}}
\label{eq:12}
\end{equation}
where for any pair of indices $x,y \in [1,..., n]$ one defines
\begin{equation}
  \label{eq:22}
  \xi^x_y=\left(\begin{array}{cc} 0_n & \mathbb I_n \\ \mathbb I_n &
    0_n \end{array}\right),\quad \eta^x_y=\left(\begin{array}{cc} \mathbb
    I_n & 0_n \\ 0_n&
   - \mathbb I_n \end{array}\right), \quad \tau^x_y=\left(\begin{array}{cc} 0_n & -\mathbb I_n \\ \mathbb I_n &
    0_n \end{array}\right).
\end{equation}
The chirality acts as $\gamma^5_E = \eta_s^t \delta_{\dot s}^{\dot t}$
on the spin indices, and as $\gamma_F = \eta^\sC_\sD\,\delta^\sI_\sJ\,\eta_\alpha^\beta $  on the internal indices:
\begin{equation}
\label{tensorJgamma}
(\Gamma\Psi)^{\partind C\colorind I}_{s\dotspinind s
  \flavind{\alpha}} = \eta_s^t \delta_{\dot s}^{\dot
t}\, \,\eta^\sC_\sD\,\delta^\sI_\sJ\,\eta_\alpha^\beta \; \Psi^{\partind D\colorind J}_{t\dotspinind t
  \flavind{\beta}}.
\end{equation}

\subsection{The grand algebra}
\label{se:tripleforsm}

Let us restart  from the most general finite algebra that satisfies all the conditions for the noncommutative space to be a manifold, Eq.~(\ref{genericalgebra}). 
The standard model coupled with gravity is described by the case $a=2$. 
The case $a=3$  
would require a 72-dimensional Hilbert space, and there is no obvious
way to build it from the particle content of the standard model. The next case, $a=4$, requires the Hilbert space to have dimension 128, which is the dimension of $\mathsf
H_F$, as defined in~\eqref{eq:10}. Said in an other way, considering together the spin and internal degrees of freedom as part of the ``grand
Hilbert space'' $\mathsf H_F$ gives precisely the number of dimension to
represent the \emph{grand algebra}
\be
{\cal A}_G=\mathbb M_4({\mathbb H})\oplus \mathbb M_8({\mathbb C}).
\ee
This means that $\cinf\otimes\mathcal A_G$ can be represented on the same
Hilbert space $\cal H$ as $\cinf\otimes{\cal A}_F$. The only difference
is that one needs to factorize $\cal H$ in~\eqref{eq:10}
as $L^2(\M)\otimes \mathsf H_F$ instead of $sp({\cal M})\otimes{\cal H}_F$. It is a remarkable ``coincidence''
that the passage from the standard model to the grand algebra, namely
from $a=2$ to $a'=4=2a$, requires to multiply the dimension
of the internal Hilbert space by $4$ (for $2(2a')^2= 2(4a)^2 =4(2(2a)^2)$) which
is precisely the dimension of spinors in a spacetime of dimension $4$. 
Once more we stress that no new particles are introduced: ${\cal A}_F$ acts on ${\cal
  H}_F={\mathbb C}^{32}$, $\mathcal A_G$ acts on
 $\mathsf H_F= {\mathbb C}^{128}$ but 
 $C^\infty(\M)\otimes {\cal A}_G$ and   $C^\infty(\M)\otimes {\cal{
   A}}_F$  acts on the same Hilbert space $\cal H$. Since the Hilbert space is not changed, either the Dirac operator will remain the same as in the standard model case, eq.~\eqref{DFbig}.

The representation of the grand algebra $\mathcal A_G$ on $\mathsf H_F$ is more
involved than the one of ${\cal A}_F$ on ${\cal H}_F$, given in~\eqref{algebraconproiettori}. 
In analogy with what was done earlier we consider an element of
${\cal A}_G$ as
two $8\times 8$ matrices, and see both of them having a block
structure of four  $4\times 4$ matrices. Thus the component $Q\in M_4(\mathbb H)$ of the grand algebra
gets two new extra indices with respect
  to the quaternionic component of $\cal A_F$, and the same is
  true for $M\in{\mathbb M}_8(\mathbb C)$. For the complex matrices we choose to identify these two new indices with the
  spinor (anti)-particles indices $\dot 0, \dot 1$; and for the quaternions with the spinor
  left-right indices $r,l$ introduced in~\eqref{eq:97}. This choice is not unique but, in all the cases, having
  both sectors diagonal on different indices ensures
  that the order zero condition is satisfied, as explained below.

We therefore have 
  \begin{equation}
    \label{eq:16}
  Q= \left(
\begin{array}{cc}
Q_{r \alpha}^{r \beta} & Q^{l \beta}_{r \alpha}\\
Q_{l \alpha}^{r \beta} &Q_{l \alpha}^{l \beta}
\end{array}\right)_{s  t}\in\mathbb M_4(\mathbb H) , \quad   M= \left(
\begin{array}{cc}
 M^{\dot 0 \sI}_{\dot 0 \sJ} &  M^{\dot 1 \sI}_{\dot 0 \sJ}\\
M^{\dot 0 \sI}_{\dot 1 \sJ} & M^{\dot 1 \sI}_{\dot 1 \sJ}
\end{array}\right)_{\dot s  \dot t}\in \mathbb M_8(\mathbb C)
  \end{equation}
where, for any
$\dot s, \dot t\in\left\{\dot 0, \dot 1\right\}$ and
$s,t\in\left\{l,r\right\}$, the matrices 
\begin{equation}
  \label{OldRepresentation}
  Q_{ s\alpha}^{t\beta }\in \mathbb M_2(\mathbb H),\quad
M^{\dot t\sI}_{\dot s \sJ} \in \mathbb M_4(\mathbb C).
\end{equation}
This means that
the representation of the element $A=(Q,M)\in{\mathcal A}_G$ is{\footnote{To
  take into account the non-diagonal action of $Q$ and $M$, we adopt the order: $C,s,I,\dot s, \alpha$.}}:
\be
A^{\sC \, t \,\sI \dot t\beta}_{\sD s \sJ \dot s\alpha} = \left(
\delta^\sC_0 \delta_{\dot s}^{\dot t} \delta^\sI_\sJ 
\,{Q}^{t \beta}_{s \alpha} +
\delta^\sC_1 M^{\dot t\sI}_{\dot s \sJ}\delta_{s}^{t}\delta_\alpha^\beta\right).\label{repa5}
\ee

As in the standard model case the quaternionic part acts on the particle sector of the
internal indices ($\delta^\sC_0$) and the complex part
on the antiparticle sector ($\delta^\sC_1$). The difference 
is that the grand algebra acts in a nondiagonal way
not only on the flavour and lepto-colour indices $\flavind\alpha,
\colorind I$, but also on the $s$ and  $\dot s$ indices. The novelty
is in this mixing of internal and spacetime indices: at the grand algebra level, the spin structure
  is somehow hidden. Specifically, the representation~\eqref{repa5} is not
  invariant under the action of the Lorentz group, or
  $Spin(4)$ since we are dealing with spin representation, in
  euclidean signature.

 The representation of $C^\infty(\M)\otimes{\cal A}_G$ is given  by~\eqref{repa5}  where the entries of $Q$ and $M$ are now functions on $\M$. Since the total Hilbert space $\cal H$ is unchanged, there is no reason to change the real structure and the grading. In
particular one easily checks that the order zero condition holds true for
the grand algebra
\begin{equation}
\label{orderzerogrand}
\left[A, {\cal J}B{\cal J}^{-1} \right]=0\quad\forall A,B\in\mathcal{A}_{G}.
\end{equation}
This is because the real structure $\cal J$ acts as the charge conjugation
operator~\eqref{eq:2}
on the spinor indices, and as $J_F$, eq.~\eqref{JF}, in $\mathcal H_F$ (where it exchanges the two blocks corresponding to particles
and antiparticles). In tensorial notations one
has
\begin{equation}
({\cal J}\Psi)^{\partind C\colorind I}_{s\dotspinind s
  \flavind{\alpha}} = - i \eta^t_s\,\tau^{\dot t}_{\dot s}\, \xi^\sC_\sD \, \delta^\sI_\sJ\,
\delta^\beta_\alpha\, \bar \Psi^{\partind D \colorind J
}_{t \dotspinind t \flavind{\beta}}
\end{equation}
where we use Einstein summation and define
\begin{equation}
  \label{eq:22}
  \xi=\left(\begin{array}{cc} 0 & 1 \\ 1 &
    0 \end{array}\right)_{\sC\sD},\quad \eta=\left(\begin{array}{cc} 1
     & 0 \\ 0&
   - 1 \end{array}\right) _{st}, \quad \tau=\left(\begin{array}{cc} 0 & -1 \\ 1 &
   0 \end{array}\right) _{\dot s \dot t}.
\end{equation}
Hence $\cal J$ preserves the indices structure in~\eqref{repa5}, apart from the exchange
$\delta^\sC_0\leftrightarrow\delta^\sC_1$: since $Q$ and $M$ act 
on different indices, the commutation~\eqref{orderzerogrand} is assured. Notice that without the
enlargement of the action of the finite dimensional algebra to the
spinorial indices, it would have been impossible to find a
representation of $\mathcal A_G$ which satisfies the order zero condition,
unless one adds more fermions. In this respect the grand algebra is not anymore an internal algebra in the usual sense, i.e.\ as acting on the matrix part of an almost commutative spin geometry.

\subsection{The Majorana coupling and the $\sigma$ field \label{grandbreaking} } 

In this section we show how the grand algebra makes possible to
have a Majorana mass giving rise to the field $\sigma$. Although the
calculations are quite involved, the principle is quite simple. Since
we have a larger algebra, the Majorana Dirac operator needs not be diagonal in
the spin indices. This added degree of freedom enables the possibility
to satisfy the order one condition in a non trivial way, namely to
still have a one form which commutes with the opposite algebra, but
that at the same time gives rise to a field.

We now show, as an example, the effects of the grading condition on the grand algebra, leading to a first reduced algebra.  In a very similar way is it possible to show how  the first order condition induced by
$D_{M_R}$  gives rise to the reduction $\mathcal A_G \to \mathcal A''_G$ of the grand algebra, as summarized in~\eqref{eq:50}. With $\mathcal A''_G$ and $D_{M_R}$ we can generate the field $\sigma$ as required by~\eqref{eq:68}. Finally the $1^\text{st}$-order condition induced by the free Dirac operator $\slashed D \otimes{\mathbb I}_F$ gives back the Standard Model.

\subsubsection{Example: reduction due to grading \label{gradingreduction} }
The grading condition imposes a reduction $\mathcal{A}_{G}\to
{\cal A}'_G$ where
 \be
{\cal A}'_G = \left(\mathbb H^l_L\oplus \mathbb H^r_L\oplus\mathbb H^l_R\oplus  \mathbb H^r_R\right)\oplus \mathbb M_8(\mathbb C) .
\label{AprimeG}
\ee
To see it, recall that the chirality $\Gamma$ in~\eqref{tensorJgamma} acts as $\gamma^5_E = \eta_s^t \delta_{\dot s}^{\dot t}$
on the spin indices, and as $\gamma_F$  on the internal indices:
\begin{equation}
(\Gamma\Psi)^{\partind C\colorind I}_{s\dotspinind s
  \flavind{\alpha}} = \eta_s^t \delta_{\dot s}^{\dot
t}\, \,\eta^\sC_\sD\,\delta^\sI_\sJ\,\eta_\alpha^\beta \; \Psi^{\partind D\colorind J}_{t\dotspinind t
  \flavind{\beta}}
\end{equation}
where $\eta^\sC_\sD$ and $\eta^\beta_\alpha$ are defined as in
\eqref{eq:22}.
Changing the order of the indices so that to match~\eqref{repa5}, one has
\begin{equation}
\label{Gammashort}
\Gamma= 
\eta^\sC_\sD \, \eta_s^t \,\delta^\sI_\sJ \,\delta_{\dot s}^{\dot t}\,\eta_\alpha^\beta.
\end{equation}
Since the representation of ${\cal A}_G$ is diagonal in the $\sC$ index, the
grading condition is satisfied if and only if it is satisfied by
both sectors - quaternionic and complex - independently.

In particular, the biggest
  subalgebra of $\cinf \otimes\A_G$ that satisfies the grading condition $[\Gamma, A]=0$
 and has bounded commutator with $\slashed D \otimes{\mathbb I}_F$ is the
  left-right algebra  $\A_{LR}$ given by
  \begin{equation}
\mathcal{A}_{LR}:=\mathbb{H}_L\oplus\mathbb{H}_R\oplus\mathbb{M}_{4}(\mathbb{C}),
\end{equation} 
In fact by~\eqref{tensorJgamma}, for
the quaternionic sector $[\Gamma, A]=0$ amounts to asking 
 $[\eta_s^t\eta_\alpha^\beta,Q_{s\alpha}^{t\beta}]=0$. This  imposes 
\begin{equation}
  Q = \left( 
    \begin{array}{cc}
      Q^r_r & 0_4 \\ 
      0_4 & Q^l_l
    \end{array}\right)_{st} 
    \label{Qdiagonal}
\end{equation}
where 
\begin{equation}
   Q^{r}_{r}=\left(
     \begin{array}{cc}
       {q^r_R}& 0_2 \\
       0_2 & {q^r_L}
     \end{array}
  \right)_{\alpha\beta}, \; Q^{l}_{l}=\left(
     \begin{array}{cc}
       {q^l_R}& 0_2 \\
       0_2 & {q^l_L}
     \end{array}
  \right)_{\alpha\beta} \quad
\text{ with } q^r_R, q^r_L, q^l_R, q^l_L\in \bb H.
\end{equation}
For complex valued matrices the requirement is $[\delta^{\dot t \sI}_{\dot s
  \sJ}, M^{\dot t \sI}_{\dot s \sJ}]=0$, which is trivially
satisfied. Hence the grading condition $[\Gamma,
A]=0$ imposes the reduction of $\A_G$ to
\begin{equation}
{\cal  A}'_{G} :=(\bb H^l_L \oplus \bb H^r_L \oplus \bb H^l_R \oplus \bb H^r_R) \oplus
M_8(\bb C). 
\end{equation}
For $A=(Q, M)\in \cinf\otimes{\cal A}'_{G}$, the boundedness of the commutator\footnote{\label{footnotenotations}To lighten notation, we omit the trivial indices in the
    product (hence in the commutators) of operators. From~\eqref{Qdiagonal} one knows that $Q$ carries the indices
    $s\alpha$ while $\gamma^\mu_E$ carries $s\dot s$, hence
    $[\slashed \del, Q]$ carries indices $s\dot s\alpha$ and should be
    written $[\delta_\alpha^\beta\slashed\del,  \delta_{\dot s}^{\dot t} Q]$.
Likewise, $[\slashed \del, M]$ carries indices $s\dot s I$ and holds for $[\delta_I^J\slashed \del, \delta_{s}^{t} M]$.}
\begin{equation}
   [\slashed D \otimes{\mathbb I}_F, A] =\left(
\begin{array}{cc}
 \delta^I_J \,[\slashed\del,  Q]&0_{64} \\ 
 0_{64} &\delta_\alpha^\beta\, [\slashed \del, M]
\end{array}
\right)_{\partind{CD}}
 \end{equation}
means that 
\begin{equation}
[\ds, Q] = i\gamma^\mu_E (\nabla_\mu^S Q)  +i [\gamma^\mu_E, Q]\nabla_\mu^S \quad\text{ and }\quad
[\ds, M] = i\gamma^\mu_E (\nabla_\mu^S M) +i[\gamma^\mu_E, M]\nabla^S_\mu
\end{equation}
are
bounded. This is obtained if and only if $Q$ and $M$ commute with all the Dirac
matrices, i.e.\ are proportional to $\delta_{s\dot s }^{t\dot t}$. For $Q$ this means 
$Q_r^r = Q_l^l$, hence the reductions
\begin{equation}
\bb H^r_R \oplus \bb H^l_R\to \bb H_R, \quad \bb H^r_L \oplus \bb H^l_L \to
\bb H_L.
\end{equation}
For $M$, this means that all the components $M_{\dot s}^{\dot t}$ in
\eqref{eq:16} are equal, i.e.\ the reduction
\begin{equation}
\label{reduc}
M_8(\bb
C)\to M_4(\bb C).
\end{equation}
Therefore ${\cal A}'_{G}$ is reduced to the usual Pati-Salam algebra $\A_{LR}$, acting diagonally on spinors.

\noindent The reduction of $\A_G$ to
  the algebra of the standard model, due to the first order condition, is summarized as follows (for details see~\cite{coldplay}):
\begin{framed}
\vspace{-.5truecm}

  \begin{eqnarray}
    \label{eq:50}
\mathcal{A}_{G}&=& M_4(\mathbb{H}) \oplus M_8(\mathbb C) \\[.25truecm]
\nonumber 
&\Downarrow& \hspace{0truecm}\text{grading condition}\\[.25truecm] 
\nonumber
{\cal A}'_G &=& \left(\mathbb H^l_L\oplus \mathbb H^r_L\oplus\mathbb H^l_R\oplus  \mathbb H^r_R\right)\oplus \mathbb M_8(\mathbb C)\\[.25truecm]  
\nonumber 
&\Downarrow& \hspace{0truecm}\text{$1^\text{st}$-order for the Majorana-Dirac operator $D_{M_R}$}\\ [.25truecm] 
   \nonumber
    \A''_G &=& (\mathbb H^l_L \oplus \mathbb H^r_L \oplus \mathbb C^l_R \oplus \mathbb C^r_R) \oplus (\mathbb C^l\oplus M_3^l(\mathbb C) \oplus \mathbb C^r \oplus M^r_3(\mathbb C))
 \text{with }\mathbb C^r_R = \mathbb C^r=\mathbb C^l\\[.25truecm] 
\nonumber &\Downarrow&\hspace{0truecm} \text{$1^\text{st}$-order for
  the free Dirac operator $\slashed D\otimes{\mathbb I}_F$}\\[.25truecm] 
\nonumber
\A_{sm} &=& \mathbb C \oplus \mathbb H \oplus M_3(\mathbb C) \\[-1truecm] 
\vspace{-1.75truecm}
\end{eqnarray}
\end{framed}

In $\mathcal{A}''_G$ with three of the four complex algebras identified, $\mathbb C^r_R = \mathbb C^r=\mathbb C^l$ the $\sigma$ field will be given by
the difference of two elements of the remaining complex algebras $\mathbb C^r_R ~\text{and}~ \mathbb{C}^l_R$:

\begin{equation}
\sigma\thicksim y_{R}(c_{R}-c\text{\textasciiacute}_{R}),\text{ with }c_{R}\in\mathbb{C}^r_{R}\text{ and }c'_{R}\in\mathbb{C}^l_{R}
\end{equation}

\subsection{Twisting the Grand Symmetry}
\label{section:twisting}

Although the first and second order conditions are satisfied,
as explained in~\cite{Devastato:2015bya}, the elements of the grand symmetry do not form a spectral triple because the commutator $[\slashed D \otimes{\mathbb I}_F, A]$ of any
of its element with the free Dirac operator is 
never bounded. This is a consequence of the noncommutativity of the Dirac's $\gamma$ with the internal algebras~\cite[Eq.\ (5.3)]{coldplay}. This is not a technical mathematical requirement, it means that it is impossible represent one-forms on the Hilbert space. 
In order to have bounded commutators, the action of $\A_G$ on spinors has to be trivial.  
In other term, to build a spectral triple with the
grand algebra ($a=4$ in~\eqref{genericalgebra}), one has to consider its subalgebra given by
$a=2$, that acts without mixing spinorial and internal indices. But then we would not have a grand algebra and a solution of the problem of the mass of the Higgs. The alternative~\cite{Devastato:2015bya} is to instead consider \emph{twisted spectral
triples}. They have been introduced in~\cite{Connes:1938fk} 
to solve the problem of the unboundedness of the commutator in a different context.
The twist  also permits to understand the breaking to the standard model as a dynamical process induced by the spectral action, as conjectured in~\cite{coldplay}. This is a spontaneous breaking from a pre-geometric Pati-Salam model to the almost-commutative geometry of the standard model, with two Higgs-like fields: scalar and vector.

We start with the definition of a \emph{twisted spectral triple} as a triple  together with an automorphism $\rho$ of $\A$ such that
\begin{equation}
  \label{eq:84}
  [{\cal D}_0, a]_\rho = {\cal D}_0a -\rho(a){\cal D}_0
\end{equation}
is bounded for any $a\in \A$.  All other conditions which do not involve commutators involving ${\cal D}_0$ and elements of the algebra are unchanged.
The question is whether to twist the commutator with ${\cal J} b^*
{\cal J}^{-1}$ as well. As explained in~\cite[Prop. 3.4]{Connes:1938fk}, the set $\Omega^1_D $ of twisted $1$-forms, namely
all the operators of the form
\begin{equation}
  \label{eq:167}
 \mathbb A_\rho = \sum_i b^i [{\cal D}_0, a_i]_\rho,
\end{equation}
is a $\A$-bimodule for the left and right  actions
\begin{equation}
  \label{eq:182}
  a\cdot\omega\cdot b := \rho(a) \omega b \quad \forall a,b\in\A,
  \omega\in \Omega^1_D
\end{equation}
therefore it is natural to twist the commutator  of the algebra with its commutant.

The bosonic fields are obtained by the \emph{twisted fluctuations of ${\cal D}_0$ by
  $\A$}, which amount to substituting ${\cal D}_0$  with~\cite{Devastato:2014bta}
\begin{equation}
{\cal D}_{{\mathbb A}_{\rho}}:={\cal D}_0+{\mathbb A}_{\rho}+\epsilon'\,{\cal J}{\mathbb A}_{\rho}{\cal J}^{-1}\label{eq:3lt}
\end{equation}
We emphasize that the fluctuation~\eqref{eq:3lt} has a different structure from the standard fluctuation~\eqref{fluctu}, consequently here we use a different notation, labelling the fluctuating Dirac operator~\eqref{eq:3lt} by ${\mathbb A}_{\rho}$, differently from~\eqref{fluctu}.

Furthermore we
assume that $\rho$ is a $*$-automorphism that commutes with the real
structure ${\cal J}$, which permits to define the twisted version of the
$1^\text{st}$-order condition as 
 \begin{equation}
   \label{eq:112}
  [ [{\cal D}_0, a]_\rho,\, {\cal J}b{\cal J}^{-1}]_{\rho_0} =    [{\cal D}_0, a]_\rho\, {\cal J}b{\cal J}^{-1} - {\cal J}
  \rho(b) {\cal J}^{-1}  [{\cal D}_0, a]_\rho =  0 \quad\quad \forall a, b\in \A.
 \end{equation}
where:
\begin{equation} 
\rho_{0}({\cal J}b{\cal J}^{-1}):={\cal J}\rho(b){\cal J}^{-1}. \label{deftwistedfirstorder}
\end{equation}

A gauge transformation for a twisted spectral
triple~ \cite{Landi:2017aa}  is implemented by
the simultaneous action on $\HH$ and ${\cal L}(\HH)$ (the space of linear operators on
$\HH$) of
the group of unitaries of $\A$,
\begin{equation}
  {\cal U}(\A):=\left\{ u\in\A, u^*u = uu^*=\mathbb I \right\}.
\end{equation}
The action on $\HH$ follows from 
the adjoint action of $\A$ (on
the left via its representation, on the right by
 $\psi a := a^\circ \psi = {\cal J}a^*{\cal J}^{-1}\psi$), i.e.
\begin{equation}
  \label{eq:24lt}
  \text{Ad}(u) \psi = u\psi u^* = u {\cal J}u{\cal J}^{-1}\psi \quad \forall
  \psi\in\HH,\, u\in {\cal U}(\A).
\end{equation}
The action on ${\cal L}(\HH)$ is defined as 
\begin{equation}
T\mapsto \text{Ad}(\rho(u)) \,T \, \text{Ad}(u^*) \qquad \forall T\in {\cal L}(\HH),
\label{eq:59lt}
\end{equation}
where
\begin{equation}
  \label{eq:34lt}
  \text{Ad}(\rho(u))= \rho(u) {\cal J}\rho(u){\cal J}^{-1}.
\end{equation}
 In particular, for $T={\cal D}_{{\mathbb A}_{\rho}}$ a twisted covariant Dirac operator
~\eqref{eq:3lt},
 one has~\cite{Landi:2017aa}
 \begin{equation}
    \label{eq:6lt}
    \text{Ad}(\rho(u))\, {\cal D}_{{\mathbb A}_{\rho}} \text{Ad}(u^*) = {\cal D}_{{\mathbb A}_{\rho}^u}
  \end{equation}
  where
  \begin{equation}
    \label{eq:7lt}
  {\mathbb A}_{\rho}^u := \rho(u){{\mathbb A}_{\rho}} u^{*}+\rho(u)\left[{\cal D},u^{*}\right]_{\rho}.
  \end{equation}
The map ${\mathbb A}_{\rho}\mapsto {\mathbb A}_{\rho}^u$
is a twisted version of the usual law of transformation of the gauge
potential in noncommutative geometry~\cite{Connes:1996fu}. 

As explained in section (\ref{gradingreduction}) the grading condition $[\Gamma,
A]=0$ imposes the reduction of $\A_G$ to
\begin{equation}
  \label{eq:62}
{\cal  B}_{LR} :=(\mathbb H^l_L \oplus \mathbb H^r_L \oplus \mathbb H^l_R \oplus \mathbb H^r_R) \oplus
M_8(\mathbb C). 
\end{equation}

For $A=(Q, M)\in \cinf\otimes{\cal B}_{LR}$,  the commutator{\footnote{\label{footnotenotations}To lighten notation, we omit the trivial indices in the
    product (hence in the commutators) of operators. From~\eqref{eq:16} one knows that $Q$ carries the indices
    $s\alpha$ while $\gamma^\mu_E$ carries $s\dot s$, hence
    $[\slashed \del, Q]$ carries indices $s\dot s\alpha$ and should be
    written $[\delta_\alpha^\beta\slashed\del,  \delta_{\dot s}^{\dot t} Q]$.
As well, $[\slashed \del, M]$ carries indices $s\dot s I$ and holds for $[\delta_I^J\slashed \del, \delta_{s}^{t} M]$.}}
\begin{equation}
\label{eq:41}
   [\slashed D\otimes \mathbb{I}_F, A] =\left(
\begin{array}{cc}
 \delta^I_J \,[\slashed\del,  Q]&0_{64} \\ 
 0_{64} &\delta_\alpha^\beta\, [\slashed \del, M]
\end{array}
\right)_{\partind{CD}}
 \end{equation}
is unbounded because in
\begin{equation}
[\ds, Q] = i\gamma^\mu_E (\nabla_\mu^S Q)  +i [\gamma^\mu_E, Q]\nabla_\mu^S \quad\text{ and }\quad
[\ds, M] = i\gamma^\mu_E (\nabla_\mu^S M) +i[\gamma^\mu_E, M]\nabla^S_\mu
\label{eq:137}
\end{equation}
$Q$ and $M$ do not commute with the Dirac'2 $\gamma$ matrices.

We will cure the unboundedness of $[\ds, Q]$ seeing the triple for the standard model as a twisted spectral triple. Unfortunately, however, the necessary twist does not reduce the complex sector, i.e.\ the condition 
\eqref{reduc} has to be imposed by hand. We will do this, for want of a better procedure.

Imposing~\eqref{reduc} on the grand algebra $\A_G$ reduced
by the grading to ${\cal B}_{LR}$ yields
\begin{equation}
  \label{eq:87}
  {\cal B}' := (\mathbb H^l_L \oplus \mathbb H^r_L \oplus \mathbb H^l_R \oplus \mathbb H^r_R) \oplus
M_4(\mathbb C).
\end{equation}
An element
$A=(Q,M)$ of $\cal B'$ is given by~\eqref{repa5} where $M$ in~\eqref{eq:16} is proportional to $\delta_{\dot s}^{\dot t}$ while $Q$ is a diagonal matrix  in  the $st$ indices:
\bea
Q&=& \left(
\begin{array}{cc}
 Q^{r}_{r} &  0_4\\
0_4 & Q^{l}_{l}
\end{array}\right)_{s t}
\nonumber\\  \label{eq:92}
  M&=& \delta_{\dot s}^{\dot t} M^\sI_\sJ \in M_4(\mathbb C).
\eea
The algebra $\cal B'$ contains the algebra of
the standard model $\A_{sm}$, and still has a part (the quaternion) 
that acts in a non-trivial way on the spin degrees of freedom.  Moreover $\cal B'$ is compatible with the twisted first-order condition in fact it is possible to show~\cite[Prop.~3.4]{Devastato:2014bta} that   $(\cinf\otimes{\cal B}', \HH, \slashed D\otimes \mathbb{I}_F, \rho)$ together with the grading $\Gamma$ and the real structure $J$ in is a graded twisted spectral triple which satisfies the twisted first-order condition of definition \ref{eq:112}.  Also~\cite[Prop.~3.4]{Devastato:2014bta} we have that, with Majorana masses, the subalgebra 
\begin{equation}
  \label{eq:96}
  {\cal B}:= \mathbb H^l_L \oplus \mathbb
  H^r_L\oplus \mathbb C^l_R \oplus \mathbb C_R^r \oplus  M_3(\mathbb C).
\end{equation}
satisfies the twisted first-order
condition
\begin{equation}
[[\slashed D\otimes{I}_F+D_{M_R}, A]_\rho, {\cal J}B{\cal J}^{-1}]_\rho = 0 
\label{eq:93}
\end{equation}

In conclusion, the twisted spectral triple
\begin{equation}
  (\cinf\otimes{\cal B},\,  L^2(\M) \otimes {\mathbb C}^{128},\,\slashed D\otimes{I}_F+ D_{M_R};\, \rho)
\label{eq:189}
  \end{equation}
solves the problem of the non-boundedness of the commutators $[\slashed D\otimes{I}_F, A]$ raised by the non-trivial action of the
 grand algebra on spinors.  The crucial point is that this twisted algebra still generates the field $\sigma$
  by a twisted fluctuation of $D_{M_R}$ as we see next.

\subsection{Twisted fluctuation and breaking to the Standard Model}
As explained in~\eqref{eq:3lt} we call
\emph{twisted fluctuation of ${\cal D}_0$ by $C^\infty(\M)\otimes\cal B$} the substitution of ${\cal D}_0$ with
   \begin{equation}
    \label{eq:143}
    {\cal D}_{\mathbb A_\rho} = {\cal D}_0 + \mathbb A_{\rho} + {\cal J}\, \mathbb A_\rho \, {\cal J}^{-1}
  \end{equation}
where $\mathbb A_\rho$ is twisted $1$-form
\begin{equation}
  \label{eq:125}
  \mathbb A_\rho = B^i [{\cal D}_0, A_i]_\rho \quad A_i, B^i\in \cinf\otimes\cal B.
\end{equation}
With diagonal $A_i=(Q_i,M_i)$ and $B^i=(R_i,N_i)$, as in~\eqref{eq:92}, $Q,R$ quaternionic matrices, $M,N$ complex valued matrices. We do not require $\mathbb A$ to be selfadjoint, we only ask that
${\cal D}_{\mathbb A_\rho}$ is selfadjoint and called it \emph{twisted-covariant Dirac operator}. It is the sum 
 ${\cal D}_{\mathbb A_\rho} = {\cal D}_X + {\cal D}_\sigma$ of the twisted-covariant free Dirac operator
\begin{equation}
  \label{eq:26}
  {\cal D}_X:= \Ds + \slashed {\mathbb A}_X + {\cal J} \slashed{\mathbb A}_X {\cal J}^{-1} \quad\quad \slashed{\mathbb A}_X :=
  B^i[\Ds, A_i]_\rho
\end{equation}
with the twisted-covariant Majorana-Dirac operator
\begin{equation}
  \label{eq:27}
  {\cal D}_\sigma := D_{M_R} + \mathbb A_{M_R} + {\cal J} \mathbb A_{M_R} {\cal J}^{-1} \quad\quad \mathbb A_{M_R}:= B^i [D_{M_R}, A_i]_\rho.
\end{equation}
In~\cite{Devastato:2014bta} ${\cal D}_X$ and ${\cal D}_\sigma$ are explicitly computed. They are parametrized by a vector field $X_\mu$ and a scalar
field $\sig$:
\begin{equation}
  \label{eq:187}
 {\cal D}_X+{\cal D}_\sigma =\slashed D - i \gamma^\mu_E \mathbb X_\mu+\sig
\end{equation}
with 
\be
{\mathbb X_\mu}:=\delta^I_J\,\rho(R^i)\,\nabla^S_\mu Q_i - \delta_\alpha^\beta
  \,\bar N^i \nabla^S_\mu \bar  M_i  \text{~~and~~} {\boldsymbol \sigma} := (\mathbb I +
\gamma^5_E \phi)
\ee
where $\phi$ is a complex scalar field given by elements of $\mathbb C^r_L$ and $\mathbb C^l_L$.
These field, and in particular ${\mathbb X}^\mu$, belong to the pre-geometric phase described by the grand symmetry, in fact they transform under the symmetry which mixes spacetime and gauge indices. Its physical consequences have not been yet investigated.

The next step is the calculation of the spectral action for the twisted-covariant
Dirac operator $ {\cal D}_{\mathbb A_\rho} = {\cal D}_X + {\cal D}_\sig$ whose
potential part is minimum when the Dirac operator $\slashed D\otimes\mathbb{I}_F \oplus
D_{M_R}$ of
the twisted spectral triple is fluctuated by a subalgebra of $ C^\infty(\M)\otimes {\cal B} $
which is invariant under the automorphism $\rho$. The maximal
such sub-algebra is precisely  the algebra $\cinf \otimes \A_{sm}$ of the standard model. 

Let us parametrize the action in term of the difference between twisted and untwisted fields as follows:
\begin{equation}
  \label{eq:171}
  \Delta(X)_\mu:= {\mathbb X_\mu} -\rho({\mathbb X_\mu}), \quad \Delta(\sig) := (\sig - \rho(\sig))D_R.
\end{equation} 
The square of the twisted-covariant Dirac operator is
\begin{equation}
\label{eq:47}
{\cal D}_{\mathbb A_\rho}^2 = -\left(\gamma^\mu_E \gamma^\nu_E \del_\mu\del_\nu + (\alpha_X^\mu+\alpha^\mu_{\sig}) \del_\mu
+\beta_X +\beta_{X\sig}+ \beta_{\sig}\right)
 \end{equation}
where
 \begin{eqnarray}\label{eq:alpha}
\alpha^{\mu}_X & := i\left\{\slashed{\mathbb X}, \gamma^\mu_E \right\} ,
\quad 
\beta_X = i\gamma^\mu_E (\del_\mu\slashed{\mathbb X}) -  \slashed{\mathbb X}\slashed{\mathbb X},
\end{eqnarray} 
while
\begin{equation}
  \label{eq:48}
\alpha^\mu_{\sig}:=  i
   \gamma^{\mu}_E\gamma^{5}_E \Delta(\sig),\quad\quad   \beta_{\sig} := -\sig^2 D_R^2,
\end{equation}
and
\begin{equation}
  \label{eq:153}
\beta_{X\sig}:= i\gamma^\mu_E \gamma^5_E \left(D_\mu (\sig D_R) + \Delta(\sig)\,\XX_\mu\right).
\end{equation}

In~\cite{Devastato:2014bta} the spectral action, i.e.\ the trace of ${\cal D}_{\mathbb A}^2$, is calculated and the main result is summarized in its potential part:
\begin{align}
V(X) &= \Lambda^2 f_2 \,\text{Tr}\, E_X^0 + \frac 12\, f_0
  \text{Tr}\,  (E_X^0)^2 \text{~~with~~} E_X^0 = \frac 12 \slashed\Delta^2(\XX) \\
 V(\sig) &= C_4 \,\phi^4   + C_2\,\phi^2 + C_0 \\
 V(X,\sig) &= \frac 12 f_0 \,\text{Tr}\,(E_{X\sig}^0)^2 +  f_0
    \,\text{Tr}\, E_X^0 E_\sig^0. 
\end{align}
with $E_{X\sig}^0= \frac 12 \gamma^5_E  \left[\slashed{\mathbb H}, \Delta(\sig)\right]$ and 
\begin{equation}
  E_X^0 E_\sig^0 = \frac 12 (3\phi^2 -1) \slashed\Delta^2(\XX) D_R^2 -
  \phi \,\slashed\Delta^2(\XX) \gamma^5_E D_R^2
\end{equation}
One obtains that the whole potential $V(X) + V(\sig) + V(X, \sig)$ is zero if and
only if both the scalar field $\sig$ and the vector field
$\Delta(X)_\mu$ are zero. Moreover an element $(Q,M)$ of $\cal B$ is invariant by
the automorphism $\rho$ if and only if
\begin{equation}
\rho(Q) = Q,
\label{eq:129}
\end{equation}
which means $\mathbb H_R^r = \mathbb H_R^l$ and $\mathbb C_L^r = \mathbb C_L^l$,
that is $(Q, M)\in \A_{sm}$.
 Therefore  the
potential part of the spectral action is minimum when the Dirac operator $\slashed D\otimes \mathbb{I}_F \oplus 
D_{M_R}$ of
the twisted spectral triple is fluctuated by a subalgebra of $ C^\infty(\M)\otimes {\cal B} $
which is invariant under the automorphism $\rho$ , i.e.\ precisely  the algebra $\cinf \otimes \A_{sm}$ of the standard
  model.
  
Summarising  the main results of grand symmetry model: starting with the enlarged algebra $\cal B$, one builds a twisted spectral triple whose fluctuations generate both an extra
scalar field $\boldsymbol \sigma$ and an additional vector field
${\mathbb X_\mu}u$. This is a Pati-Salam like model - the unitary of $\cal
B$ yields both an $SU(2)_R$ and an $SU(2)_L$, together with an extra $U(1)$ - but in a pre-geometric
phase since the Lorentz symmetry (in our case: the Euclidean $SO(n)$ symmetry) is
not explicit. The spectral action spontaneously breaks this model to
the standard model, in which the Lorentz symmetry is explicit, with the scalar and the vector fields playing
a role similar as the one of Higgs field. We thus have a dynamical model of emergent geometry.

\section{Beyond the Standard Model without the first order condition}

In this section we describe another extension of the inner fluctuations~\cite{CCVPati-Salam} based on spectral triples that do not fulfil the first-order condition,  involving the addition of a quadratic term to the usual linear terms. Without the restriction of the first-order condition, invariance under inner automorphisms requires the inner fluctuations of the Dirac operator to contain a quadratic piece expressed in terms of the linear part. This leads immediately to a Pati-Salam $SU(2)_{R} \times SU(2)_{L} \times SU(4)$ type model which unifies leptons and quarks in four colours. Besides the gauge fields, there are 16 fermions in the $(2, 1, 4)+(1,2,4)$ representation, fundamental Higgs fields in the (2, 2, 1), (2, 1, 4) and $(1,\ 1,\ 1+15)$ representations. Depending on the precise form of the initial Dirac operator there are additional Higgs fields which are either composite depending on the fundamental Higgs fields listed above, or are fundamental themselves. As in the case of the Grand Symmetry of the previous section, these additional fields break spontaneously the Pati-Salam symmetries at high energies to those of the Standard Model and make the model compatible with the experimental Higgs mass.


The first order condition is one of the conditions for a noncommutative geometry to be a manifold. Nevertheless,  in~\cite{CCvS} it was suggested that this condition can be violated, on phenomenological grounds. This has various consequences. The  Dirac operator~\{fluctu} with these generalised transformation fails to be gauge invariant in the usual sense. The authors of~\cite{CCvS} considered the more general, ``universal'', one form
so that the fluctuations of the Dirac operator have now three contributions. Two of them re the usual ones and satisfy the first order condition:
\begin{align}
{\mathbb A}_{(1)} & =\sum_{j}a_{j}[{\cal D}_0,\ b_{j}]\\
\tilde{{\mathbb A}}_{(1)} & =\sum_{j}\hat{a}_{j}[{\cal D}_0,\ \hat{b}_{j}] \ \text{with}\ \hat{a}_{i}={\cal J}a_{i}{\cal J}^{-1},\ \hat{b}_{i}={\cal J}b_{i}{\cal J}^{-1}
\end{align}
while the last one is absent if the first order condition is imposed, it is:
\be
{\mathbb A}_{(2)}  =\sum_{j}\hat{a}_{j}[{\mathbb A}_{(1)},\ \hat{b}_{j}]=\sum_{j,k}\hat{a}_{j}a_{k}[[{\cal D}_0,\ b_{k}],\ \hat{b}_{j}]
\ee
it is a \emph{nonlinear} correction in the terms of the algebra.
This defines a novel fluctuated operator
\be
{\cal D}_{\mathbb{A}}={\cal D}_0+{\mathbb A}_{(1)}+\tilde{{\mathbb A}}_{(1)}+{\mathbb A}_{(2)} \label{Dsecondorder}
\ee
We clarify that  here, as well as in~\eqref{eq:3lt}, the fluctuation~\eqref{Dsecondorder} has a different structure from both~\eqref{eq:3lt} and~\eqref{fluctu} therefore it is indicated with a different subscript notation, ${\cal D}_{\mathbb{A}}$.

The gauge transformations (for details see~\cite{CCvS}) are
\bea
{\mathbb A}_{(1)}&\mapsto& u{\mathbb A}_{(1)}u^{*}+u[{\cal D}_0,\ u^{*}]
\nonumber\\
{\mathbb A}_{(2)}&\mapsto& {\cal J}u{\cal J}^{-1}{\mathbb A}_{(2)}{\cal J}u^{*}{\cal J}^{-1}+{\cal J}u{\cal J}^{-1}[u[{\cal D}_0,\ u^{*}],\ {\cal J}u^{*}J^{-1}]
\eea
these transformations have to be made in order, i.e.\ the ${\mathbb A}_{(1)}$ of the second line is the one obtained from the the first line. The transformation of $\tilde{\mathbb A}_{(1)}$ is a straigthforward generalization.
In~\cite{CCvS} a semi-group structure, Pert($\mathcal{A}$), is given to the inner fluctuations. This means that 

an inner fluctuation of $D$ by $\displaystyle \sum_{j}a_{j}\otimes b_{j}^{\mathrm{o}}\in\mbox{Pert}(\mathcal{A})$, is now simply given by
$
{\cal D}_0\mapsto\sum_{j}a_{j}{\cal D}_0 b_{j}.
$

As explained after~\eqref{genericalgebra} the simplest and most general finite algebra satisfying all the conditions for a noncommutative space to be a manifold - except for the {\it order one condition} - equipped with a real structure and a grading  is the algebra $\mathcal A_{LR}=\mathbb{H}_{R}\oplus \mathbb{H}_{L}\oplus M_{4}(\mathbb{C})$ defined in~\eqref{bigalgebra}. 
The order one condition restricts the above algebra further to the subalgebra $\mathcal A_{sm}=
\mathbb{C}\oplus \mathbb{H}\oplus M_{3}(\mathbb{C})$ defined in~\eqref{smalgebra}.

The new Dirac operator ${\cal D}_{\mathbb{A}}$ of~\eqref{Dsecondorder} has now nonlinear contributions, but the Hilbert space is still the same, as is the algebra $\mathcal A_{LR}$. The consequence of the new term is a change of the representation of the Higgs, which now can be seen as a \emph{composite} particle, due to the presence of $\mathbb a_2$. Consider the representations of the group of unitaries of the algebra $\mathcal A_{LR}$, i.e.$
SU(2)_{R}\times SU(2)_{L}\times SU(4)
$
where $SU(4)$ represents the color group with the lepton number as the fourth color The Higgs fields appearing in $A_{(2)}$ are composite, they depend quadratically on those appearing in $A_{(1)}$ and are in the representations $(2_{R},\ 2_{L},\ 1)$ , $(2_{R},\ 1_{L},\ 4)$ and $(1_{R},\ 1_{L},\ 1+15)$ of $SU(2)_{R} \times SU(2)_{L} \times SU(4)$. Otherwise, there will be additional fundamental Higgs fields in the $(3_{R},\ 1_{L},\ 10)$ , $(1_{R},\ 1_{L},\ 6)$ and $(2_{R},\ 2_{L},\ 1+15)$ representations, but not the $(2_{R},\ 1_{L},\ 4)$.

The calculations of the spectral action are not dissimilar from the ones of the Grand symmetry, but it is convenient to split the Hilbert space in a different way. We will leave the details out and refer to~\cite{CCvS, CCVPati-Salam, Chamseddine:2015ata} for details and give the final result for the calculation of the spectral action: 
\begin{align}
S_{CCvS} & =\frac{24}{\pi^{2}}f_{0}\Lambda^{4}\int d^{4}x\sqrt{g}\nonumber \\
 & \,\,\,-\frac{2}{\pi^{2}}f_{2}\Lambda^{2}\int d^{4}x\sqrt{g}\left(R+\frac{1}{4}\left(H_{\mathrm{\dot{a}}I\dot{c}K}H^{\mathrm{\dot{c}}K\dot{a}I}+2\Sigma_{\dot{a}I}^{cK}\Sigma_{cK}^{\dot{a}I}\right)\right)\nonumber \\
 & {\displaystyle \,\,\,+\frac{1}{2\pi^{2}}f_{40}\int d^{4}x\sqrt{g}\left[-{\displaystyle \frac{3}{5}C_{\mu\nu\rho\sigma}^{2}+{\displaystyle \frac{11}{30}R^{*}R^{*}+g_{L}^{2}(W_{\mu\nu L}^{\alpha})^{2}+g_{R}^{2}(W_{\mu\nu R}^{\alpha})^{2}+g^{2}(V_{\mu\nu}^{m})^{2}}}\right.}\nonumber \\
 & \,\,\,+\nabla_{\mu}\Sigma{}_{aI}^{\dot{c}K}\nabla^{\mu}\Sigma{}_{\dot{c}K}^{aI}+\frac{1}{2}\nabla_{\mu}H_{\dot{a}I\dot{b}J}\nabla^{\mu}H^{\dot{\mathrm{a}}I\dot{b}J}+\frac{1}{12}R\left(H_{\dot{a}I\mathrm{\dot{c}}K}H^{\dot{c}K\dot{a}I}+2\Sigma{}_{\dot{a}I}^{cK}\Sigma{}_{cK}^{\dot{a}I}\right)\nonumber \\
 & \,\,\,\left.+\frac{1}{2}\left|H_{\dot{a}I\dot{c}K}H^{\mathrm{\dot{c}}K\dot{b}J}\right|^{2}+2H_{\dot{a}I\dot{c}K}{\displaystyle \Sigma_{bJ}^{\dot{c}K}H^{\dot{a}I\dot{d}L}\Sigma_{\dot{d}L}^{bJ}+\Sigma_{aI}^{\dot{c}K} \Sigma_{\dot{c}K}^{bJ}\Sigma_{bJ}^{\dot{d}L}\Sigma_{\dot{d}L}^{aI}}\right].\label{PatiSalamAction}
\end{align}
Apart from the usual terms already present in the other cases, the new terms are the composite fields $\Sigma_{\dot{a}I}^{bJ}$ and $H_{\dot{a}I\dot{b}J}$ which transform under the $(2_{R},\ 2_{L},\ 1+15)$ and $(3_{R},\ 1_{L},\ 10)+(1_{R},\ 1_{L},\ 6)$ representations of $ SU(2)_{R}\times SU(2)_{L}\times SU(4)$ . Moreover, for generic Dirac operator one also generates the fundamental field $(1_{R},\  3_L,\ 10)$.

It is possible to see that this model contains, as particular solution, the Standard Model, as coming froma  Pati-Salam model. It is the extra fields of this unified model which play the role of lowering the mass of the Higgs, making it compatible with the measure one. The classical approach to grand unified theories suffers the presence of an arbitrary and complicated Higgs representations. This problem is solved in the noncommutative spectral model by having minimal representations of the Higgs fields. Remarkably, a very close model to the one deduced here is the one considered by Marshak and Mohapatra~\cite{Mohapatra:1980qe} where the $U(1)$ of the left-right model is identified with the $B-L$ symmetry. In addition this type of model arises in the first phase of breaking of $SO(10)$ to $SU(2)_{R}\times SU(2)_{L}\times SU(4)$  (see for example~\cite{Slansky:1981yr, Altarelli:2013aqa}).

\section{Twist and Lorentz Structure}
\label{sec:twistlorentz}

In this section we connect the twist introduced in Sect.~\ref{section:twisting} with a Lorentz structure of the theory. 
We show that when the automorphism $\rho$ in a
twisted spectral triple $(\A, \HH, D; \rho)$ is inner,
then there exists a natural $\rho$-twisted inner product on
$\HH$. Furthermore, for the twisted geometry of the
Standard Model \ref{section:twisting}, this
inner-product is a Krein product of  Lorentzian spinors~\cite{Devastato:2017rlo}.  

The twisted fluctuations of the
Dirac operator, which were initially introduced in analogy with the non twisted case~\cite{Devastato:2014bta}, have been placed~\cite{Landi:2017aa} on the same rigorous footing as
Connes' original ``fluctuations of the metric''~\cite{Connes:1996fu},
namely as a way to export a real twisted spec\-tral triple to a Morita
equivalent algebra. In particular, in  case of self-Morita equivalence,
one obtains formula~\eqref{eq:3lt}.
Additionally, a gauge
transformation is implemented as in the non-twisted case, namely as a
change of connection in the bimodule that implements Morita
equivalence. This yields formula~\eqref{eq:6lt}, which is our starting point in this section.

There is an important difference between the twisted and the
non-twisted cases: while usual fluctuations preserve the
selfadjointness of the Dirac operator, twisted-fluctuations may not. In~\cite{Landi:2017aa} this issue was addressed working out the necessary and sufficient
conditions, such that the unitary $u$ which implements the twisting automorphism (in
case the latter lifts to an inner automorphism of ${\cal B}(\HH)$) must satisfy in order to preserve selfadjointness.  Interestingly, there are other solutions beyond the obvious ones 
(i.e.\  $u$ invariant under the twist). 

In this section we provide an alternative solution: instead of trying
to preserve selfadjointness, we investigate whether there is a ``more natural''
property  preserved under a twisted
fluctuation. We find one: selfadjointness with
respect to the inner product induced by the twist. Unexpectedly, in
the
case of the twisted spectral triple of the Standard Model, the
induced product is the
Krein product of Lorentz spinors.
The Lorentz
structure emerges, \emph{uniquely} (if fermions are untouched by the twist) from the algebraic properties of the twisted Euclidean spectral triple
~\cite{Landi:2017aa}).

\subsection{Twisted inner product}
\label{subsec:twistinner}
The first step is the definition of an inner product which takes the twist $\rho$ defined in Sect.~\ref{section:twisting} into account. This is provided by:
\begin{df}
\label{de:twist}
A  $\rho$-twisted inner product, in short a $\rho$-product,
  $\langle\cdot ,\, \cdot \rangle_\rho$ is an inner product on $\HH$ such
  that
   \begin{equation}
    \langle\Psi,\mathcal{O}\Phi\rangle_{\rho}=\langle\rho(\mathcal{O})^\dag\Psi,\Phi\rangle_{\rho}\qquad
    \forall {\cal O}\in{\cal B}(\HH),\; \Psi,\,\Phi\in\HH,
    \label{inner product}
  \end{equation}
  where $\rho({\cal
    O})^\dag$ is the adjoint of $\rho({\cal
    O})$ with respect to the initial Hilbert inner product
  $\langle\cdot,\cdot  \rangle$.
\end{df}
\noindent  We denote 
\begin{equation}
{\cal O}^+:= \rho({\cal O})^\dag
\label{rhoadjoint}
\end{equation}
the  adjoint of a bounded operator ${\cal O}$ with respect to the $\rho$-twisted
inner product.We call ${\cal
  O}^+$ the $\rho$-adjoint. Analogously we define $\rho$-hermitean ($\mathcal O=\rho(\mathcal O)^+$) and $\rho$-unitary ($U\rho(U)^\dagger=\rho(U)^\dagger U=\mathbb 1 $) operators.

A particular class of twists is the one implemented by a unitary operator $R$:
\begin{equation}
  \rho({\cal O}) = R{\cal O} R^\dag \qquad \forall {\cal O}\in{\cal B}(\HH),
\end{equation}
It can be checked\cite{Devastato:2017rlo} that this defines a good $\rho$-product:
\begin{equation}
\langle\Psi,\Phi\rangle_\rho=\langle\Psi, R\Phi\rangle=\langle R^\dag\Psi, \Phi\rangle.
\label{eq:28lt}
\end{equation}
With this product $R$ is both unitary (by definition) and $\rho$-unitary. The $\rho$-product defined in~\ref{de:twist} need not be positive definite, but the unitarity of $R$ ensures that in this case it is not degenerate.

The extension of an inner automorphism $a\to uau^*$ of $\A$ to an automorphism of ${\cal B}(\HH)$ is not unique (just consider
two distinct unitaries $R_1$, $R_2$ in ${\cal B}(\HH)$ such that $R_1
a R_1^\dag = R_2 a
R_2^\dag$ for any $a\in \A$). Any such extension defines 
an automorphism of $\A_\circ$: 
\begin{equation}
  \label{automorphismA0}
  \rho({\cal J} a^*{\cal J}^{-1}) = R\, {\cal J} a^* {\cal J}^{-1} R^\dag\quad \forall a\in\A.
\end{equation}
We say that an inner automorphism is compatible with
  $\cal J$ if it admits an extension such that~\eqref{automorphismA0}
  agrees with $\rho_\circ\in\text{Aut}(\A_\circ)$ defined in
  \ref{deftwistedfirstorder}. 
  More precisely:
\begin{df}   
\label{def:compreal}
 Given a real spectral triple $(\A, \HH, D)$,  an inner automorphism $\rho$ of $\A$ is compatible with
  the real structure ${\cal J}$ if there exists a unitary $R\in\cal B(\HH)$
  such that
  \begin{align}
\rho(a) = RaR^\dag \quad \text{ and }\quad
    \label{eq:39bis}
     {\cal J}\,R\, a^*\,R^\dag {\cal J}^{-1} = R {\cal J} a^* {\cal J}^{-1} R^\dag \qquad \forall a\in\A.
  \end{align}
\end{df}

\noindent This condition  is verified in particular when the inner
automorphism can be implemented by a unitary $R$ such that

\begin{equation}
  \label{eq:40lt}
  {\cal J} R = \pm  R {\cal J} .
\end{equation}

\subsection{Lorentzian signature and Krein space} 
\label{subsec:lorentzkrein}

If $R$ is selfadjoint (and not the
identity), it will have eigenvalues $\pm 1$, and it splits $\HH$ in two eigenspaces
$\HH_+, \HH_-$. The $\rho$-product  is positive (resp.\  negative) definite on $\HH_+$ (resp.\ $\HH_-$.  this is  a \emph{Krein space}\footnote{An excellent introduction to Krein spaces in the context of NCG and particle physics is~\cite{Bizi:2018psj}. This reference also contain a very good bibliography if one wishes to further study the mathematical aspects.}. For this $R$ is a fundamental symmetry, i.e.\  it satisfies
$R^2=\I$ and the inner product $\langle\cdot , R\cdot\rangle_\rho$
is positive definite  on $\HH$ (in our case, this is simply the
Hilbert product one started with).

For the model of Sect.~\ref{section:twisting} turn out that the flip $R$ is one of the $\gamma$ matrices: $\gamma^0_E$ which in the usual basis is the same for the Euclidean and Lorentzian cases:
\begin{equation}
R=\begin{pmatrix} 0 &\mathbb 1_2\\
\mathbb 1_2 & \mathbb 0 \end{pmatrix} =\gamma^0_E=\gamma^0_M
\end{equation}
With this twist $\HH$ has become a Krein space.The $\rho$-product~\eqref{inner product} is now the usual inner product
of quantum field theory in Lorentz signature, where instead of
$\psi^\dagger$ it appears $\bar\psi=\psi^\dagger \gamma_E^0$:
\be \langle\psi,\phi\rangle_\rho=\langle \psi, \gamma_E^0\phi\rangle
= \int d^{4}x\psi^{\dagger}\gamma_E^{0}\phi:=\int
d^{4}x\bar\psi\phi.
\label{eq:Minkosky} \ee
These statements are basis independent, che choice of which $\gamma$ we chose for the flip is tantamount to the choice of one of Euclidean direction to be ``time''. It now makes sense to define the integral on a
 time slice and have fields normalized only for the space integral, which is what is commonly done. 
However, the $\rho(\gamma^i_E)$'s are
  not the Lorentzian signature (i.e.\ Minkowskian) gamma matrices,
  \begin{equation}
    \label{eq:63lt}
    \gamma^{0}_M =  \gamma^{0}_E\;,\quad \gamma_{M}^{j}= i\gamma_{E}^{j}
    \quad j=1,2,3.
  \end{equation}
 In Sect.~\ref{se:Wicketc} we considered the effect of change of the change of signature (and Wick rotation) for the $\gamma$'s 
$W:\gamma^\mu_E\to\gamma^\mu_M$, that is
\begin{equation}
 W(\gamma^0_E)=\gamma^0_E,\quad
 W(\gamma^j_E)=i\gamma^j_E,
 \end{equation}
one has that the twist is in some sense the
 square of the Wick rotation
 \begin{equation}
   \rho(\gamma^0_E)= W(W(\gamma^0_E)),\quad  \rho(\gamma^j_E)= W(W(\gamma^j_E)).
 \end{equation}

The Euclidean Dirac matrices are selfadjoint for the Hilbert product
of $\mathrm{sp}(\mathcal{M})$, but (except for $\gamma^0_E$) not $\rho$-hermitian since from
\eqref{rhoadjoint} one has
\begin{equation}
  (\gamma^j_E)^+ = \rho(\gamma^j_E)^\dag= -{\gamma^j_E}^\dag. \label{gammaEM}
\end{equation}
On the contrary, the Minkowskian gamma matrices (except $\gamma^0_M$)  are not selfadjoint for
the Hilbert product since~\eqref{eq:63lt} yields 
$(\gamma^j_M)^\dag = - \gamma^j_M$; but they are $\rho$-hermitian since
\begin{equation}
  \rho(\gamma^j_M) = i\rho(\gamma^j_E) = -i\gamma^j_E= -\gamma^j_M,
\end{equation}
so that
\begin{align}
(\gamma_M^j)^+= \rho(\gamma^j_M)^\dag  &=(\gamma_0 \gamma_M^j \gamma_0)^\dag
 =\gamma_0 (\gamma_M^j)^\dag\gamma_0 = -\gamma_0 \gamma_M^j\gamma_0 =-\rho(\gamma_M^j)=\gamma^j_M.
\label{eq:68lt}
  \end{align}
The ``temporal'' gamma matrix $\gamma^0:=\gamma^0_E=\gamma^0_M$ is both
selfadjoint and $\rho$-hermitian.
 
The twist naturally defines a Krein structure, while maintaining in
     the background the Euclidean structure.  Applications of
     Krein spaces to noncommutative
 geometry framework  have been recently
    studied in\cite{VanDenDungen,Brouder:2015qoa} as well as in~\cite{Franco:2012fk,Franco:2014fk,Franco:2012fkbis} (see reference
    therein for earlier attempts of adapting Connes noncommutative
    geometry to the Minkowskian signature). In~\cite{PaschkeSitarz, Verch,  Bochniak:2018ucd, Bochniak:2019aik} mathematical discussions on Lorentzian triples is present.

\subsection{Fermionic Action}
\label{subsec:fermact}

For a twisted spectral triple, both fermionic~~\eqref{SFE} and bosonic actions~\eqref{BSAdef} are well defined, but their invariance under a gauge
 transformation, i.e.\ 
\begin{equation}
  \label{eq:62lt}
  {\cal D}_{\mathbb{A}_\rho}\mapsto {\cal D}_{\mathbb{A}_\rho^u}:=\text{Ad}(\rho(u)) \, {\cal D}_{\mathbb{A}_\rho}\, \text{Ad}(u^*) \quad \text{ and } \quad  
\psi\longmapsto  \text{Ad}(u)\psi,
\end{equation}
is not ensured. On the one side $S^F( {\cal D}_{\mathbb{A}_\rho^u})$ is invariant for $u=\rho(u)$. 
On the other $S_B( {\cal D}_{\mathbb{A}_\rho^u})$ is defined only if 
    $ {\cal D}_{\mathbb{A}_\rho^u}$ is selfadjoint, or at least normal. But for an arbitrary unitary $u$, the operator $ {\cal D}_{\mathbb{A}_\rho^u}$ need not be selfadjoint. But it still has compact resolvent. Therefore if the operator is at least normal, the trace in~\eqref{BSAdef} is finite for any value of the cutoff
    $\Lambda$ and the bosonic action $S^B({\cal D}_{\mathbb{A}_\rho^u})$ is then well defined. 

In this section we will modify the fermionic action to make it invariant under~\eqref{eq:62lt} for \emph{any} unitary~$u$. The bosonic action will be considered in the next section. We will consider a Lorentzian twisted fermionic variant of the fermionic action of Sect.~\ref{se:Wicketc}, another variation based on the Euclidean version can be found in~\cite{Devastato:2017rlo}. We choose the former because it is related to Krein spaces.
\be
\label{kreinfermionaction}
S_{F_\rho}=\langle {\cal J} \psi,
    {\cal D}^M_{\mathbb{A}_\rho}\psi\rangle_\rho
\ee
 where ${\cal D}^M_{\mathbb{A}_\rho}$ is a
    $\rho$-hermitian twisted fluctuation of the Minkowskian operator~ \eqref{eq:78lt}.
The Lorentzian action has been considered in~\cite{Barrett,VanDenDungen}.

Our prescription is based on the substitution of the inner product in~\eqref{SFL} with the $\rho$-product. This ensures invariance under~\eqref{eq:62lt}.
The Krein structure induced by
the twist suggests a way to define a fermionic
action which is antisymmetric on the whole of $\HH$: by assuming that $\cal D$ is $\rho$-hermitian. We will not discuss here the construction of Krein or twisted spectral triples, which is performed for example in~\cite{VanDenDungen, Brouder:2015qoa, Bizi:2018psj}, it will suffice for us to note that if
 $(\A, \HH, {\cal D};\rho)$ is a real twisted spectral triple with
  $\rho$ an inner automorphism of ${\cal B}(\HH)$ compatible with the
  real structure in the sense of Def.~\ref{def:compreal}, then 
$\langle {\cal J} \psi, {\cal D}\phi\rangle_\rho\quad
  \forall \psi, \phi\in\text{Dom} \,{\cal D}$
is a bilinear form invariant under the simultaneous transformations
of~ \eqref{eq:62lt}. In other words
 \begin{equation}
\langle{\cal J}\psi, {\cal D}\phi\rangle_\rho=  \langle {\cal J}\text{Ad}(\rho(u))\psi, \text{Ad}(\rho(u)){\cal D}\text{Ad}(\rho(u^*)) \text{Ad}(\rho(u))\phi\rangle_\rho   \quad
  \forall \psi, \phi\in\text{Dom} \,{\cal D},\; u\in{\cal U}(\A). \label{vecchiaprop}
\end{equation}

 As an illustration, consider the twisted spectral triple of the
Standard Model~\eqref{almostspectraltriple} with $\ds$, $\cal J$ and $\gamma_E$
substituted with their  Lorentzian version,
\begin{equation}
\ds_M  := i\gamma^\mu_M\partial_\mu,\quad  J_M := -i\gamma^2_M \, cc,\quad   \gamma_M:=\gamma^0_M\gamma^1_M\gamma^2_M\gamma^3_M= i^3
  \gamma^0_M\gamma^1_E\gamma^2_E\gamma^3_E = -i\gamma_E,
\label{eq:23lt}
\end{equation}
where $\gamma_M^\mu$
are the Minkowskian Dirac matrices~\eqref{eq:63lt}. 

The Minkowskian Dirac operator
\be
\label{eq:78lt}
{\cal D}^M_0:=   \ds_M\otimes \I_{96} + \gamma_M\otimes D_F
\ee
 is $\rho$-hermitian since
  \begin{align*}
    ({\cal D}^M_0)^+ = \rho(({\cal D}^M_0)^\dag) &= \gamma^0 (\ds_M)^\dag\gamma^0 \otimes
    \I_{96} + \gamma^0_M\gamma_M^\dag\gamma^0_M\otimes D_F,\\
&=-i\gamma^0_M(\gamma^\mu_M)^\dag \gamma^0_M \partial_\mu +
  \gamma_M\otimes D_F= -i\gamma^\mu_M\partial_\mu +
  \gamma_M\otimes D_F={\cal D}^M_0,
  \end{align*}
where we used
$\ds_M^\dag=i(\gamma^\mu_M)^\dag\partial_\mu$ and~\eqref{gammaEM}.\\
\\
Moreover, it is possible to show the antisymmetry of the fermionic action~\eqref{kreinfermionaction} :
\begin{equation}
\label{antisymmetry}
  \langle {\cal J}_M\psi,
    {\cal D}^M_{\mathbb{A}_\rho}\psi\rangle_\rho = -  \langle {\cal D}^M_{\mathbb{A}_\rho}\psi,{\cal J}_M\psi\rangle_\rho
\end{equation}
using the fact that
\begin{equation}
    \overline{\gamma^0_M} =\gamma^0_M,\quad   \overline{\gamma^1_M} =\gamma^1_M,\quad   \overline{\gamma^2_M} =-\gamma^2_M,\quad   \overline{\gamma^3_M} =\gamma^3_M,
 \end{equation}
so that on a  Lorentzian four dimensional manifold, the real structure satisfies
\begin{align*}
(J_M)^2& = (-i
 \gamma_M^2\,cc)^2=\gamma_M^2\overline{\gamma_M^2}=
 -(\gamma^2_M)^2=\I;\\
  J_M \ds_M &= \ds_M{J_M}\quad \text{ for }\quad J_M \ds_M - \ds_M{J_M}= \left(\gamma^\mu_M\gamma^2_M +
    \gamma^2_M\overline{\gamma^\mu_M}\right)\partial_\mu \, cc=0;\\
J_M \gamma_M & =-i\gamma_M^2(\overline{\gamma^0_M}\,\overline{\gamma^1_M}\,\overline{\gamma^2_M}\,\overline{\gamma^3_M})cc
 = i\gamma_M^2(\gamma^0_M\gamma^1_M\gamma^2_M\gamma^3_M)cc  =-i
  (\gamma^0_M\gamma^1_M\gamma^2_M\gamma^3_M) \gamma_M^2 cc\\[4pt] &= \gamma_M J_M.
\end{align*}
Since $J_F^2=\I$, the first equation yields $(J_M\otimes J_F)^2=\I$. The second 
and third equations, together with $D_FJ_F=J_F D_F$ (coming from the
$KO$ dimension $6$) yield
\begin{align}
\nonumber
  {\cal D}^M_0 (J_M\otimes J_F) &= \ds_M J_M \otimes J_F +
  \gamma_M{J}_M\otimes D_F J_F = {J}_M \ds_M\otimes J_F +
  J_M\gamma_M\otimes J_FD_F \\ &=  (J_M\otimes J_F) {\cal D}^M_0,
\end{align}
 Finally one has $\gamma^0 J_M =
 -J_M\gamma^0$, showing that~ \eqref{kreinfermionaction} is antisymmetric as expected.

The gauge invariance in Prop.~\eqref{vecchiaprop} does not
rely on the selfadjointness of the Dirac operator, and thus is still valid for
$\rho$-hermitean Dirac operator. What must be checked, however, for
\eqref{antisymmetry} to make sense is that a twisted perturbation
of a $\rho$-hermitean operator is still $\rho$-hermitean, and that
this property is preserved under gauge transformation. A direct calculation~\cite[Prop.~4.4]{Devastato:2017rlo} shows this.

\subsection{Bosonic action}
\label{subsec:bosonact}

For the usual (hermitean) Dirac operator the bosonic action~\eqref{BSAdef} is trivially invariant for the transformation ${\cal D}\to U{\cal D}U^\dagger$, and it has been computed in~~\cite{Devastato:2014bta} for a selfadjoint twisted fluctuation
${\cal D}_{\mathbb{A}_\rho}$ of the Dirac operator of the Standard Model. 
The substitution of ${\cal D}^\dagger$ with the Krein adjoint ${\cal D}^+$ may seem natural, but it quickly runs into problems because the heath kernel technique, by its nature, works only for elliptic operators, while  ${\cal D}^+{\cal D}$ is hyperbolic operator.
Considers instead the following $\rho$-hermitian Dirac operator:
\begin{equation}
{ D}={D}^+ = \rho({D})^\dag,
\end{equation}
it has compact resolvent and it is possible to write~\eqref{BSAdef} in a twisted form  as
\begin{equation}
\text{Tr} \,\chi\left(\frac{\rho({D}) {D}}{\Lambda^2}\right).
\end{equation}
This has no reference to Hilbert adjointness. Taking for $D$ the $\rho$-Hermitian Minkowskian Dirac
operator $i\gamma^\mu_M\partial_\mu$ (which has locally
  compact resolvent, see~\cite [Prop.\ 4.2]{DungenRennie} and reference therein), it returns the Euclidean
action: by cyclicity of the trace, one
can substitute $\rho(D) D$ with $\frac 12\left(\rho(D) D +D\rho(D)\right)$, which is nothing but the Euclidean Laplacian  (up to a sign):
\begin{align}
\frac{1}{2}\left(\rho(D)D+D\rho(D)\right) & =\frac{1}{2}\left(i\gamma^{\mu\dagger}_M\partial_{\mu}i\gamma^{\nu}_M\partial_{\nu}+i\gamma^{\mu}_M\partial_{\mu}i\gamma^{\nu\dagger}_M\partial_{\nu}\right)\\
 & =-\frac{1}{2}\left(\gamma^{\mu\dagger}_M\gamma^{\nu}_M\partial_{\mu}\partial_{\nu}+\gamma^{\mu}_M\gamma^{\nu\dagger}_M\partial_{\mu}\partial_{\nu}\right)\\
 & =-\frac{1}{2}\left(\gamma^{\mu\dagger}_M\gamma^{\nu}_M+\gamma^{\mu}_M\gamma^{\nu\dagger}_M\right)\partial_{\mu}\partial_{\nu}\\
 & =-g_{E}^{\mu\nu}\partial_{\mu}\partial_{\nu}
\end{align}
where $g_E$ is the Euclidean metric.\\
\\
The modifications of the bosonic spectral action that we have shown here do not yield the bosonic action in a Lorentzian signature, which is a well-known and difficult problem. However, twists could shed a new light on the problem, suggesting the traditional approach of quantum field theory:  to start with a Lorentzian signature, for which a twist
is adapted, then to perform a Wick rotation yielding the Einstein-Hilbert action in Euclidean signature and Wick rotating back to the Lorentzian physical model. The added value of the twist should be thus to prescribe a geometry upon which to “Wick rotate back”.

\section{Conclusions and Outlook}

There are several direction in which this research must go. First  there is the comparison with experiments. Unlike other theories based on sophisticated mathematics, this approach reaches physically testable predictions immediately, in some sense it has its roots in an analysis of the standard model. There are however many points to solve, we cited already the Euclidean-Lorentz problematic.
The present RG analysis of the spectral action is carried out with the one-loop precision, but if one wants to calculate more precise numbers it will be necessary to go beyond the one-loop approximation.

What is central in the NCG approach to the particle physics is the spectral aspect of the action, but there is still much to be understood for the expansion and its physical role.
As we have said in Sec.~\ref{BLMapprox}, resuming the heat kernel expansion one can capture the UV behaviour~\eqref{HIGH} of the spectral action.
It is clear, that this ultraviolet behaviour 
 leads to nonrenormalizable QFT  and some UV
completion is needed, nevertheless nothing forbids to impose this UV completion at the energy
scale $\tilde \Lambda$ which is greater than $\Lambda\sim 10^{16}$ GeV itself, e.g. $\tilde\Lambda\sim M_{Pl}\sim 10^{19}$ GeV.
In such a setup there appears an energy region between $\Lambda$ and $\tilde\Lambda$, where the behaviour of bosons
is governed by Eq.~\eqref{HIGH}, see the discussion in \cite{Vassilevich:2015fwa}. It is clear, that the asymptotic~\eqref{HIGH} will have nontrivial physical consequences for example
 Coulomb's electrostatic law will change at short distances; it is also obvious that Eq.~\eqref{HIGH} will also have nontrivial
 thermodynamical consequences, which may be relevant for a dynamics of the early Universe.
Therefore we find it very interesting to figure out physical consequences of this high momenta regime.

Another important point is the further study of the possible pre-geometric phase hinted by the Grand Symmetry. This is pointing to a genuine noncommutative space, not just the product of ordinary spacetime times a finite noncommutative algebra. It is likely that going beyond the standard model will require a new vision of spacetime. This is pointing in the direction of quantum gravity. The scales are not too different. Unification of the three gauge interactions suggests a scale of the order of $10^{16}$~GeV, while the common wisdom indicates the onset of quantum gravitational effects at a scale of $10^{19}$~GeV. This does take into account possible numerical factors, $16\pi^2$ is more than two orders of magnitude! Moreover, the onset of ``new physics'' will alter the running, and possibly alter the unification and quantum gravity scales. A common scale may be the scale at which a quantum spacetime has to be considered. For such an object the tool of noncommutative geometry, applied to the standard model in this review, may be the most appropriate.

An alternative definition of the spectral action via the zeta function regularization~\cite{zeta}
also opens interesting perspectives.
In contrast to the cutoff based spectral action, this formulation does not exploit the cutoff scale $\Lambda$, therefore it is natural to raise a question, whether a scale-invariant formulation of the spectral action exists. In such a formulation all the physical scales are supposed to be generated dynamically.  The scale invariant extensions of the standard model received significant attention in particle physics~\cite{AlexanderNunneley:2010nw,Ferreira:2018itt,Hill:2018qcb} and  cosmology~\cite{Ferreira:2018qss,Vicentini:2019etr}.  Modern theoretical physics offers various mechanisms of the dynamical scale generation in both scalar~\cite{Coleman:1973jx,Jones:2017ejm,Kurkov:2016zpd} and gravitational~\cite{Adler:1982ri,Sakharov:1967pk,Visser:2002ew} sectors. 

It would be very interesting to go ahead in this direction and to construct a scale-invariant and scalar-less formulation of the spectral NCG, where not just scales, but the scalar fields are generated dynamically from fermions, like it happens in the technicolor scenario. Such a formulation would be very interesting from the gravitational point of view. Indeed, in the zeta-function-based spectral Lagrangian the only allowed gravitational term
would be the Weyl tensor square, what is nothing
but a classical Lagrangian of conformal gravity\footnote{In general, the zeta-based  scale-invariant spectral Lagrangian may contain $\phi^2 R$ and $\phi^4$ terms, where $\phi$ is a scalar field. A presence of the scalar curvature $R$ breaks the PT symmetry, what prohibits to apply the formalism of~\cite{Mannheim:2006rd,Bender:2007wu}.}.  This higher derivative gravitational model is renormalizable by power counting, 
and, what is also important, PT-symmetric. Generally speaking the higher-derivative theories are non unitary,
however, a presence of the PT-symmetry allows to quantise it in a unitary way~\cite{Mannheim:2011ds}, using the PT-symmetry Bender-Mannheim formalism~\cite{Mannheim:2006rd,Bender:2007wu}, see also~\cite{Andrianov:2016ffj, Novikov:2017dsl, Novikov:2018ybq, Novikov:2019ntb} for the recent progress PT-symmetric studies in particular in the context of the noncommutative geometry~\cite{Novikov:2019}.
We emphasize, that once one succeeds to build such a model, it will be UV complete, and then one can study quantum gravitational
corrections in a systematic way.

\subsection*{Acknowledgments}
We thank U.~Aydemir, L.~Boyle, S.~Farnsworth, A.~Sitarz and D.~Vassilievich for useful suggestions and comments on the draft.
FL would like to thank Thierry Masson, Roland Tray, George Zoupanos and Biswajit Chakraborty for organising conferences and series of lectures in Vietnam, Greece and India.
This work has been supported by the INFN Iniziativa Specifica GeoSymQFT and Spanish
MINECO under project MDM-2014-0369 of ICCUB (Unidad de Excelencia `Maria de Maeztu').


\end{document}